\documentclass[aps,prl, twocolumn,amsmath,amssymb,longbibliography,superscriptaddress]{revtex4-1}
\usepackage{amsmath,amssymb,amsfonts,bm}
\usepackage{graphicx}
\usepackage{xcolor}
\usepackage{bbold}
\usepackage{graphicx}
\usepackage{dcolumn}
\usepackage{epstopdf}
\usepackage{epsfig}
\usepackage{url}
\usepackage{hyperref}
\usepackage{verbatim}
\usepackage[export]{adjustbox}
\usepackage{scalerel}
\usepackage{braket}
\usepackage[normalem]{ulem}
\usepackage{cancel}

\usepackage{booktabs}

\usepackage{svg}

\usepackage{lipsum}
\usepackage{enumitem}

\usepackage{float}

\usepackage{IEEEtrantools}

\newcommand{\kcgin}{K_{\mathrm{c}, \mathrm{Gin}}}

\newcommand{\1}{\mathrm{I}}
\newcommand{\be}{\begin{equation}}
\newcommand{\ee}{\end{equation}}
\newcommand{\ba}{\begin{aligned}}
\newcommand{\ea}{\end{aligned}}

\newcommand{\tauplateau}{\tau_{\mathrm{plat}}}

\newcommand{\fk}{\tilde{K}}
\newcommand{\fkc}{\tilde{K}_{\mathrm{c}}}

\newcommand{\HSYK}{H_{\mathrm{SYK}}}

\newcommand{\rhor}{\rho}

\newcommand{\tauhei}{\tau_{\mathrm{Hei}}}
\newcommand{\tauth}{\tau_{\mathrm{Th}}}
\newcommand{\tauedge}{\tau_{\mathrm{edge}}}

\newcommand{\TCUE}{\mathrm{TCUE}}

\newcommand{\red}[1]{{\color{red} #1}}
\newcommand{\brown}[1]{{\color{brown}#1}}

\newcommand{\syan}[1]{{\color{cyan} #1}}

\newcommand{\Kdc}{K_{\mathrm{dc}}}
\newcommand{\Kcon}{K_{\mathrm{c}}}
\newcommand{\Kconf}{\tilde{K}_{\mathrm{c}}}

\newcommand{\ff}{f}
\newcommand{\valpha}{\vec{\alpha}}

\newcommand{\ree}{\mathrm{Re}}
\newcommand{\imm}{\mathrm{Im}} 

\usepackage{amsfonts}
\usepackage{booktabs}
\usepackage{siunitx}

\usepackage{multirow}
\usepackage{booktabs}
\usepackage{pifont}

\usepackage{colortbl}
\usepackage[toc,page,header]{appendix}

\newcommand{\otimess}{\otimes_{\mathrm{s}}}

\newcommand{\id}{\mathbb{1}}

\newcommand{\Phirkc}{\Phi_{\mathrm{1D}}}
\newcommand{\Phirko}{\Phi_{\mathrm{0D}}}

\newcommand{\GUE}{\mathrm{GUE}}
\newcommand{\Gin}{\mathrm{GinUE}}

\newcommand{\taudev}{\tau_\mathrm{dev}}

\newcommand{\talpha }{{\tilde{\alpha}}}

\newcommand{\rhofluc}{\rho_{\mathrm{fluc}}}
\newcommand{\rhoav}{\rho_{\mathrm{av}}}

\newcommand{\DeltaD}{\Delta_{\mathrm{D}}}
\newcommand{\Deltahei}{\Delta_{\mathrm{Hei}}}
\newcommand{\Deltath}{\Delta_{\mathrm{Th}}}

\newcommand{\mS}{\mathcal{S}}

\newcommand{\titleinfo}{Spectral form factor in chaotic, localized, and integrable \\
open quantum many-body systems}

\graphicspath{{Figures/}}
\graphicspath{{Figures/Stephen/}}
\graphicspath{{Figures/Stephen/Final_Figs/}}

\begin{document}


\title{Spectral form factor in chaotic, localized, and integrable \\
open quantum many-body systems}
\author{Jiachen Li$^*$
}
\affiliation{Department  of  Physics,  Princeton  University,  Princeton,  NJ  08544,  USA}

\author{Stephen Yan$^*$}
\affiliation{Department of Physics, University of California, Santa Barbara, CA 93106, USA}
\thanks{J.L. and S.Y. contributed equally to this work.}
\affiliation{Department  of  Physics,  Princeton  University,  Princeton,  NJ  08544,  USA}
\author{Toma\v{z} Prosen}
\affiliation{Department of Physics, Faculty of Mathematics and Physics, University of Ljubljana, Jadranska 19, SI-1000 Ljubljana, Slovenia}
\affiliation{Institute of Mathematics, Physics and Mechanics, Jadranska 19, SI-1000, Ljubljana, Slovenia}
\author{Amos Chan}
\affiliation{Physics Department, Lancaster University, Lancaster
LA1 4YB, United Kingdom}
\affiliation{Princeton Center for Theoretical Science, Princeton University, Princeton, NJ 08544, USA}
\date{\today}

\begin{abstract}
We numerically study the spectral statistics of open quantum many-body systems (OQMBS)  as signatures of quantum chaos (or the lack thereof), using the dissipative spectral form factor (DSFF), a generalization of the spectral form factor to complex spectra.
%
We show that the DSFF of
chaotic OQMBS generically displays the \textit{quadratic} ramp-plateau behaviour of the Ginibre ensemble from random matrix theory, in contrast to the \textit{linear} ramp-plateau behaviour of the Gaussian ensemble in closed quantum systems. 
Furthermore, in the presence of many-body interactions,
such RMT behaviour emerges only after a time scale $\taudev$, which generally increases with system size for sufficiently large system size, and can be identified as the non-Hermitian analogue of the \textit{many-body Thouless time}.
%
%
The universality of the random matrix theory behavior is demonstrated by surveying twelve models of OQMBS, including random Kraus circuits (quantum channels) and random Lindbladians (Liouvillians) in several symmetry classes, as well as Lindbladians of paradigmatic models such as the Sachdev-Ye-Kitaev (SYK), XXZ, and the transverse field Ising models. 
%
%
We devise an unfolding and filtering procedure to remove variations of the averaged density of states which would otherwise hide the universal RMT-like signatures in the DSFF for chaotic OQMBS.
Beyond chaotic OQMBS, we study the spectral statistics of non-chaotic OQMBS, specifically the integrable XX model and a system in the many-body localized (MBL) regime in the presence of dissipation, which exhibit DSFF behaviours distinct from the ramp-plateau behaviour of random matrix theory. 
Lastly, we study the DSFF of  Lindbladians with the Hamiltonian term set to zero, i.e. only the jump operators are present, and demonstrate that the results of RMT universality and scaling of many-body Thouless time survive even without coherent evolution.
As side results, we compute the density of states, nearest-neighbour spacing distribution, and complex spacing ratio for the studied models. 
\end{abstract}

\maketitle

\textbf{Introduction. --} 
The study of open quantum systems is of importance since realistic systems cannot be perfectly isolated from their environment. 
Such open systems, represented by density matrices, have dynamics described by quantum channels under the Markovian approximation, where a macroscopic number of observables can be treated as quantum noise~\cite{gardiner2004quantum}.
Two prominent descriptions are the Kraus operator formulism~\cite{CHOI1975, kraus1983}, and the quantum master equation known as the Lindbladian~\cite{gorini1976, lindblad1976}.
%
%
%
Spectral statistics have historically served as a robust diagnostic of quantum chaos in closed systems~\cite{bohigas1984characterization}, and have contributed to the recent revival of the study of many-body quantum chaos~\cite{Cotler_2017, complexity2017, kos_sff_prx_2018, cdc1, cdc2, bertini2018exact, saad2019semiclassical}. 
%
%
For open systems, the spectral statistics has been explored much less, with recent investigations focusing on the scale of mean-level spacing \cite{akemann2019, sa2019a, ueda2019nonhermMBL, sa2020b, luitz2020, Shklovskii2020, Tzortzakakis2020, Peron2020, rubio2021, garcia2022symmetry, xiao2022level, sa2022lindbladian, sa2023symmetry, kawabata2023symmetry}.
With rapid advancements in non-Hermitian physics \cite{Ashida2020review} and the recent discovery that the late-time relaxation of open systems is not solely governed by the spectral gap~\cite{shirai2020lindbladgap, ueda2020lindbladgap, twostep2022znidaric}, it is pertinent to ask what are the universal features of spectral correlations at \textit{all} scales in OQMBS.
%
%
%
%
%

\begin{figure*}[t]
\centering
\includegraphics[width=0.96\textwidth]{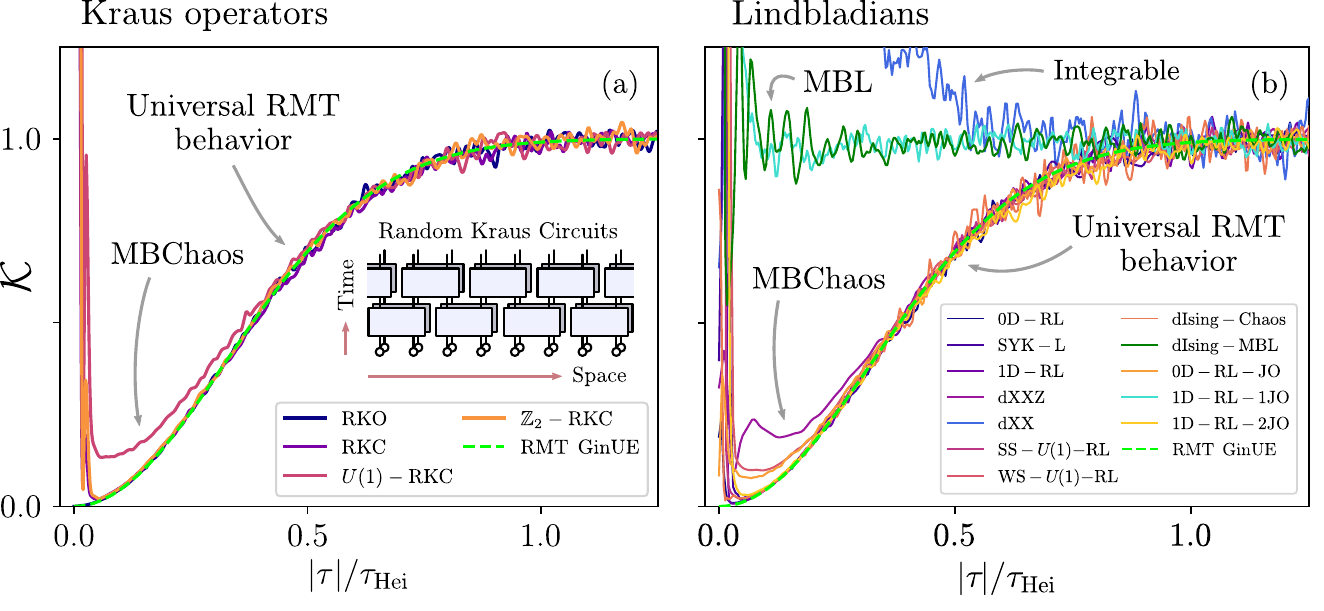}
    \caption{
    The DSFF for (a)
    the Kraus operator models RKO ($\log n = 6$), RKC ($L=6$), $U(1)$-RKC ($L=8$)
    and $\mathbb{Z}_2$-RKC ($L=7$);
    and (b) the random Lindbladian models $0$D-RL ($\log n=7$), SYK-L ($L=12$), $1$D-RL ($L=7$), dXXZ ($L=7$), dXX ($L=7$), SS-$U(1)$-RL ($L=6$), WS-$U(1)$-RL ($L=7$), dIsing-Chaos ($L=7$), dIsing-MBL ($L=7$), 0D-RL-JO ($L=6$), 1D-RL-1JO ($L=6$), and 1D-RL-2JO ($L=6$).
    The DSFF is computed along $\theta = \pi/4$ with 5000 samples. 
For RKC models, $0$D-RL, SYK-L and dXXZ, we apply unfolding and Gaussian filtering procedures. For RKO, we directly compute the DSFF.  For all other models, we only apply the filtering procedure, see details in SM~\cite{supplementary_paper2}. Note that only the models, dIsing-MBL, dXX and 1D-RL-1JO, do not exhibit RMT GinUE behaviour.
}\label{fig:many_model_same_L}
\end{figure*}

\begin{figure*}[t]
\centering
\includegraphics[width=1 \textwidth  ]{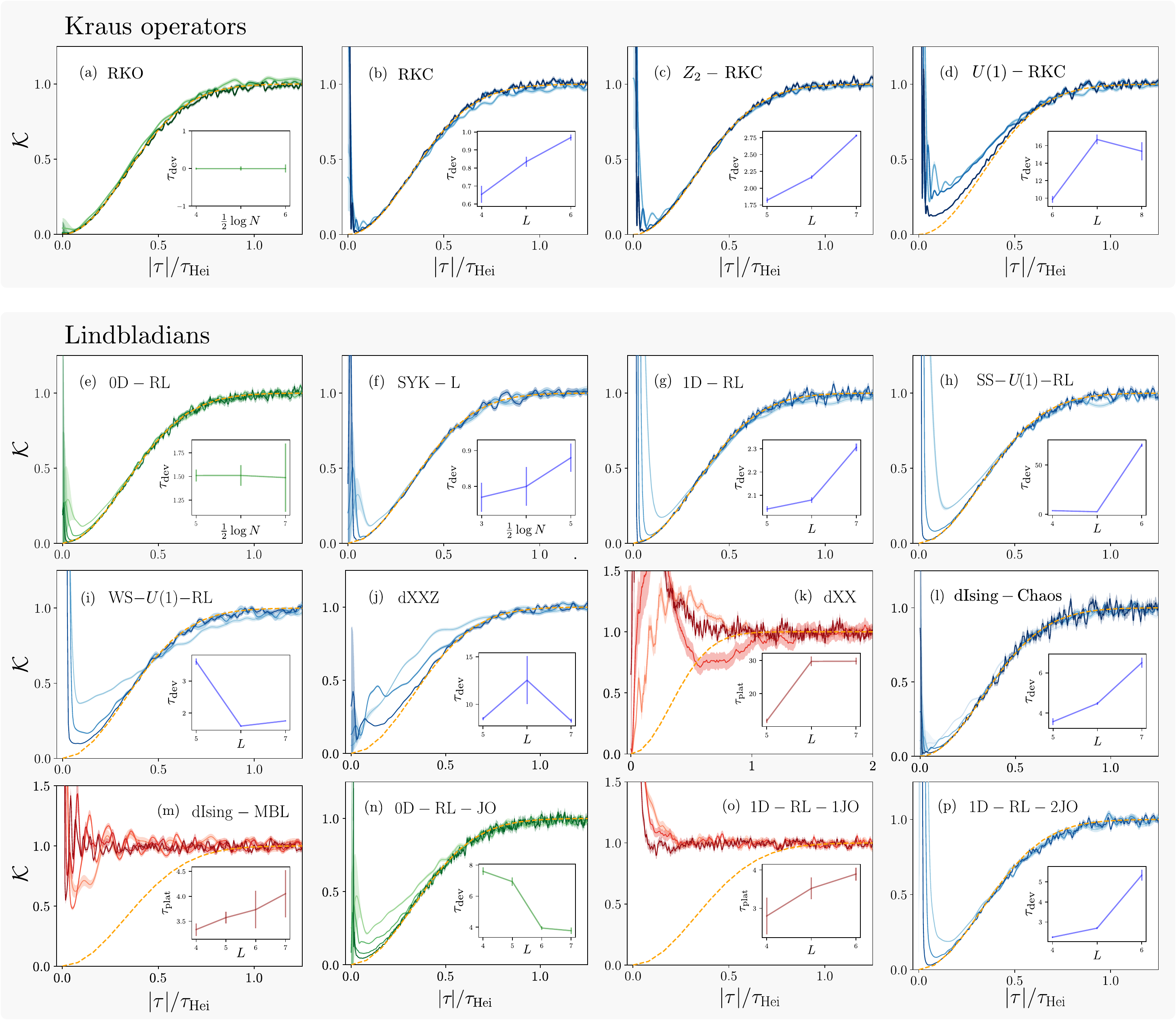}
    \caption{DSFF and the onset of RMT quadratic ramp for structureless models (green), many-body chaotic models (blue), and MBL and integrable models (red) of OQMBS for different system sizes. The DSFF data is taken along $\theta = \pi/4$ with 5000 samples. 
    DSFF of larger (smaller) system sizes are labelled with darker (lighter) colors. The theoretical curve of GinUE from RMT, $\kcgin$, is in orange.  The unfolding and filtering procedures are as in Fig.~\ref{fig:many_model_same_L}. 
    In the insets, we plot $\taudev$ against $L$. In contrast to the structure-less models (RKO, 0D-RL, 0D-RK-JO), we observe an increase in $\taudev$ in $L$ for all models with many-body interaction, with the exceptions of $U(1)$-RKC, $U(1)$-RL and dXXZ, whose dependence of $\taudev$ in $L$ has not stabilized.
    \label{fig:dsffpanel}}
\end{figure*}

In this paper, by surveying a wide range of paradigmatic models, we showed that the DSFF of chaotic OQMBS generically display the universal quadratic ramp-plateau behaviour from RMT (Fig.~\ref{fig:many_model_same_L}).
Further, We show that in the presence of many-body interaction, the DSFF of chaotic OQMBS shows an early-time deviation from the RMT until a time scale, $\taudev$, which diverges with system size, and which we identify as the analogue of the many-body Thouless time~\cite{cdc2, Kos_2018} in OQMBS (Fig.~\ref{fig:dsffpanel}).
%
%
%
%
 We devise a procedure using unfolding transformation and filtering to remove variations of density of states (DOS) which would otherwise hide the RMT-like signatures in DSFF for chaotic OQMBS.
%
Lastly, we study the spectral correlation of non-chaotic OQMBS, specifically integrable spin chains and (prethermal) many-body localized (MBL) systems in the presence of dissipation, 
showing a DSFF behaviour distinctive from the ramp-plateau behaviour of GinUE, and thereby demonstrating that DSFF diagnoses (the lack of) chaos in OQMBS. 

\vspace{0.1cm}
\textbf{Dissipative spectral form factor. --}
The spectral form factor (SFF) is one of the simplest non-trivial and analytically tractable diagnostic of quantum chaos~\cite{Berry, Sieber_2001, muller_2004, muller_2005}. 
The SFF captures correlations between eigenlevels at all scales, including the level repulsion and spectral rigidity 
%
%
%
%
and has recently been shown to capture novel signatures particularly in quantum many-body physics~\cite{kos_sff_prx_2018, cdc1, cdc2, bertini2018exact, friedman2019, moudgalya2020spectral, Roy2020random, flack2020statistics, liaogalitski2020, chan2021trans,  garratt2020manybody, chan2020lyap, swingle2020hydrodynamic, bertini2021random, mblexistence}, and in the studies of black holes and holography~\cite{Cotler_2017, complexity2017, del_Campo_2017, ALTLAND201845, saad2019semiclassical, Gharibyan_2018}. 
%
For a non-Hermitian matrix with complex spectra, however, the SFF is exponentially growing or decaying in time due to the imaginary parts of the complex eigenvalues. 
%
%
The DSFF has been introduced to circumvent this problem by considering the complex spectrum as a two-dimensional gas and probing the correlation therein~\cite{li2021spectral, fyodorov1997}.
Specifically, given a $N\times N$ matrix with spectrum $\{z_n = x_n + i y_n \}_{n=1}^N$, consider a generalized two-parameter partition function 
\be\label{eq:pfun}
Z(t,s) = \sum_{n  } e^{\mathrm{i}x_n t + \mathrm{i}y_n s} \;, 
\ee
%
%
%
and the connected DSFF~\cite{li2021spectral, fyodorov1997} defined by
\begin{equation} \label{eq:dsff_def_a}
\Kcon(t,s):= 
\big\langle
\left|
Z(t,s)
\right|^2
\big\rangle
-
\left|
\left\langle
Z(t,s)
\right\rangle
\right|^2
\;.
\end{equation}
%
%
%
%
%
%
We define the complex time $\tau \equiv |\tau | \, e^{i \theta} \equiv t+ is$, and will abusively use the radial coordinates $(|\tau|,\theta)$ as the arguments of $K$.
%
%
%
 [\onlinecite{li2021spectral}] derived the exact analytical solution of DSFF for the GinUE for all $N$, which simplifies in large-$N$ as
%
%
\be \label{eq:dsff_ginue_largeN}
\kcgin(|\tau| , \theta) = N 
\left( 1
-  e^{ -|\tau|^2/ 4N} \right) \,.
\ee
The role of the GinUE solution of DSFF is analogous to that of the Gaussian unitary ensemble solution of SFF, namely and loosely that the DSFF behaviour of sufficiently generic or  ``chaotic'' non-Hermitian matrices display universal behaviour from the GinUE of RMT \cite{li2021spectral, dsff2022kulkarni, 
Shivam_2023ginibre, Garc_a_Garc_a_2023dsff, cipolloni2023dissipative}. Alternative approaches in quantifying spectral correlation in non-Hermitian matrices beyond the scale of mean-level spacing have been explored in \cite{can2019a, kawabata2023dynamical, matsoukas2023non, matsoukas2023unitarity, yoshimura2023robustness, zhou2023universal}.
%
%
%
%
%
%
A few comments are in order: 
Firstly, note that $\kcgin (|\tau| , \theta)$ only depends on the absolute value of $\tau$, i.e. the DSFF of chaotic systems are rotationally symmetric in complex time, at least after the onset of RMT behaviour, and for this reason we opt to focus on $K(|\tau|, \theta=\pi/4)$ in this paper.
%
Secondly, as a function of $|\tau|$, the DSFF shows a \textit{(dip)-ramp-plateau} behaviour~\footnote{At early time $|\tau | \lesssim \tauedge \ll \tauhei$, 
DSFF dips from $K(0,\theta)= N^2$ with a form described by the non-universal disconnected DSFF, $|\left\langle Z(t,s)\right\rangle|^2$, discussed in \cite{li2021spectral}, but excluded in \eqref{eq:dsff_def_a} for simplicity},
analogous to the SFF for closed quantum systems: 
%
%
%
%
At $ |\tau| \lesssim \tauhei$,
DSFF increases \textit{quadratically} $\kcgin \simeq |\tau|^2/4$ in large $N$
%
until it reaches late time $ |\tau| \gtrsim \tauhei $ with $ \tauhei \sim \sqrt{N}$, where DSFF reaches a plateau at $N$. 
Crucially, the DSFF GinUE ramp behaviour is drastically different from the corresponding SFF GUE behaviour, which is \textit{linear} in time. 
%

%
%
%
%

\vspace{0.1cm}

\textbf{Models. --} We will introduce two classes of models. 
The first class consists of 
zero- and one-dimensional Kraus operators $\Phi$, i.e. $\Phi[\rho] = \sum_{a}  M_a \rho M^\dagger_a$ with density matrix $\rho$ and operators $M_a$ satisfying $\sum_{a} M_a^\dagger M_a = \mathbb{1}$~\cite{nielsen_chuang_2010}. 
For one-dimensional $L$-site models  with on-site Hilbert space dimension $q$, we define random Kraus circuits (RKC)  with brick-wall geometry in the superoperator representation [Fig.~\ref{fig:many_model_same_L} (a) inset] as
\begin{align}\label{eq:rkc}
\begin{split}
\Phi = 
&
\left( 
\,
\bigotimes_{i \in 2 \mathbb{Z}}  V_{i} 
\right)
\left(
\,
\bigotimes_{i \in 2 \mathbb{Z} + 1 } V_{i}
\right)
\end{split}
\,,
\end{align}
where $V_i$ is a set of Kraus operators acting on supersite $i$ and $i+1$  as
\begin{equation}\label{eqn:brick_def}
V_i = 
\sum_a  v_{i,a} \otimess v^*_{i,a}  
 \,.
\end{equation}
Here $v_{i,a}$ are $q^2$-by-$q^2$ Kraus operators satisfying $\sum_a v^{\dagger }_{i, a} v_{i, a}= \id $ for each $i$. 
$\otimess$ denotes the tensor product between operators acting to the left and right of a density matrix.
%
%
%
To incorporate symmetries, 
consider the conserved quantity $I$, its associated symmetry operator $S=\exp(i \varphi I)$ with real parameter $\varphi$, and the adjoint representation of $S$ given by $\mS [\rho] = S\rho S^{-1}$.  
A Kraus operator which respects our symmetry admits the block-diagonal decomposition:

\be\label{eqn:block_diagonal}
v_{i} = \sum_{\mu} P_{\mu} v_{i}^{\mu} P_{\mu}\, ,
\;\;\;  \text{s.t.} \;\;\;  [\mS, \Phi] = [S, v_{i}]=0
\,,
\ee
where $P_\mu$ is a projector to the symmetry sector labelled by $\mu$.
%
    %
The $v^\mu_a$ is a Kraus operator generated by a protocol using truncated random unitary matrices \cite{truncue2000, wojciech2009}:
    We take a $nd$-by-$nd$ unitary $T$ from the circular unitary ensemble (CUE) with $d^2$ blocks of size $n$-by-$n$ denoted $t_{ab}$, $a,b=1,\dots, d$, such that  $T_{(a\alpha),(b\beta)}= [t_{ab}]_{\alpha \beta}$, where $\alpha, \beta=1,\dots, n$ and $(a\alpha)$ is a composite index.
    We take 
    $v_a = t_{1a}$ 
    such that the Kraus condition is satisfied via the unitarity of $T$. 
We consider zero- and one-dimensional variations of RKC \eqref{eq:rkc} listed below, and written more explicitly in the supplementary material (SM)~\cite{supplementary_paper2}. 
%
%
\begin{enumerate}
    \item[1.] \textbf{0D Random Kraus operator (RKO)} is our prototypical zero-dimensional model without symmetries, where we take $\Phirko = V$ defined in \eqref{eqn:brick_def} with $n$-by-$n$ matrices and $v_a$.
    %
    \item[2.] \textbf{1D RKC without symmetries (RKC)} is a one-dimensional model without symmetries, i.e. the sum~\eqref{eqn:block_diagonal} for $v_i$ is trivial. 
    \item[3.] \textbf{$\mathbb{Z}_2$-symmetric RKC ($\mathbb{Z}_2$-RKC)} with $q=2$ is a RKC  that conserves $I= \prod_{i=1}^L \sigma^z_i$, the parity operator. 
    \item[4.] \textbf{$U(1)$-symmetric RKC ($U(1)$-RKC)} with $q=2$ is a RKC that conserves $I=\sigma^z_{\mathrm{tot}} = \sum_{i=1}^L \sigma^z_i$, the total magnetization. 
\end{enumerate}

The second class of models are the Liouvillian of quantum master equations $\frac{d\rho (t) }{dt} = \mathcal{L}[\rho]$ in the Lindblad form~\cite{lindblad1976, gorini1976}. In the superoperator representation, we have
\be
\begin{aligned} \label{eq:lindblad_model}
\mathcal{L}=& - \mathrm{i} ( H \otimess \id -  \id \otimess H^*) 
\\
&+ \sum_{a=1}^d  (  2 F_a \otimess F_a^* 
- F_a^\dagger F_a \otimess \id
- \id \otimess F_a^\dagger F_a ) \, ,
\end{aligned} 
\ee
where the Hamiltonian $H$ governs the unitary time evolution, the second term describes the coupling between the system and the environment via jump operators $F_a$. 
We study the Lindblad dynamics with variations given below.
%
%
%
%
%
\begin{itemize}
    \item[5.] \textbf{0D Random Lindbladian (0D-RL)} is our prototypical zero-dimensional model, where $H\in \GUE(n)$ and $F_a\in \Gin(n)$ are respectively taken from the Gaussian unitary and complex Ginibre ensembles of $n$-by-$n$ matrices.
    \item[6.] \textbf{Sachdev-Ye-Kitaev Lindbladian (SYK-L):} The Sachdev-Ye-Kitaev Hamiltonian is given by $\HSYK= \mathrm{i} \sum^L_{i_1 < i_2 < i_3< i_4} J_{i_1 i_2 i_3 i_4} \psi_{i_1} \psi_{i_2} \psi_{i_3} \psi_{i_4}$ with Majorana fermion operators $\psi_i = \psi_i^\dagger$ satisfying $\{\psi_{i}, \psi_{j} \} = 2 \delta_{ij}$.  We consider jump operator 
    $F_a =\sum_{1\leq i < j \leq L} K^a_{ij} \psi_i \psi_{j}$.
    $J_{ijkl}$ and $K^{a}_{ij}$ are independently drawn from the normal distribution with variance $J$ and $k$ respectively.
    %
    %
    \item[7.] \textbf{1D Random Lindbladian (1D-RL)} is a one-dimensional chain with length $L$ and on-site dimension $q$. We choose $H = \sum_{i=1}^L h_{i, i+1}$ with two-site gates $h_{i, i+1} \in \GUE(q^2)$, and two-site jump operators $F_i \equiv \ell_{i, i+1} \in \Gin(q^2)$ acting on site $i$ and $i+1$.
    %
    \item[8.] \textbf{Strongly-$U(1)$-symmetric 1D-RL (SS-$U(1)$-RL)} 
is defined identically to 1DRL, except that 
  $h_{i, i+1}  = \bigoplus_{\mu=-q+1}^{q-1}
    h^{(\mu)}_{i, i+1}$ 
    and  
    $\ell_{i, i+1} = 
    \bigoplus_{\mu=-q+1}^{q-1}
    \ell_{i, i+1}^{(\mu)} $
 where  
    $h^{(\mu)}_{i, i+1}  \in \GUE(q-|\mu|)$ and $ \ell^{(\mu)}_{i, i+1}   \in \Gin(q-|\mu|)$ \
    such that  the model is strongly symmetric, conserving the total magnetization $I=\sigma^z_{\mathrm{tot}}$, i.e. 
    \be
 [ \mS, \mathcal{L}] =   [I, H]= [I, \ell_{i, i+1}] =0. 
    \ee
    \item[9.] \textbf{Weakly-$U(1)$-symmetric 1D-RL (WS-$U(1)$-RL)} is defined similar to SS-$U(1)$-RL with the same Hamiltonian, but with different random jump operators $\ell_{i, i+1}$ such that they satisfy the following commutation relations
    \be
 [ \mS, \mathcal{L}]=[I,h_{i, i+1}] =0  \, ,  \;\;  [I, \ell_{i, i+1}^{(m)}] = m\ell_{i, i+1}^{(m)},
\ee
where  again $\mS [\rho] = S\rho S^{-1}$.
%
The explicit forms of $\ell_{i, i+1}^{(m)}$ are given in the SM~\cite{supplementary_paper2}.
\end{itemize}
Furthermore, we study two spin chains with dissipation that display chaotic and non-chaotic behaviours: 

\begin{itemize}
    \item[10.] \textbf{Dissipative XXZ and dissipative XX models } are defined through 
$H=J\sum_{l=1}^{N-1}(\sigma_{l}^{x}\sigma_{l+1}^{x}+\sigma_{l}^{y}\sigma_{l+1}^{y}+\Delta\sigma_{l}^{z}\sigma_{l+1}^{z}) +J'\sum_{l=1}^{N-2}(\sigma_{l}^{x}\sigma_{l+2}^{x}+\sigma_{l}^{y}\sigma_{l+2}^{y}+\Delta'\sigma_{l}^{z}\sigma_{l+2}^{z})$
and 
$F_{\mathrm L}^{+}=\sqrt{\gamma_{\mathrm{L}}^{+}}\sigma_{1}^{+}$, $F_{\mathrm L}^{-}=\sqrt{\gamma_{\mathrm L}^{-}}\sigma_{1}^{-}$, $F_{\mathrm R}^{+}=\sqrt{\gamma_{\mathrm R}^{+}}\sigma_{L}^{+}$, $F_{\mathrm R}^{-}=\sqrt{\gamma_{\mathrm R}^{-}}\sigma_{L}^{-}$, $F_{l}=\sqrt{\gamma} \sigma_{l}^{z}$
where  $\sigma_{l}^{\alpha}, \alpha=x,y,z$ are Pauli matrices and $\sigma^{\pm}_{l}=\sigma_{l}^{x}\pm \mathrm{i}\sigma_{l}^{y}$.
Following \cite{akemann2019}, we consider two choices of parameters: 
(a) Chaotic dissipative XXZ model \textbf{(dXXZ)}, with  parameters sampled from normal distributions $J,J'\in\mathcal{N}(1,0.09)$, $\Delta\in \mathcal{N}(0.5, 0.0225)$, $\Delta'\in\mathcal{N}(1.5,0.2025)$, $\gamma=0$, $\gamma_{L}^{+}=0.5$, $\gamma_{L}^{-}=0.3$, $\gamma_{R}^{+}=0.3$, and $\gamma_{R}^{-}=0.9$; 
(b) Integrable dissipative XX model \textbf{(dXX)}, with $J\in\mathcal{N}(1,0.09)$, $J'=0$, $\Delta=0$, $\Delta'=0$, $\gamma=1$, $\gamma_{L}^{+}=0.5$, $\gamma_{L}^{-}=1.2$, $\gamma_{R}^{+}=1$, and $\gamma_{R}^{-}=0.8$. 
    %
    \item[11.] \textbf{Dissipative transverse field Ising model in the chaotic and MBL phases}
    ~\cite{hamazaki2022lindbladian}
    are defined with $H = J \sum_{l=1}^L \sigma^z_l \sigma^z_{l+1} + g \sum_{l=1}^L \sigma^x_l + \sum_{l=1}^L h_l \sigma^z_l,$ where the on-site disorder $h_l \in [-W, W]$ is drawn from a flat distribution of width $W$. We take 
    $F_a =  \sqrt{\frac{\gamma}{2}} (\sigma^x_a - \mathrm{i} \sigma^y_a)$ with $a = 1, \dots, L$. 
    Without dissipation, in finite system sizes, $H$ displays many-body localized phenomenology for sufficiently large $W$. In this paper 
    we focus on $J=1$, $g=-0.9$, $\gamma = 0.5$ and two choices of $W$: 
    (a) $W=.6$ \textbf{(dIsing-Chaos)} and (b) $W=5$ \textbf{(dIsing-MBL)}.
\end{itemize}
Lastly, we study Lindbladians with the Hamiltonian term turned off:
\begin{itemize}
    \item[12.] \textbf{Jump-operator-only random Lindbladians} are defined  with $H=0$ with three variations: (a)  0D Random Lindbladian with jump operators only \textbf{(0D-RL-JO)} which coincides with 0D-RL (model 5) with $H=0$. (b)  1D Random Lindbladian with 1-site jump operators only \textbf{(1D-RL-1JO)} with  $F_i = \ell_{i} \in \text{GinUE}(q=2)$ acting as one-site operator on site $i$ with $d=L$. (c)   1D Random Lindbladian with 2-site jump operators only \textbf{(1D-RL-2JO)} which coincides with 1D-RL (model 7) with $H=0$.
\end{itemize}
For each model above with symmetries, we analyse the eigenvalues of the superoperator after projection into the largest symmetry sector.
In addition to the study of the DSFF in Fig.~\ref{fig:many_model_same_L} and~\ref{fig:dsffpanel}, we provide numerical data for the single-realization spectra, DOS, nearest-neighbour spacing distribution, and complex spacing ratio in the SM~\cite{supplementary_paper2}.

\begin{figure*}[t]
\centering
\includegraphics[width=1 \textwidth  ]{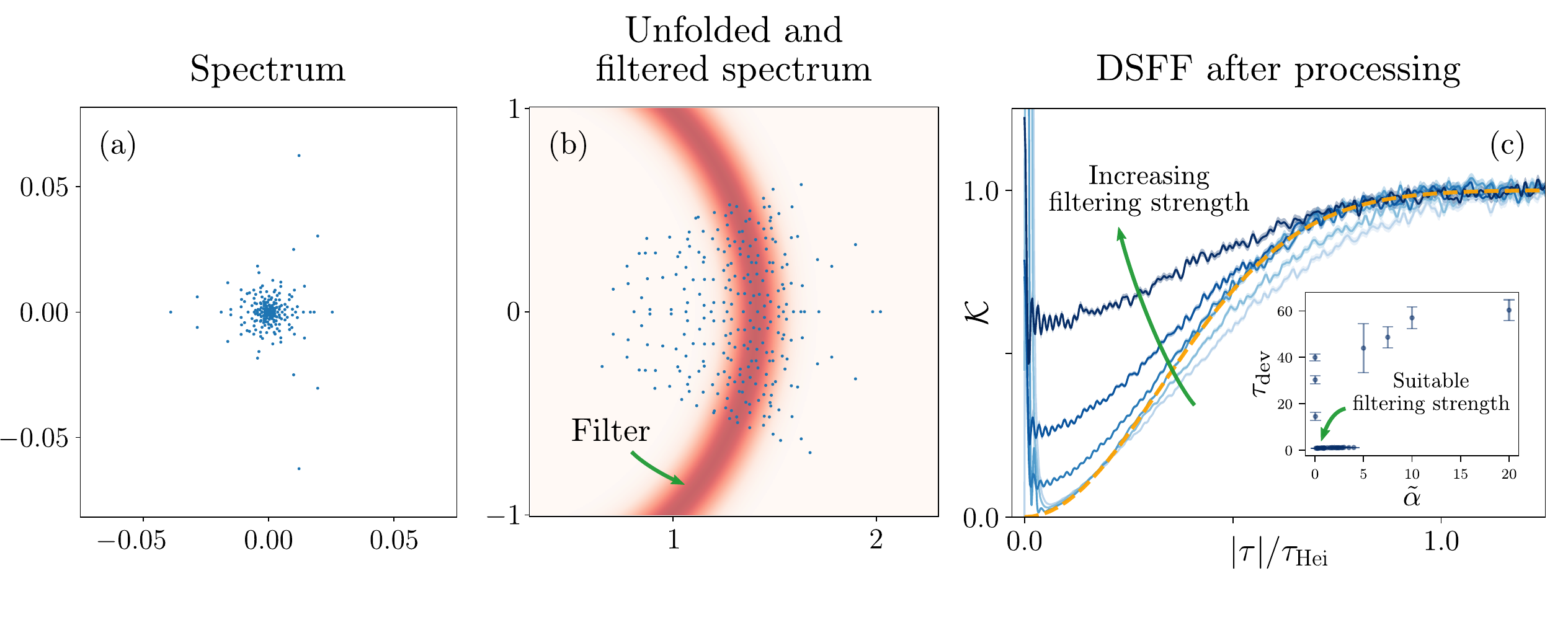}
    \caption{{Unfolding and filtering.} (a) A single-realization spectrum $\{z\}$ of RKC for $L=6$. (b) The unfolded spectrum $\{ z^{1/\log_2 N}\}$ (of the spectrum in (a)) is subsequently filtered  with the filtering function $f(z)= e^{-\alpha(|z|-\mu)^2}$. (c) DSFF of the RKC for varying filtering strength $\tilde{\alpha} \in [1/100, 20]$ from light to dark blue. Inset: $\taudev$ against $\tilde{\alpha}$ shows a trough. The suitable range of filtering strength is given at the lower range of $\tilde{\alpha}$ at the trough. 
    Note that the artificial edge of the unfolded spectrum in (b) is due to the branch cut in the unfolding function. We perform sanity checks by unfolding the GinUE spectrum, and find that such edges do not qualitatively distort the DSFF even in the absence of a filtering scheme, see Appendices E and G.
    }
    \label{fig:unfold}
\end{figure*}

\textbf{Unfolding and filtering. --} 
The density of states (DOS) can be written as a sum of two terms, $\rho_{\mathrm{av}}$, accounting for the smooth, averaged behavior, and $\rho_{\mathrm{fluc}}$, for the fluctuations about the average, i.e. $\rho(z) = \rho_{\mathrm{av}}(z) + \rho_{\mathrm{fluc}}(z)$.
The expected quadratic ramp behaviour for chaotic OQMBS means that the variation in $\rho_{\mathrm{av}}$ has to be removed before comparing the level fluctuations $\rhofluc$ in different regions. This process is called ``unfolding.''
Without unfolding, regions of the spectrum with varying $\rho_{\mathrm{av}}(z)$ give \textit{quadratic} ramps with different values of $\tauhei$, summing to a smeared-out DSFF behaviour (such problem does not arise for the \textit{linear} ramp behaviour of the SFF in closed chaotic quantum systems).  
Ideally, we would like to find an ``unfolding" function transforming $z \mapsto g(z)$ such that the resulting $\rhoav(z)$ is perfectly uniform and local relationships between eigenvalues are preserved, as reflected in, e.g., the nearest neighbour spacing distribution and the complex spacing ratio~\cite{sa2019a}.  
However, unfolding for complex spectra is considerably more challenging than for the real counterpart, and such function $g$ might not even exist in principle.
%
%
We remedy this problem in two steps. Firstly, for each model, we empirically find a conformal transformation as the (imperfect) unfolding function $g$ such that the resulting $\rhoav$ is closer to uniform over a subregion of the complex plane~\cite{supplementary_paper2}. 
%
%
%
Secondly, we employ a filtering procedure which favours the DSFF contributions from the flatter region of the unfolded spectrum $\{g(z_n) \equiv \tilde{z}_n = \tilde{x}_n + i \tilde{y}_n \}$, analogous to the filtering procedure for SFF~\cite{Gharibyan_2018}.
Therefore, we define a filtered partition function,
%
%
%
\be\label{eq:filtered_pfun}
\tilde{Z}_f(t,s; \{ \alpha\}) = \sum_{n} e^{\mathrm{i}\tilde{x}_n t + \mathrm{i} \tilde{y}_n s} 
f
        \left(\tilde{x}_n,\tilde{y}_n ; \{ \alpha \} \right)
        \, ,
\ee 
where $\{ \alpha\}$ parameterize the filtering function $f(\tilde{x}_n, \tilde{y}_n; \{\alpha\})$. 
To be concrete, we focus on the Gaussian filter $f(x; \alpha, \mu) = e^{-\alpha(x-\mu)^2}$ in either  Cartesian or radial  coordinates, with $\mu$ chosen to be the center of the flatter region in the unfolded spectrum. 
The determination of filtering strength $\alpha$ is discussed below. 
An example of the DSFF for the unfolded and filtered spectrum for the RKC is provided in Fig.~\ref{fig:unfold}.
The filtered connected DSFF, $\Kconf(t,s;\{{\alpha} \})$, is defined as in \eqref{eq:dsff_def_a}, with $Z$ replaced by $\tilde{Z}$. 
%
%
%

\textbf{Onset of RMT quadratic ramp. --} 
Based on the SFF behaviour in closed chaotic many-body systems, we expect three energy scales in the chaotic OQMBS: (i) $\DeltaD$, the energy scale over which the DOS varies, defined more precisely in SM~\cite{supplementary_paper2}; (ii) $\Deltath$, the scale within which the RMT behaviour emerges in the spectral correlation of the OQMBS (discussed below); and (iii) $\Deltahei$, the mean level spacing in the complex plane.
Again, analogous to the closed chaotic systems~\cite{cdc2, Kos_2018}, we consider an OQMBS chaotic if it exhibits RMT DSFF behaviour at sufficiently large time scale. 
For such systems, and given filtering strength, we define this time scale as
\begin{align}\label{eq:tth_def_main}
\taudev(\{\alpha\}) = \frac{\sqrt{N}}{\tauhei} \times 
\min
&
\Bigg\{
\tau: \frac{\tilde{K}_{\mathrm{c}}(|\tau|, \theta; \{\alpha \} )}{  \tilde{K}_{\mathrm{c}}( |\tau| \to \infty, \theta; \{\alpha \})  }
\nonumber
\\
&- \frac{ \kcgin(|\tau|, \theta ) }{N}
\leq \epsilon 
\Bigg\}
\,,
\end{align}
where here $\tauhei$ is the time scale at which the unfolded and filtered DSFF reaches the plateau, and consistently, $\tauhei$ can be estimated in terms of the average level-spacing weighted by the filter (see~\cite{supplementary_paper2}). 
Given that the mean level-spacing of a spectra with flat $\rhoav(z)$ scales as $1/\sqrt{N}$, the factor $\sqrt{N}/ \tauhei$ above compensates for the artificial effect of filtering. See e.g. Fig.~\ref{fig:unfold}(d) inset.
The early-time deviation from RMT behaviour can be due to non-physical effects like filtering and non-universal behaviour like finite-size effects and the ``dip'' in the disconnected DSFF~\cite{Gharibyan_2018}. Below, we argue that $\taudev$ can indeed be identified as the many-body Thouless time $\tau_{\mathrm{Th}} \propto \Deltath^{-1}$ for appropriate choices of filtering strength.
%

%

\textbf{Choice of filtering strength. --}  
In order to choose the appropriate filtering strength $\alpha$, it is useful to define a dimensionless filtering strength $\tilde{\alpha} = \alpha \DeltaD^2$.
We choose $\alpha$ to satisfy $\DeltaD^{-2} \ll \alpha \ll \Deltath^{-2} < \Deltahei^{-2}$, such that our filtering function is strong enough to reduce the effect of inhomogeneity in the DOS, but weak enough to preserve the physics at the scale $\Deltath$.  
%

The typical dependence  of $\taudev$ on $\tilde{\alpha}$ for chaotic OQMBS exhibits a trough-like behaviour, see e.g. the RKC model in Fig.~\ref{fig:unfold}(c) inset.
For weak filter strengths $\tilde \alpha$, $\taudev$ is large because the DSFF is distorted by the non-uniform DOS and does not exhibit universal RMT behavior.  
For strong filter, $\taudev$ is large because the filter removes the signatures of level repulsion in RMT at scale $\Deltahei$.
For intermediate filter strengths, we observe a trough in $\taudev(\tilde\alpha)$ 
where OQMBS displays prominent RMT DSFF behaviour. In practice, we pick smaller values of $\taudev$ in the trough to ensure that $\alpha \ll \Deltath^{-2}$. 
For all studied chaotic OQMBS, the curve of $\taudev$ versus $\tilde \alpha$ displays a trough, whose width increases in system size, see Fig.~\ref{fig:unfold}(c) and \cite{supplementary}. This trough is relatively flat, and hence we do not expect the value of $\taudev$ to be sensitive to the precise choice of filtering strength within the trough.
%
%

Finally, we benchmark our filtering procedure by computing the DSFF of the spectrum of GinUE, after applying three conformal transformations to introduce inhomogeneities to the originally flat DOS.  The form of the DSFF of GinUE before the transformation is known exactly~\eqref{eq:dsff_ginue_largeN}. 
For suitably chosen filter strengths, we recover the GinUE RMT behavior of the flat spectrum from the filtered DSFF of the transformed spectrum for all three transformations we considered: (i) $\{z_n \}$, (ii) $\{\log z_n \}$, and  (iii) $\{\log z^p \}$. 

\textbf{Chaotic OQMBS and many-body Thouless time. --} In Fig.~\ref{fig:many_model_same_L} and Fig.~\ref{fig:dsffpanel},
we compute the connected DSFF for OQMBS of models 1 to 9, the dXXZ (model 10) and dIsing-Chaos (model 11). Additionally, we compute the nearest-neighbour spacing distribution, and complex spacing ratio in the SM~\cite{supplementary_paper2}.
The DSFF, spacing distribution and ratio of all models converge to RMT GinUE behaviour in increasingly large system size at sufficiently large times (as parameterised by $\taudev$). 
Remarkably, chaotic OQMBS generically display the RMT quadratic ramp, i.e. signatures of GinUE eigenvalue correlation well beyond the scale of mean-level spacing, or equivalently, at times well before the Heisenberg time.

%
%
%


Now we investigate the early-time deviation from RMT DSFF behaviour, which we refer to as the ``bump.''
 Note that the bump should not be confused with the decaying behaviour from $N^2$ to the ramp  due to the \textit{disconnected} part of DSFF~\cite{li2021spectral} or SFF~\cite{Gharibyan_2018}. 
 We demonstrate that for OQMBS, 
(i) a sensible definition of the height of the bump in DSFF increases in system size, see SM~\cite{supplementary_paper2}; and 
(ii) the region of the bump in $|\tau|$, as parameterised by $\taudev$,
increases in system size for sufficiently large system size. In Fig.~\ref{fig:dsffpanel}, we present the DSFF for increasing $L$ for most models.
We see a notable difference 
between  models without structures
(RKO and 0D-RL) and those with many-body interactions (e.g. SYK) or spatial structure (e.g. RKC, 1D-RL, and dIsing-Chaos). 
For the models without structure, $\taudev$ does not change appreciably with system size, while for models with many-body interaction, it increases for sufficiently large $L$.
In other words, due to locality, increasingly long time is required for the many-body systems to be indistinguishable from structureless random matrices, as far as the spectral statistics is concerned.
Note that several models (e.g. $U(1)$-RKC, $U(1)$-RL, dXXZ), display RMT DSFF behaviour for the largest system size, the dependence of $\taudev$ with system size has not stabilized due to finite-size effects.
Note that, for the collapsed curves in the main panels of Fig.~\ref{fig:dsffpanel}, $\taudev/\tauhei$ is decreasing for larger $L$, i.e., the bump appears to move towards the left.  However, this is due to the exponential growth of the Heisenberg time; 
as seen in the insets of Fig.~\ref{fig:dsffpanel}, $\taudev$ itself increases sub-exponentially for OQMBS, unlike the case for 0D systems.
(i) and (ii) suggest that the deviation from RMT in DSFF is a genuine many-body effect in OQMBS, not due to finite-size effect. 
Therefore, we identify $\tau_{\mathrm{dev}}$ as the many-body Thouless time 
for open quantum many-body systems.  

%
%
%

\textbf{Many-body localized and integrable systems. --}
In Fig.~\ref{fig:many_model_same_L} and Fig.~\ref{fig:dsffpanel},
we present the DSFF of dIsing-MBL and dXX after applying the filtering protocol, as examples of MBL and integrable systems, respectively.
For both systems, the DSFF increases rapidly to (and above) the plateau, fluctuates wildly even after ensemble averaging, and does not display the ramp-plateau signature of RMT. 
In insets of Fig.~\ref{fig:dsffpanel}, we compute the time when the DSFF increases to the plateau, defined as
\begin{align}\label{eq:tplat_def_main}
\tauplateau = \frac{\sqrt{N}}{\tauhei} \min
\Bigg\{
\tau: \Big|\frac{\tilde{K}_{\mathrm{c}}(|\tau|, \theta; \{\alpha \} )}{  \tilde{K}_{\mathrm{c}}( |\tau| \to \infty, \theta; \{\alpha \})  }
\nonumber- 1\Big|
\leq \epsilon 
\Bigg\}
\,,
\end{align}
where an additional time average has been applied to $\tilde{K}_\mathrm{c}$, and where $\tauhei$ is defined by the mean level spacing weighted by the filter~\cite{supplementary_paper2}.
We observe that $\tauplateau$ in the MBL and integrable systems does not scale with the inverse mean level spacing, i.e. $\tauplateau$ does not scale exponentially in system size $L$ (although still increasing in $L$ for the accessible system sizes), in contrast to the chaotic case, and analogous to the behaviour of SFF in MBL and certain integrable closed systems \cite{mblexistence, Garratt_2021, sffmbl_2021_prakash}.
We therefore conclude that the DSFF clearly distinguishes chaotic and non-chaotic systems, including MBL and integrable systems.
%

\textbf{Jump-operator-only Lindbladians. --} We study the DSFF of  Lindbladians with the Hamiltonian term set to zero, i.e. only the jump operators are present. 
In Fig.~\ref{fig:dsffpanel} (n-p), we plot the DSFF for Lindbladians with structureless, 1-site and 2-site jump-operators, and see that they display behaviours similar to the MBL, structureless chaotic and spatially-extended chaotic models respectively.  
These results demonstrate that the RMT universality and scaling of many-body Thouless time can survive even with only driven-dissipative evolution (without coherent evolution). 
%

\textbf{Conclusions. --} In this paper, by surveying a wide range of paradigmatic models, we showed that chaotic OQMBS generically display the quadratic ramp-plateau behaviour of the Ginibre ensemble from RMT after a time scale $\taudev$, which we identify as the non-Hermitian analogue of the many-body Thouless time.
We further show that the long-range spectral statistics of MBL and integrable OQMBS behave distinctly from the chaotic OQMBS.
Recently, experimental protocols have been proposed to measure the SFF in closed quantum many-body  systems~\cite{Vasilyev_2020, sffexp2022, Dag2023}.
However, in practice, decoherence is an inherent feature of current noisy intermediate-scale era quantum (NISQ) simulators, i.e. such systems are naturally open. Therefore, it would be beneficial to devise a protocol to probe quantum many-body dynamics using DSFF.
Such protocol requires one to overcome certain technical obstacles, e.g. treating the real and imaginary eigenvalues separately as in Eq.~\eqref{eq:pfun}. 
%
For similar reasons, it is an interesting but non-trivial task to gain a better analytic handle on the DSFF in OQMBS
, whose closed quantum many-body system analogues have only recently been understood~\cite{kos_sff_prx_2018, cdc1, cdc2, bertini2018exact}. 

%

\textit{Acknowledgement.}
We are thankful to Shinsei Ryu and Yan Fyodorov for helpful discussions.
A.C. acknowledges support from the Royal Society grant RGS{$\backslash$}R1{$\backslash$}231444, and the fellowships from EPSRC EP/X042812/1, the Croucher foundation and the PCTS at the Princeton University.
T.P. acknowledges ERC Advanced grant 101096208 -- QUEST,
and ARIS research program P1-0402 and grant N1-0219.
%

%


\onecolumngrid
\newpage


\newpage

\onecolumngrid
\newpage 

\appendix
\setcounter{equation}{0}
\setcounter{figure}{0}
\renewcommand{\thetable}{S\arabic{table}}
\renewcommand{\theequation}{S\thesection.\arabic{equation}}
\renewcommand{\thefigure}{S\arabic{figure}}
\setcounter{secnumdepth}{2}

\begin{center}
{\Large Supplementary Material \\
\vspace{0.2cm}
\titleinfo 
}
\end{center}
In this supplementary material we provide additional details about: 
\begin{enumerate}[label=\Alph*]
\item Models
\begin{itemize}
    \item[---] Random Kraus circuits
\begin{itemize}
    \item[1.] Random Kraus operators (RKO)
    \item[2.] Random Kraus circuits (RKC)
    \item[3.] $\mathbb{Z}_2$-symmetric Random Kraus circuits ($\mathbb{Z}_2$-RKC)
    \item[4.] $U(1)$-symmetric Random Kraus circuits ($U(1)$-RKC)
\end{itemize}
    \item[---] Random Lindbladians
\begin{itemize}
    \item[5.]  0D Random Lindbladian (0D-RL)
    \item[6.] Sachdev-Ye-Kitaev Lindbladian (SYK-L)
    \item[7.] 1D Random Lindbladian (1D-RL)
    \item[8.] Strongly-$U(1)$-symmetric 1D Random Lindbladian (SS-$U(1)$-RL)
    \item[9.] Weakly-$U(1)$-symmetric 1D random Lindbladian (WS-$U(1)$-RL)
    \item [10.] Dissipative XXZ and XX models 
    \begin{itemize}
        \item[(a)] Chaotic dissipative XXZ model (dXXZ)
        \item[(b)] Integrable dissipative XX model (dXX)
    \end{itemize}
    \item[11.] Dissipative transverse field Ising model
    \begin{itemize}
        \item[(a)] Dissipative transverse field Ising model in the chaotic phase (dIsing-Chaos)
        \item[(b)] Dissipative transverse field Ising model in the many-body localized phase (dIsing-MBL)
    \end{itemize}
    \item[12.] Jump-operator-only Lindbladian
    \begin{itemize}
        \item[(a)] 0D Random Lindbladian with jump operator only (0D-RL-JO)
        \item[(b)] 1D Random Lindbladian with 1-site jump operators only (1D-RL-1JO)
        \item[(c)] 1D Random Lindbladian with 2-site jump operators only (1D-RL-2JO)
    \end{itemize}
\end{itemize}
\end{itemize}
\item Spectral properties
\item Nearest neighbour spacing distribution
\item Complex spacing ratio 
\item Unfolding
\item Filtering
\begin{itemize}
    \item[---] Definition
    \item[---] Early time effect of filtering
    \item[---] Late time effect of filtering
    \item[---] Definition of Heisenberg time after filtering
    \item[---] Choice of filtering strength $\alpha$
\end{itemize}
\item Sanity checks
\begin{itemize}
    \item[---] Sanity check 1: DSFF of $\{z \} $ of GinUE with filtering 
    \item[---] Sanity check 2: DSFF of $\{\log z \} $ of GinUE with filtering
    \item[---] Sanity check 3: DSFF of $\{z^p \} $ of GinUE with filtering and $p = \frac{1}{4}$
    \item[---] Example: DSFF of RKO with filtering
\end{itemize}

\item Summary table and additional numerics for dissipative spectral form factor (DSFF) 
\item Height of Bumps
\end{enumerate}


\section{Models }\label{app:models}
In this section, we write explicitly the content of the gates used to construct the zero-dimensional RKO, the one-dimensional RKC, and its variants.
We use the convention where the matrix representation of the superoperators is of size $N$.  
For many-body systems, we use $L$ to denote the system size in one-dimensional systems, or the number of particles in zero-dimensional systems (see the SYK-L model below).
We use $I$ for conserved quantities, $S=\exp(i \varphi I)$ for its associated symmetry operator  with real parameter $\varphi$, and $\mS$ for the adjoint representation of $S$ defined by $\mS [\rho] = S\rho S^{-1}$.  

\subsection{Random Kraus circuits}
\begin{itemize}
    \item[1.] \textbf{Random Kraus operator (RKO)} -- In the superoperator representation, the 0D-RKO acting on a Hilbert space with dimension $n$ is given by 
    %
    \begin{equation}
        \Phirko =
        \sum_{a=1}^d v_a \otimess v^*_a
        \,,
    \end{equation}
    where  
     $v_a$ is a  $n$-by-$n$  Kraus operator satisfying $\sum_a v_a v^\dagger_a = \id$ and is generated by a protocol using truncated random unitary matrices \cite{truncue2000, wojciech2009}:
    We take a $nd$-by-$nd$ Haar-random unitary $T \in U(dn)$ with $d^2$ blocks of size $n$-by-$n$ denoted by $t_{ab}$, $a,b=1,\dots, d$, such that  $T_{(a\alpha),(b\beta)}= [t_{ab}]_{\alpha \beta}$ where $\alpha, \beta=1,\dots, n$ and $(a \alpha)$ is a composite index. 
    We take 
    $v_a = t_{1a}$ for $a= 1,\dots,d$ such that the Kraus operator condition is satisfied via the unitarity of $T$. 
    %
    %
   For simplicity, to represent Kraus operators sampled in this manner, we will use the notation
    \be
    \{ v_a  \} \in \TCUE(n,dn) \, .
    \ee
    \item[2.] \textbf{Random Kraus circuit (RKC)} -- The RKC without any conserved quantities, $\Phirkc$, is defined in the main text by~\eqref{eq:rkc} and~\eqref{eqn:brick_def}, which we reproduce below,
\be
\Phirkc = 
\left( 
\,
\bigotimes_{i \in 2 \mathbb{Z}}  V_{i} 
\right)
\left(
\,
\bigotimes_{i \in 2 \mathbb{Z} + 1 } V_{i}
\right)
\, , \qquad \qquad 
V_i = 
\sum_a  v_{i,a} \otimess v^*_{i,a}  
 \,.
\ee
 For RKC without conserved quantities, we take 
    %
        \be
   \left\{ v_{i,a} \right\} \in \TCUE\left(q^2,dq^2 \right) \, ,
    \ee
    i.e., 
    $\{ v_{i, a} \}$ are taken from the truncated CUE.
    \item[3.] \textbf{$\mathbb{Z}_2$-symmetric random Kraus circuit ($\mathbb{Z}_2$-RKC)} --  Consider the parity operator given by $I = \prod_{i=1}^L \sigma^z_i$, where $\sigma^z_i$ is the Pauli-$z$ operator acting on site $i$.
   To construct a $\mathbb{Z}_2$-RKC preserving the parity of the circuit with on-site dimension $q=2$, we consider $2$-site
    gates  with the block structure in the basis of $\sigma^z$ as follows,
    \begin{equation}
     v_{i,a} = 
    \bigoplus_{\mu=0}^1
    v_{i,a}^{(\mu)}  
    \;, 
    \end{equation}
 where  
            \be
   \left\{ v^{(\mu)}_{i,a} 
   \right\} \in \TCUE\left(2,2d \right) \, .
    \ee
for $\mu=1,2$, such that
\be
[ \mS, \Phi] =[I, v_{i,a}] =0 
\ee
We focus on the symmetry sector $I=1$ in this paper, although for $I=-1$, we expect our observables to behave identically.
    \item[4.] \textbf{$U(1)$-symmetric random Kraus circuit ($U(1)$-RKC)}: Consider the $U(1)$ charge, $ I = \sigma^z_{\mathrm{tot}} = \sum_{i=1}^L \sigma^z_i$. 
   To construct a $U_1$-RKC with on-site dimension $q=2$, we consider $2$-site
    gates  with the block structure in the basis of $\sigma^z$ as follows,
    \begin{equation}
     v_{i,a} = 
    \bigoplus_{\mu=-1}^1
    v_{i,a}^{(\mu)}       
    \end{equation}
 where  
\begin{IEEEeqnarray}{rlrl}
   \left\{ v^{(\mu=\pm 1)}_{i,a}  \right\} &\in \TCUE\left(1,d \right)  
  \qquad \qquad   
     \left\{ v^{(0)}_{i,a}  \right\}& \in \TCUE\left(2,2d \right) \,.
\end{IEEEeqnarray}
such that
\be
[\mS, \Phi]=[I, v_{i,a}] =0 \;.
\ee
%
    %
 We focus on the symmetry sector $I=0$ in this paper.
\end{itemize}

\subsection{Random Lindbladians}
To make this section self-contained, we reproduce the Lindbladian \eqref{eq:lindblad_model} in the superoperator representation, 
\be
\begin{aligned} \label{eq:lindblad_model2}
\mathcal{L}=& - \mathrm{i} ( H \otimess \id -  \id \otimess H^*) 
+ \sum_{a=1}^d  (  2 F_a \otimess F_a^* 
- F_a^\dagger F_a \otimess \id
- \id \otimess F_a^\dagger F_a ) \, ,
\end{aligned} 
\ee
where $H$ and $F_a$ are the Hamiltonian and jump operators specified by the variation of the models below.
Again, we take the dimension of the Hilbert space to be $n$, and the size of the matrix representation of the superoperator $\mathcal{L}$ to be $N=n^2$.  The choice of $d$ will depend on the specific model considered below.
%
%
\begin{itemize}
    \item[5. ] \textbf{0D Random Lindbladian (0D-RL)} -- On the superspace of a Hilbert space of size $n$, the Hamiltonian and jump operators are given by
            \be
    H \in \GUE(n), \qquad \qquad 
  F_a \in \Gin(n) \, ,
    \ee
   where $\GUE(n)$ and $\Gin(n)$ are the Gaussian and complex Ginibre unitary ensemble of $n$-by-$n$ random matrices respectively. We take  $d=1$ for this model.
  
  \item[6. ] \textbf{Sachdev-Ye-Kitaev Lindbladian (SYK-L)} -- The Sachdev-Ye-Kitaev Hamiltonian is given by \be
  \HSYK= \mathrm{i} \sum^L_{i_1 < i_2 < i_3< i_4} J_{i_1 i_2 i_3 i_4} \psi_{i_1} \psi_{i_2} \psi_{i_3} \psi_{i_4}  \;, 
  \ee
  with Majorana fermion operators $\psi_i = \psi_i^\dagger$ satisfying $\{\psi_{i}, \psi_{j} \} = 2 \delta_{ij}$, and  real Gaussian random variables $J_{i_1 i_2 i_3 i_4}\in \mathcal{N}(0,J)$. We consider jump operators 
  \be
  \ba
  F_a &=\sum_{1\leq i < j \leq L} K^a_{ij} \psi_i \psi_{j} \;,\; \;a=1,2,\dots, d \;,
  \ea
  \ee
  with complex Gaussian random variables $K^a_{ij}\in \mathcal{N}(0,k) + i\mathcal{N}(0,k)$~\cite{Kulkarni_2022, Sa_syklindblad_2022}. From~\cite{kawabata2023symmetry}, SYK-L has two parity symmetries $(-1)^{\cal F^{+}}$ and $(-1)^{\cal F^{-}}$ where $\left( -1 \right)^{F^{\pm}} = \left[ \prod_{i=1}^{\left( L-1 \right)/2} \left( 2\mathrm{i} \psi_{2i-1}^{\mp} \psi_{2i}^{\mp} \right) \right]\,( \sqrt{2} \psi_{L}^{\mp} )$. In this paper, we take parameters $d=L$, $J=\frac{3!}{L^3}$ and $k=0.04$, and we focus on the symmetry section $(-1)^{\cal F^{\pm}}=1$.

    \item[7. ] \textbf{1D Random Lindbladian (1D-RL)} -- On the superspace of a one-dimensional system    
    with on-site Hilbert space dimension $q$, we take the Hamiltonian and jump operators to be:
\be\label{eq:hl}
H = \sum_{i=1}^{L-1} h_{i, i+1} \,,  \qquad F_i = \ell_{ i,i+1} \,,\qquad i = 1,2,\dots, L-1.
\ee
$h_{i,i+1}$ and $\ell_{i, i+1}$ are two-site random operators acting on site $i$ and $i+1$, and independently drawn from the GUE and GinUE respectively, i.e.
\be
   h_{i, i+1} \in \GUE(q^2), \qquad \qquad 
   \ell_{i, i+1}
   \in \Gin(q^2)
   \,.
\ee
The number of jump operators is just $d=L-1$. In this paper we focus on the case $q=2$, with open boundary conditions. 



    \item[8. ] \textbf{Strongly-$U(1)$-symmetric 1D Random Lindbladian (SS-$U(1)$-RL)} -- Consider the $U(1)$ charge $ I = \sigma^z_{\mathrm{tot}} = \sum_{i=1}^L \sigma^z_i$, 
    where $\sigma^z_i$ has eigenvalues $\{-\frac{q-1}{2}, -\frac{q-1}{2}+1, \dots, \frac{q-1}{2}\}$ measuring the magnetization along the $z$-axis on site $i$. 
     SS-$U(1)$-RL is defined similarly as the 1D-RL, except that 
\be\label{eq:U1hl1}
    h_{i, i+1}  =\bigoplus_{\mu=-q+1}^{q-1}
    h^{(\mu)}_{i, i+1} , \qquad \qquad 
       \ell_{i, i+1}=\bigoplus_{\mu=-q+1}^{q-1}
    \ell_{i,i+1}^{(\mu)} , \qquad i = 1,2,\dots, L-1,
\ee
 where in the local symmetry sector 
 $\mu = \sigma^{z}_{i} + \sigma^{z}_{i+1}$, we choose 
    \be\label{eq:U1hl2}
     h^{(\mu)}_{i,i+1} \in \GUE(q-|\mu|),
         \qquad \qquad    
    \ell^{(\mu)}_{i,i+1}   \in \Gin(q-|\mu|).
    \ee
    such that  
    \be
     [\mS, \mathcal{L}] = 
    [I, H]=    [I, h_{i,i+1}]= [I, F_i] = [I, \ell_{i,i+1}]=0 
    \ee
i.e.  this model is a ``strongly symmetric'' Lindbladian with respect to $I$ since individual terms in \eqref{eq:lindblad_model2} commute with the symmetry operator~\cite{buca2012,
albert2014}. 
%
%
%
%
We find that for $q=2$, the SS-$U(1)$-RL is not chaotic, i.e. probes of the spectral 
correlations, e.g. the DSFF and NNSD, do not converge to Ginibre RMT behaviour. The lack of chaotic behaviour is due to the reduction in degrees of freedom for the Hamiltonian and jump operators after imposing $U(1)$ symmetry.
%
%
We focus on  SS-$U(1)$-RL with $q=3$, the minimum chaotic case, and the symmetry sector $I=0$ in this paper.
%
%
    \item[9. ] \textbf{Weakly-$U(1)$-symmetric 1D Random Lindbladian (WS-$U(1)$-RL)} -- 
    Consider again the  $U(1)$ charge, $ I = \sigma^z_{\mathrm{tot}} = \sum_{i=1}^L \sigma^z_i$. 
As mentioned above, we find that the strongly symmetric SS-$U(1)$-RL is only chaotic when $q$ is taken to be 3. 
Another way to impose symmetry while retaining chaotic behaviour with $q=2$ is to relax the strong symmetry of Lindbladian to a weak symmetry.
 To this end, we consider a WS-$U(1)$-RL that acts on the 
 superspace of a one-dimensional system with on-site Hilbert space dimension $q=2$. 
 It has the same Hamiltonian as the SS-$U(1)$-RL,

    \be
    h_{i,i+1}  = \bigoplus_{\mu=-1}^1
    h^{(\mu)}_{i,i+1} , \qquad 
    h^{(\pm 1)}_{i,i+1}, \in \GUE(1), \qquad 
    h^{(0)}_{i,i+1} \in \GUE(2)\, ,\qquad i = 1,2,\dots, L-1
    \ee
 with the jump operators 
    \be
    F_i = \ell_{i,i+1}^{(m=0,\pm{1},\pm{2})}, \qquad i = 1,2,\dots, L-1
    \ee
where
\begin{align}
    \ell_{i,i+1}^{(\pm{2})}=& \qquad \beta_{i}^{\pm} \sigma_{i}^{\pm}\sigma_{i+1}^{\pm}, \\
    \ell_{i,i+1}^{(1)}=&
        \begin{pmatrix}
        0&\mu_{i}^{+}&\nu_{i}^{+}&0
        \\
        0&0&0&\sigma_{i}^{+}
        \\
        0&0&0&\rho_{i}^{+}
        \\
        0&0&0&0
        \end{pmatrix},
    \\
    \ell_{i,i+1}^{(-1)}=&
        \begin{pmatrix}
        0&0&0&0
        \\
        \mu_{i}^{-}&0&0&0
        \\
        \nu_{i}^{-}&0&0&0
        \\
        0&\sigma_{i}^{-}&\rho_{i}^{-}&0
        \end{pmatrix}
        \;.
\end{align}
Here $\sigma_{i}^{+}$ and $\sigma_{i}^{-}$ are the ladder operators on site $i$, 
$\beta_{i}^{\pm}, \mu_{i}^{\pm}, \nu_{i}^{\pm}, \sigma_{i}^{\pm}, \rho_{i}^{\pm} \in \Gin(1)$ 
and $\ell_{i, i+1}^{(m=0)}$ is just the charge conserving jump operators defined in \eqref{eq:U1hl1} and \eqref{eq:U1hl2}. 
The above Hamiltonians and jump operators are chosen to satisfy
\be\label{eq:ws_eq1}
\qquad [I, H] = [I,h_{i, i+1}] =0  \, , \qquad \qquad  [I , \ell_{i, i+1}^{(m)}] = m\ell_{i, i+1}^{(m)},
\ee
such that the Lindbladian superoperater satisfies
\be\label{eq:ws_eq2}
[{\mathcal{L}} , \mathcal{S}] =0 \;.
\ee
%
By \eqref{eq:ws_eq1} and \eqref{eq:ws_eq2}, this model is considered a ``weakly symmetric'' Lindbladian~\cite{buca2012,albert2014}. We focus on the largest symmetry section $\mathcal{S} = 0$ in this paper.

\item[10.] \textbf{Dissipative XXZ and XX models} -- 
Here we consider a one-dimensional dissipative XXZ model with nearest neighbor and next-to-nearest neighbor interactions.
The Hamiltonian and jump operators are defined as
\be\label{app:eq:xxz}
\begin{split}
&H=J\sum_{l=1}^{N-1}(\sigma_{l}^{x}\sigma_{l+1}^{x}+\sigma_{l}^{y}\sigma_{l+1}^{y}+\Delta\sigma_{l}^{z}\sigma_{l+1}^{z})\\
&\quad \quad \quad +J'\sum_{l=1}^{N-2}(\sigma_{l}^{x}\sigma_{l+2}^{x}+\sigma_{l}^{y}\sigma_{l+2}^{y}+\Delta'\sigma_{l}^{z}\sigma_{l+2}^{z})\\
&F_{\mathrm L}^{+}=\sqrt{\gamma_{\mathrm{L}}^{+}}\sigma_{1}^{+}\\
&F_{\mathrm L}^{-}=\sqrt{\gamma_{\mathrm L}^{-}}\sigma_{1}^{-}\\
&F_{\mathrm R}^{+}=\sqrt{\gamma_{\mathrm R}^{+}}\sigma_{L}^{+}\\
&F_{\mathrm R}^{-}=\sqrt{\gamma_{\mathrm R}^{-}}\sigma_{L}^{-} \,
\\
&F_{l}=\sqrt{\gamma} \sigma_{l}^{z} \;,
\end{split}
\ee
where  $\sigma_{l}^{\alpha}$ with $ \alpha=x,y,z$ are the Pauli matrices and $\sigma^{\pm}_{l}=\sigma_{l}^{x}\pm i\sigma_{l}^{y}$.
Like in WS-$U(1)$-RL, this model has the $U(1)$ weak symmetry, and we focus on the $\mathcal{S}=0$ sector.
Following \cite{akemann2019}, we consider two choices of parameters: 
\begin{itemize}
    \item[(a)] \textbf{Chaotic dissipative XXZ model (dXXZ)} is defined by \eqref{app:eq:xxz} with parameters drawn from normal distributions $J,J'\in\mathcal{N}(1,0.09)$, $\Delta\in \mathcal{N}(0.5, 0.0225)$, $\Delta'\in\mathcal{N}(1.5,0.2025)$, $\gamma=0$, $\gamma_{L}^{+}=0.5$, $\gamma_{L}^{-}=0.3$, $\gamma_{R}^{+}=0.3$, and $\gamma_{R}^{-}=0.9$. 
    
    \item[(b)] \textbf{Integrable dissipative XX model (dXX)} is defined by \eqref{app:eq:xxz} with $J\in\mathcal{N}(1,0.09)$, $J'=0$, $\Delta=0$, $\Delta'=0$, $\gamma=1$, $\gamma_{L}^{+}=0.5$, $\gamma_{L}^{-}=1.2$, $\gamma_{R}^{+}=1$, and $\gamma_{R}^{-}=0.8$. Note that this model is Bethe ansatz integrable with a mapping to the Fermi-Hubbard chain with imaginary interaction.
\end{itemize}

   \item[11.] \textbf{Dissipative transverse field Ising model 
    } -- 
    We consider the Hamiltonian
    \be\label{app:eq:tfim}
    H = J \sum_{l=1}^L \sigma^z_l \sigma^z_{l+1} + g \sum_{l=1}^L \sigma^x_l + \sum_{i=1}^L h_l
    \sigma^z_l \; ,
    \ee
    where the on-site disorder $h_l \in [-W, W]$ is drawn from a flat distribution of width $W$. We take the jump operators to be 
    \be\label{app:eq:tfim_jump}
F_a =  \sqrt{\frac{\gamma}{2}} \sigma^-_a \;,
\ee 
with $a = 1, \dots , L$.
    Without dissipation, in finite system sizes, $H$ displays many-body localized phenomenology for sufficiently large $W$. In this paper we take $J=1$, $g=-0.9$ and $\gamma = 0.5$. Following~\cite{hamazaki2022lindbladian}, we consider the model in two different phases:
    \begin{itemize}
    \item[(a)] \textbf{Dissipative transverse field Ising model in the chaotic phase (dIsing-Chaos)} defined by \eqref{app:eq:tfim} and \eqref{app:eq:tfim_jump} with $W=0.6$.
    
    \item[(b)] \textbf{Dissipative transverse field Ising model in the many-body localized phase (dIsing-MBL)} defined by \eqref{app:eq:tfim} and \eqref{app:eq:tfim_jump} with $W=5$.
\end{itemize}

\item[12.] \textbf{Jump-operator-only Lindbladian. } We consider the  Lindbladians 
acting on the superspace of a Hilbert space of size $n$, where the Hamiltonian is set to be zero, i.e. $H=0$, and the jump operators are drawn from the GinUE with varying size of operator support. Again the number of jump operator is set to be $d$.
\begin{itemize}
    \item[(a)] \textbf{0D Random Lindbladian with jump operator only (0D-RL-JO):} This model is a variation of the model 0D-RL with $H=0$. We take only one jump operator drawn from the GinUE of $n$-by-$n$ matrices, i.e. $d=1$ and $F_1 \in \text{GinUE}(n)$. 
    \item[(b)] \textbf{1D Random Lindbladian with 1-site jump operators only (1D-RL-1JO):} The jump operators are taken to be $F_i = l_{i}$ with $i=1,2,\dots, L$, where $l_{i} \in \text{GinUE}(q)$ are one-site random operators acting on site $i$. The number of jump operator is $d=L$, and the on-site dimension is $q=2$.
        
        
    \item[(c)] \textbf{1D Random Lindbladian with 2-site jump operators only (1D-RL-2JO):} This model is a variation of the model 1D-RL with $H=0$. The jump operators are taken to be $F_i = l_{i,i+1}$ with $i=1,2,\dots, L-1$, where $l_{i,i+1} \in \text{GinUE}(q^2)$ are two-site random operators acting on site $i$ and $i+1$. The number of jump operator is $d=L-1$, and the on-site dimension is $q=2$. 
\end{itemize}

\end{itemize}

\section{Spectral properties}\label{app:spectra}
In this section, we provide the plots for the single realization spectra and heat maps of the
DOS for Krauss operators and Lindbladian models. 
%
%

%

\subsection{Random Kraus circuits}
\begin{figure}[H]
\begin{minipage}[t]{0.39\textwidth}
\includegraphics[width=\linewidth,keepaspectratio=true]{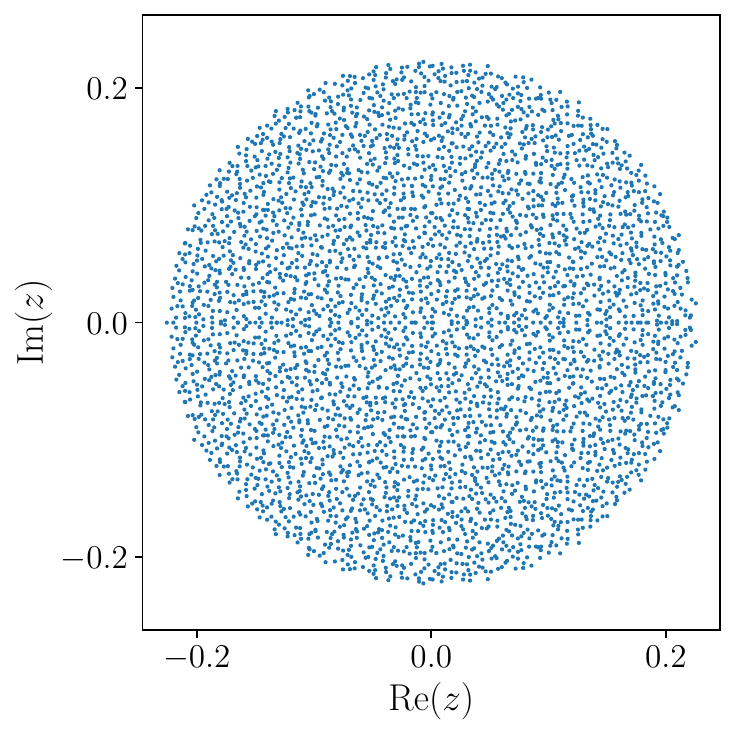}
\end{minipage}
\hspace*{\fill} 
\begin{minipage}[t]{0.475\textwidth}
\includegraphics[width=\linewidth,keepaspectratio=true]{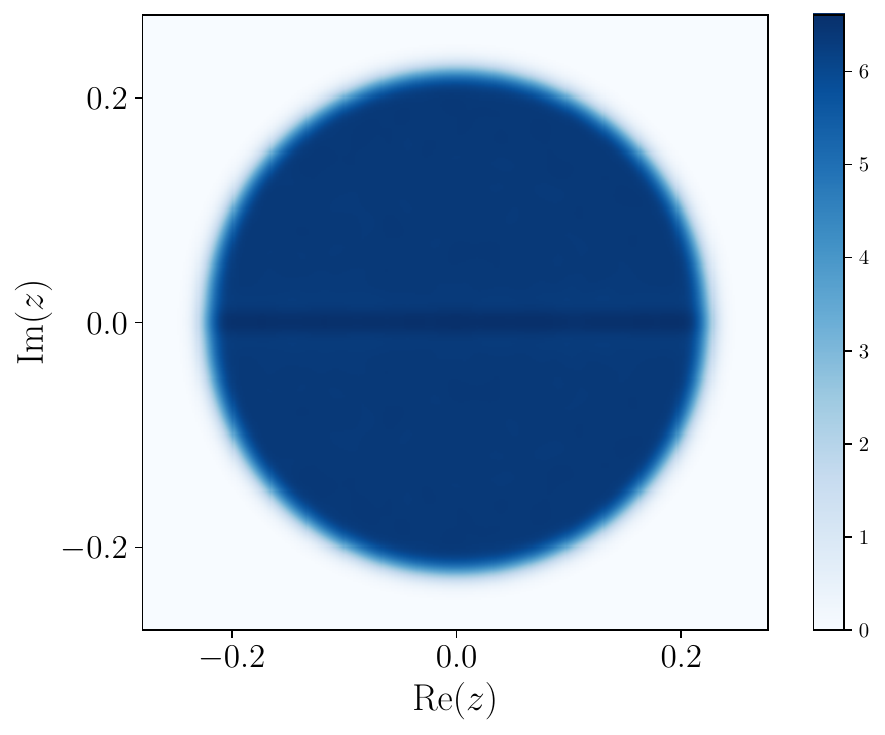}
\end{minipage}
    \caption{The spectrum of a single realization (left) and the heat map of the DOS (right)
    of the RKO
    with dimension $N=4096$.
    Note that there is a single leading eigenvalue at $z=1$, which we have removed for clarity of presentation.  
    The spectrum is roughly uniform, displays reflection symmetry across the real axis, 
    exhibits level repulsion, and lies within a circular region surrounding the origin of 
    the complex plane.
    }
    \label{fig:single_spectra_0d}
\end{figure}

\begin{figure}[H]
\begin{minipage}[t]{0.22\textwidth}
\includegraphics[width=\linewidth,keepaspectratio=true]{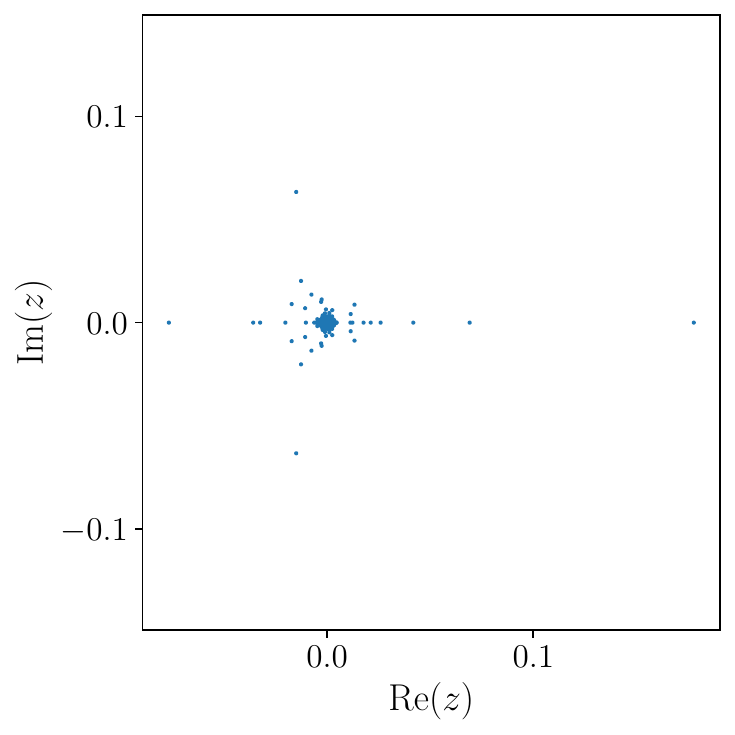}
\end{minipage}
\hspace*{\fill} 
\begin{minipage}[t]{0.258\textwidth}
\includegraphics[width=\linewidth,keepaspectratio=true]{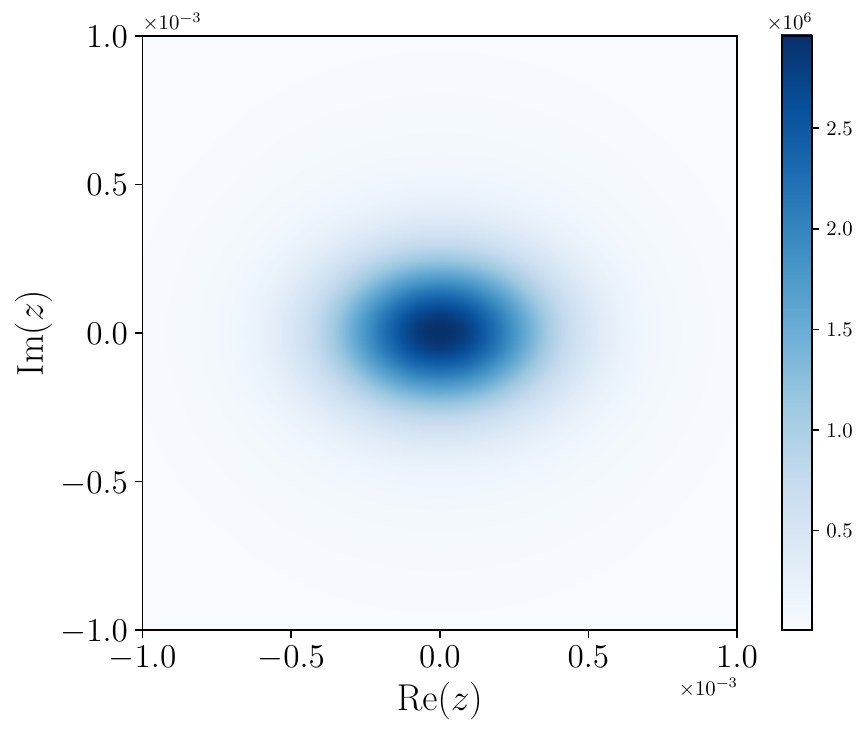}
\end{minipage}
\hspace*{\fill} 
\begin{minipage}[t]{0.22\textwidth}
\includegraphics[width=\linewidth,keepaspectratio=true]{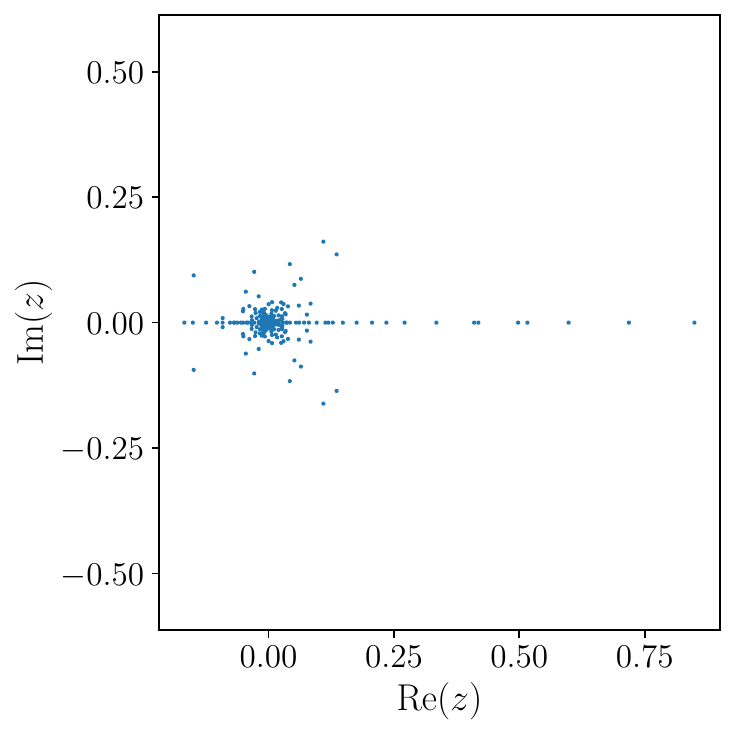}
\end{minipage}
\hspace*{\fill} 
\begin{minipage}[t]{0.258\textwidth}
\includegraphics[width=\linewidth,keepaspectratio=true]{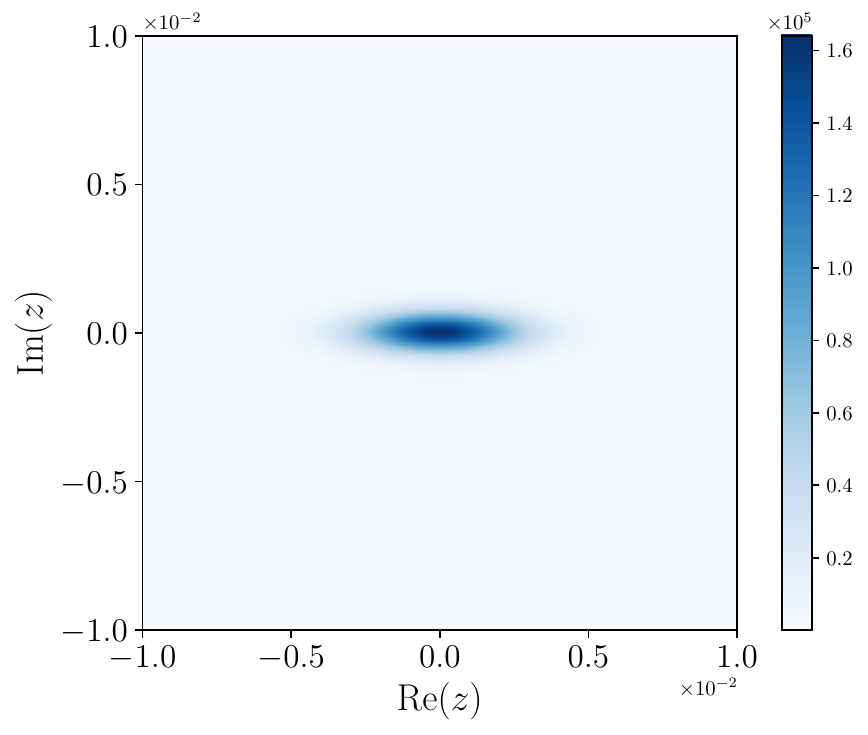}
\end{minipage}
    \caption{Spectrum of a single realization (first from left) and DOS (second from left) of the RKC model with $L=6$.  Again, we remove the eigenvalue at $z=1$.
    %
    %
    %
    Spectrum of a single realization (third from left) and DOS (forth from left) of the $U(1)$-RKC model with $L=8, S_z=4$.
    Both of the spectra are sharply peaked at the
    origin (note the scale) with tails 
    extending outwards.
    }
    \label{fig:single_spectra_1DRKC}
\end{figure}

For Kraus circuits, we include the representative examples of the RKO (\autoref{fig:single_spectra_0d}),
the RKC
and $U(1)$-RKC (\autoref{fig:single_spectra_1DRKC}).

\subsection{Random Lindbladians}\label{app:dos_lindblad}
\begin{figure}[H]
\begin{minipage}[t]{0.24\textwidth}
\includegraphics[width=\linewidth,keepaspectratio=true]{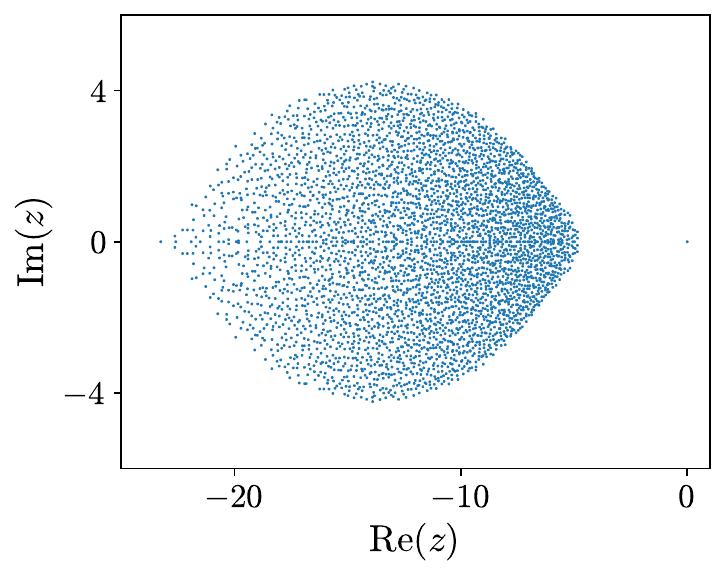}
\end{minipage}
\hspace*{\fill} 
\begin{minipage}[t]{0.24\textwidth}
\includegraphics[width=\linewidth,keepaspectratio=true]{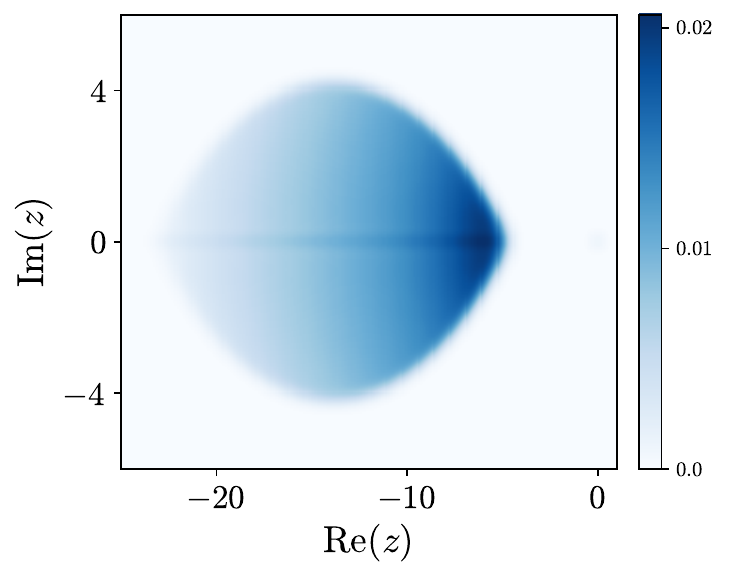}
\end{minipage}
\hspace*{\fill} 
\begin{minipage}[t]{0.24\textwidth}
\includegraphics[width=\linewidth,keepaspectratio=true]{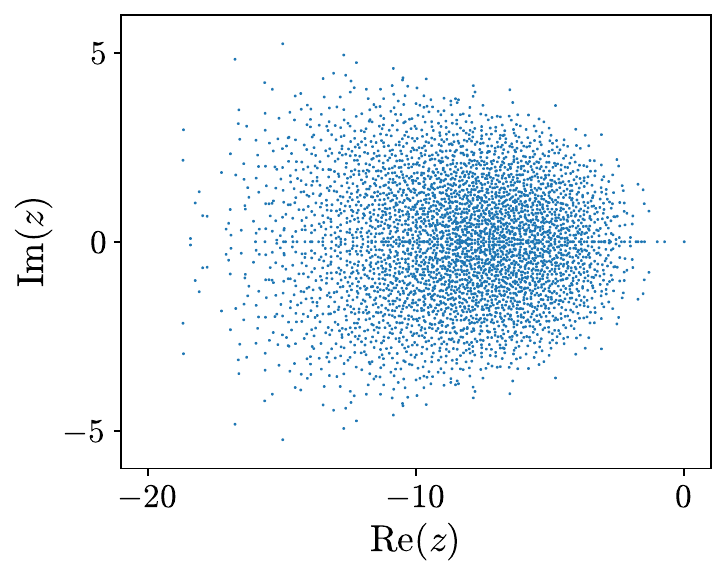}
\end{minipage}
\hspace*{\fill} 
\begin{minipage}[t]{0.24\textwidth}
\includegraphics[width=\linewidth,keepaspectratio=true]{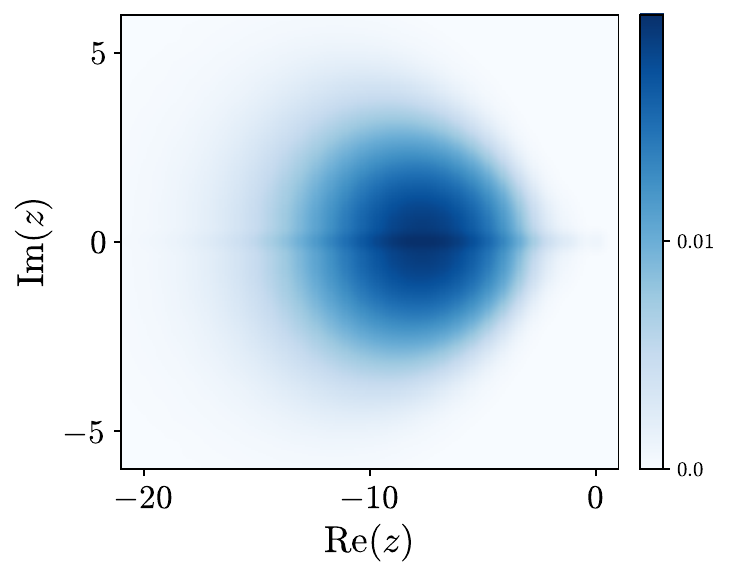}
\end{minipage}
    \caption{(a) Spectrum of a single realization and (b) the heat map of the DOS of the 0D-RL. (c) Spectrum of a single realization and (d) the heat map of the DOS of the 1D-RL model with dimension $N=4096$ and $L=6$ respectively.
    The spectra of the 0D-RL has  a shape of a lemon (c.f. \cite{denisov2019}) and gets denser to the origin. The DOS of the 1D-RL has a smooth peak at around $z=-7.5$.
    }
    \label{fig:spectra_0DRL_1DRL}
\end{figure}

\begin{figure}[H]
\begin{minipage}[t]{0.24\textwidth}
\includegraphics[width=\linewidth,keepaspectratio=true]{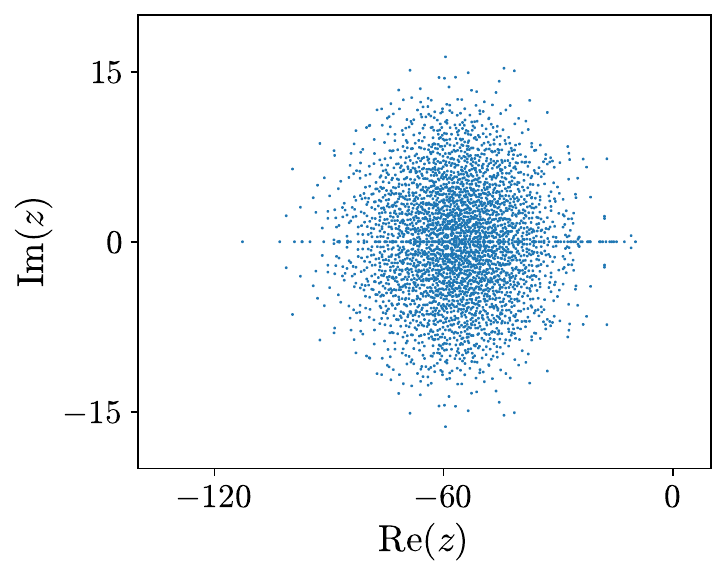}
\end{minipage}
\hspace*{\fill} 
\begin{minipage}[t]{0.24\textwidth}
\includegraphics[width=\linewidth,keepaspectratio=true]{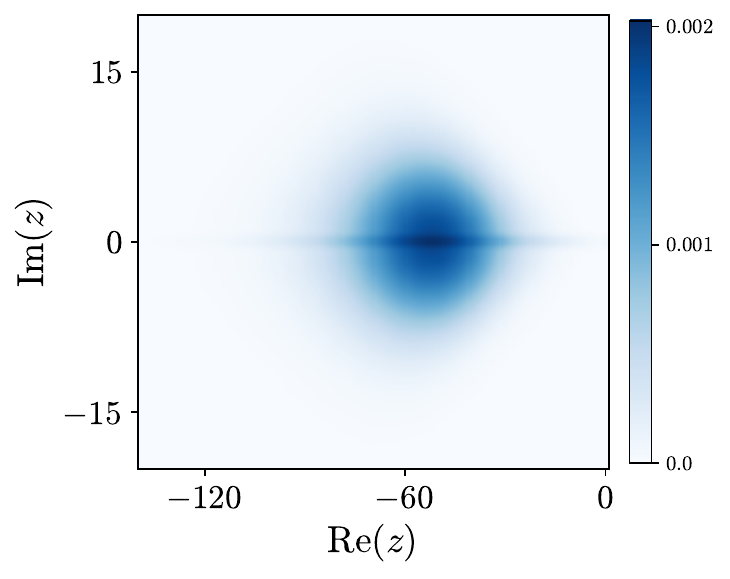}
\end{minipage}
\hspace*{\fill} 
\begin{minipage}[t]{0.24\textwidth}
\includegraphics[width=\linewidth,keepaspectratio=true]{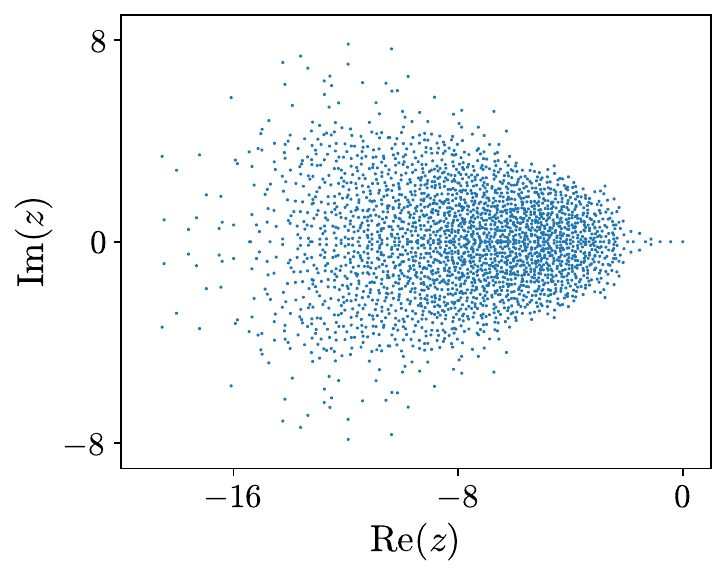}
\end{minipage}
\hspace*{\fill} 
\begin{minipage}[t]{0.24\textwidth}
\includegraphics[width=\linewidth,keepaspectratio=true]{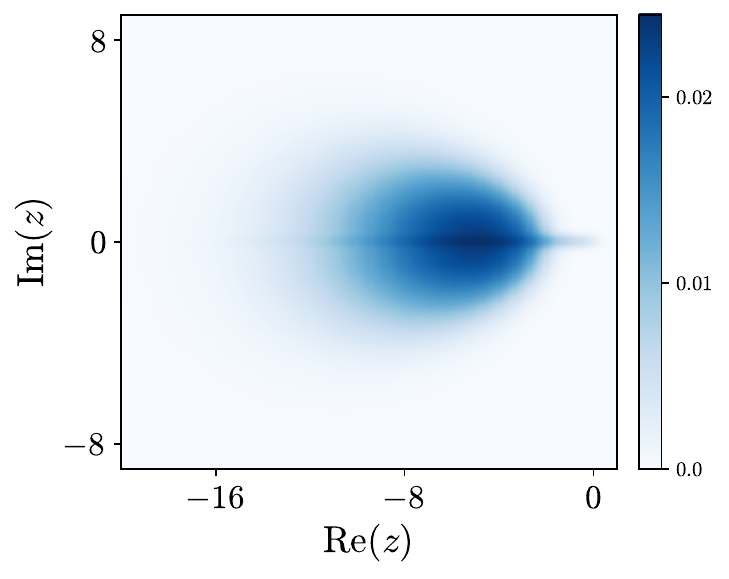}
\end{minipage}
    \caption{
    Spectrum of a single realization (first) and  the heat map of the DOS (second) of the WS-$U(1)$-RL model with weak symmetry and physical dimension $L=7$.
    The spectra has a smooth peak at around $z=-50$. Notice that for the weak symmetry case, there is no steady state, i.e. there is not an eigenvalue with the value $z=0$.
    Spectrum of a single realization (third) and  the heat map of the DOS (fourth) of the SS-$U(1)$-RL model with weak symmetry and physical dimension $L=5$.
    The spectra has a smooth peak at around $z=-5$.
    }
    \label{fig:single_spectra_1d_lindblad}
\end{figure}

\begin{figure}[H]
\begin{minipage}[t]{0.19\textwidth}
\includegraphics[width=\linewidth,keepaspectratio=true]{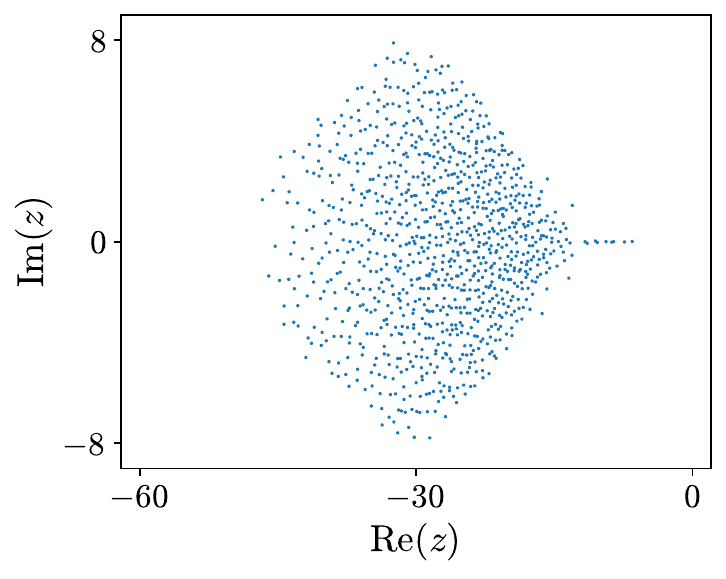}
\end{minipage}
\hspace*{\fill} 
\begin{minipage}[t]{0.19\textwidth}
\includegraphics[width=\linewidth,keepaspectratio=true]{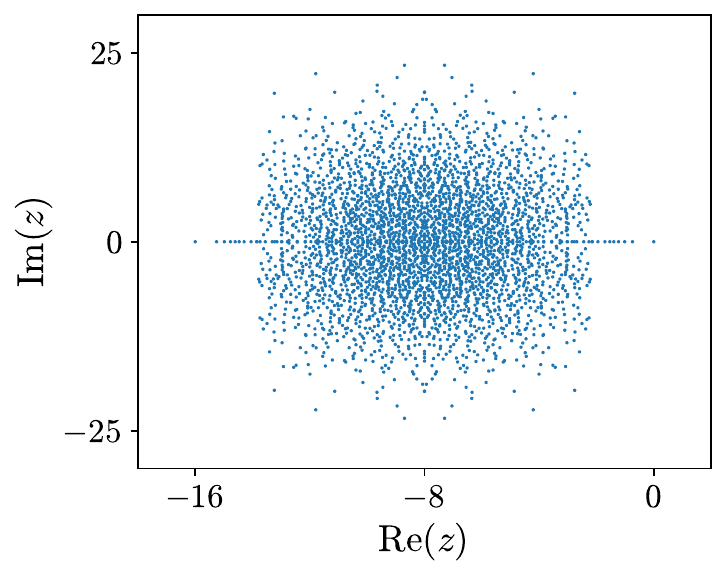}
\end{minipage}
\hspace*{\fill} 
\begin{minipage}[t]{0.19\textwidth}
\includegraphics[width=\linewidth,keepaspectratio=true]{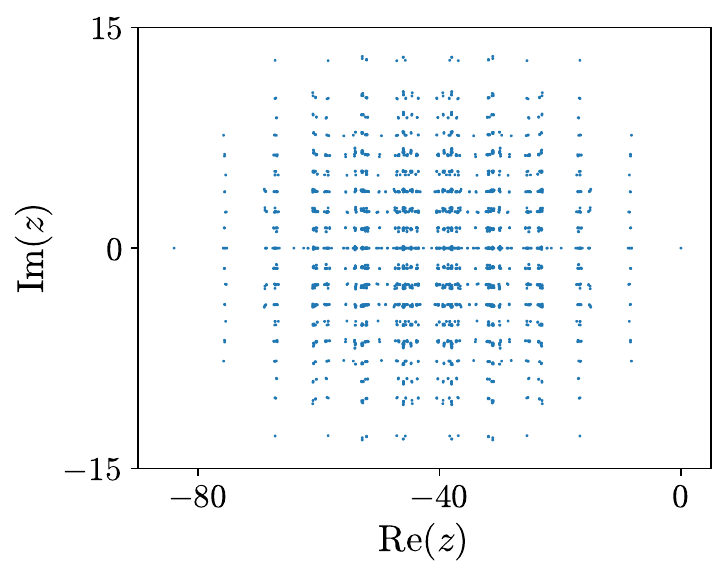}
\end{minipage}
\hspace*{\fill} 
\begin{minipage}[t]{0.19\textwidth}
\includegraphics[width=\linewidth,keepaspectratio=true]{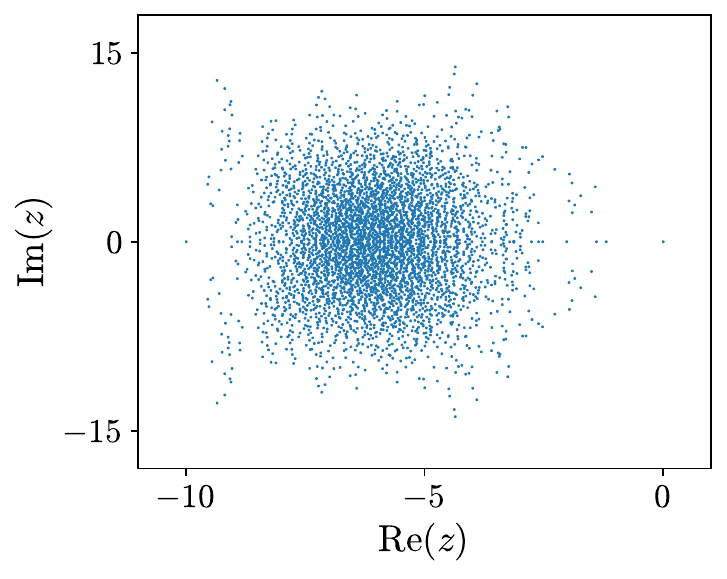}
\end{minipage}
\hspace*{\fill} 
\begin{minipage}[t]{0.19\textwidth}
\includegraphics[width=\linewidth,keepaspectratio=true]{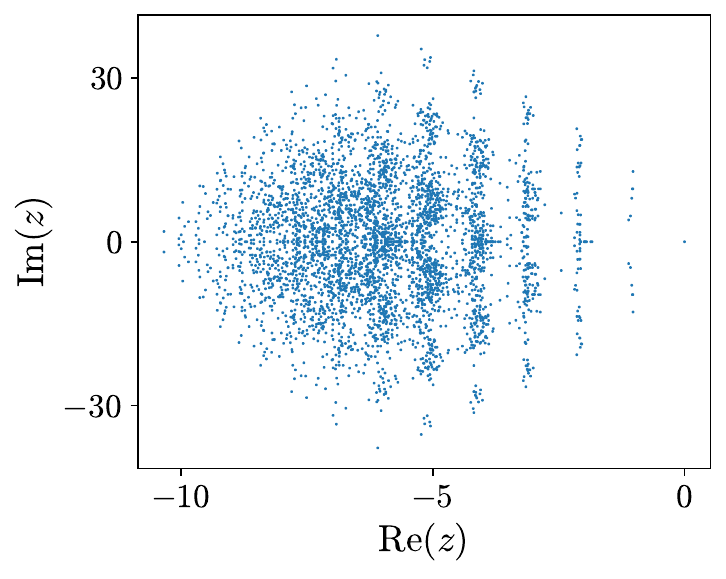}
\end{minipage}
\hspace*{\fill} 
\begin{minipage}[t]{0.19\textwidth}
\includegraphics[width=\linewidth,keepaspectratio=true]{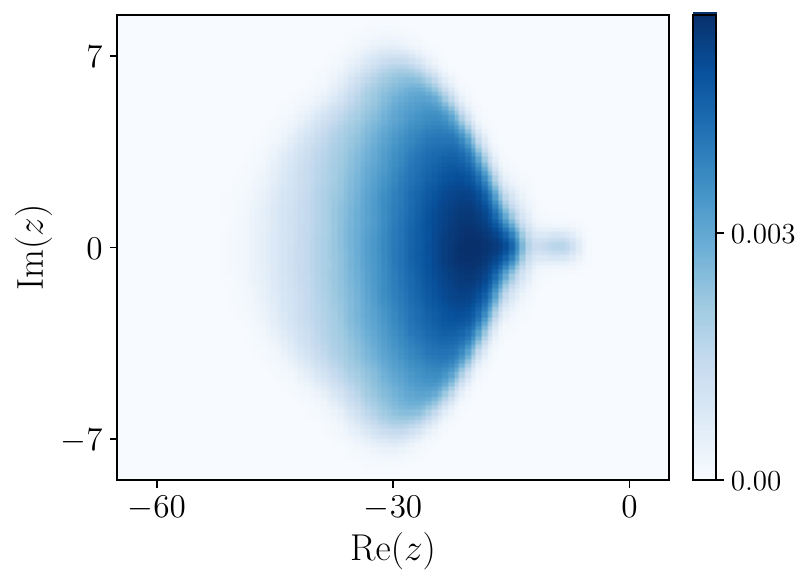}
\end{minipage}
\hspace*{\fill} 
\begin{minipage}[t]{0.19\textwidth}
\includegraphics[width=\linewidth,keepaspectratio=true]{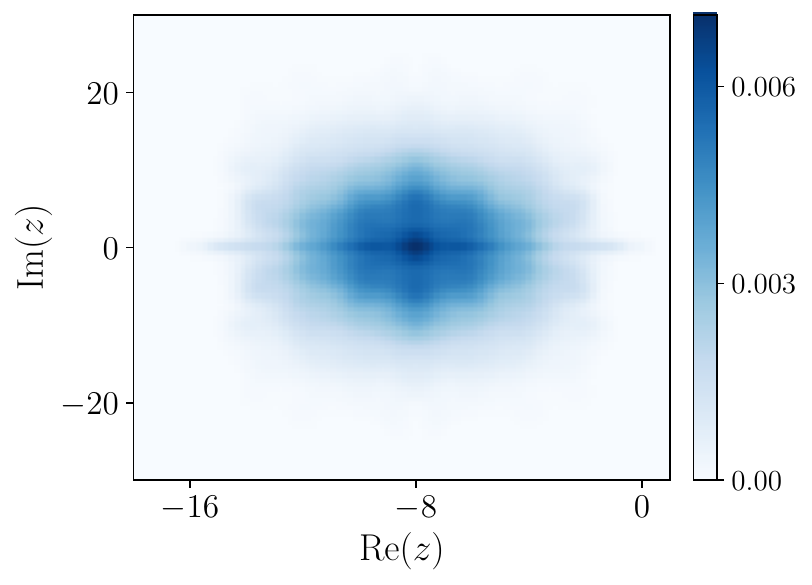}
\end{minipage}
\hspace*{\fill} 
\begin{minipage}[t]{0.19\textwidth}
\includegraphics[width=\linewidth,keepaspectratio=true]{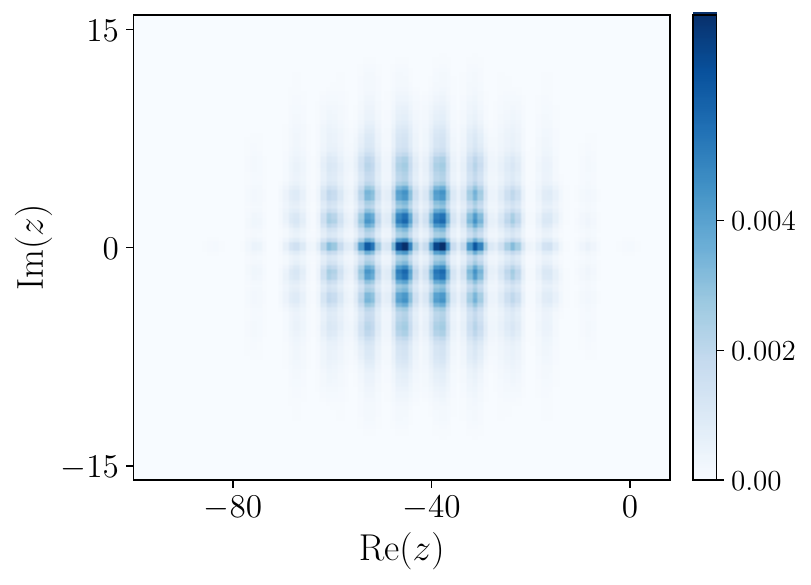}
\end{minipage}
\hspace*{\fill} 
\begin{minipage}[t]{0.19\textwidth}
\includegraphics[width=\linewidth,keepaspectratio=true]{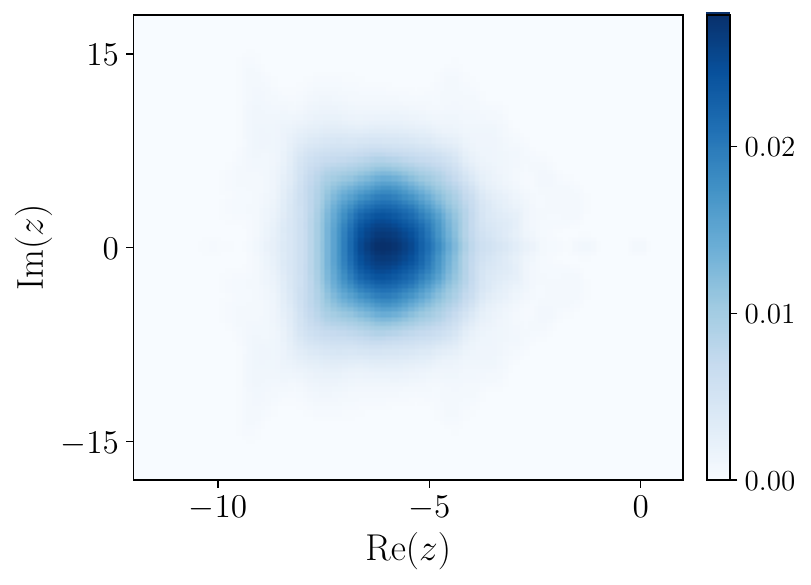}
\end{minipage}
\hspace*{\fill} 
\begin{minipage}[t]{0.19\textwidth}
\includegraphics[width=\linewidth,keepaspectratio=true]{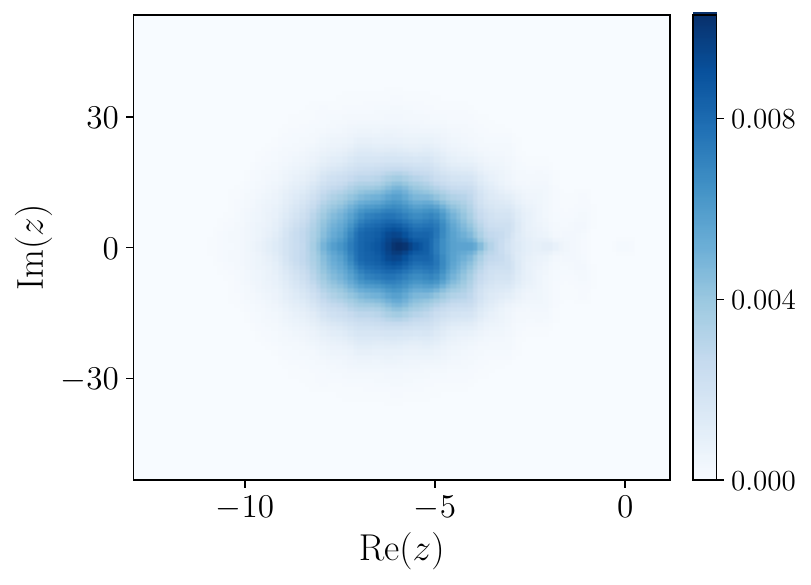}
\end{minipage}
    \caption{
    Spectrum of a single realization (first row) and  the heat map of the DOS (second row) of, from left to right, the SYK-L with $L=12$, dXXZ with $L=7$, dXX with $L=7$, dIsing-Chaos with $L=6$ and dIsing-MBL with $L=6$.
    }
    \label{fig:single_spectra_lindblad_models}
\end{figure}

\begin{figure}[H]
\begin{minipage}[t]{0.3\textwidth}
\includegraphics[width=\linewidth,keepaspectratio=true]{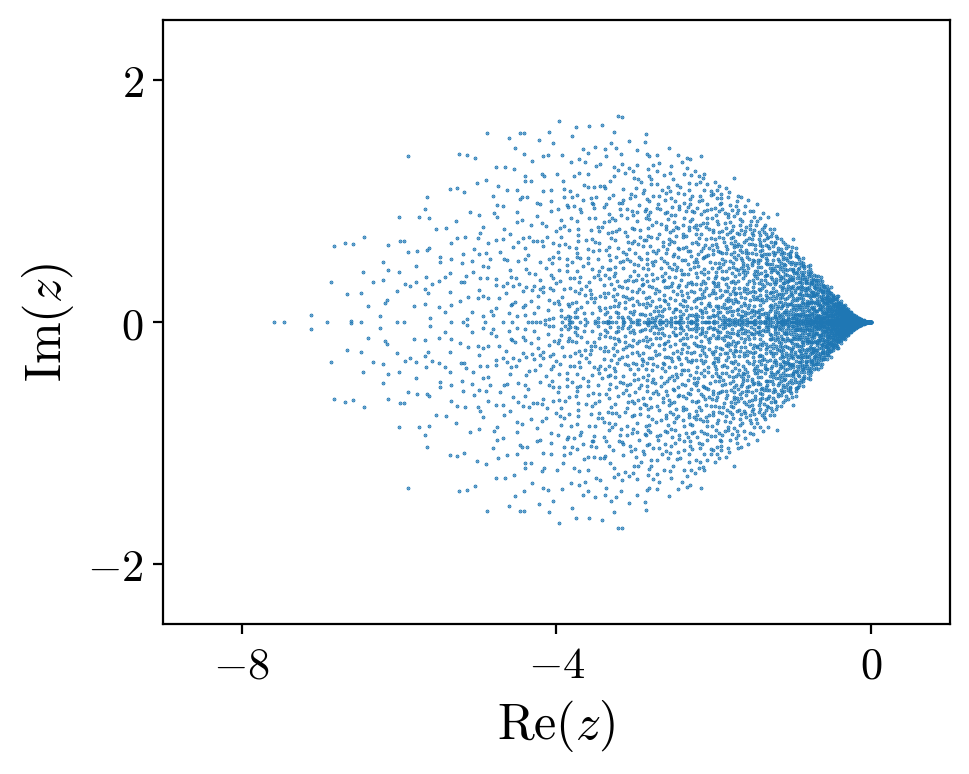}
\end{minipage}
\hspace*{\fill} 
\begin{minipage}[t]{0.3\textwidth}
\includegraphics[width=\linewidth,keepaspectratio=true]{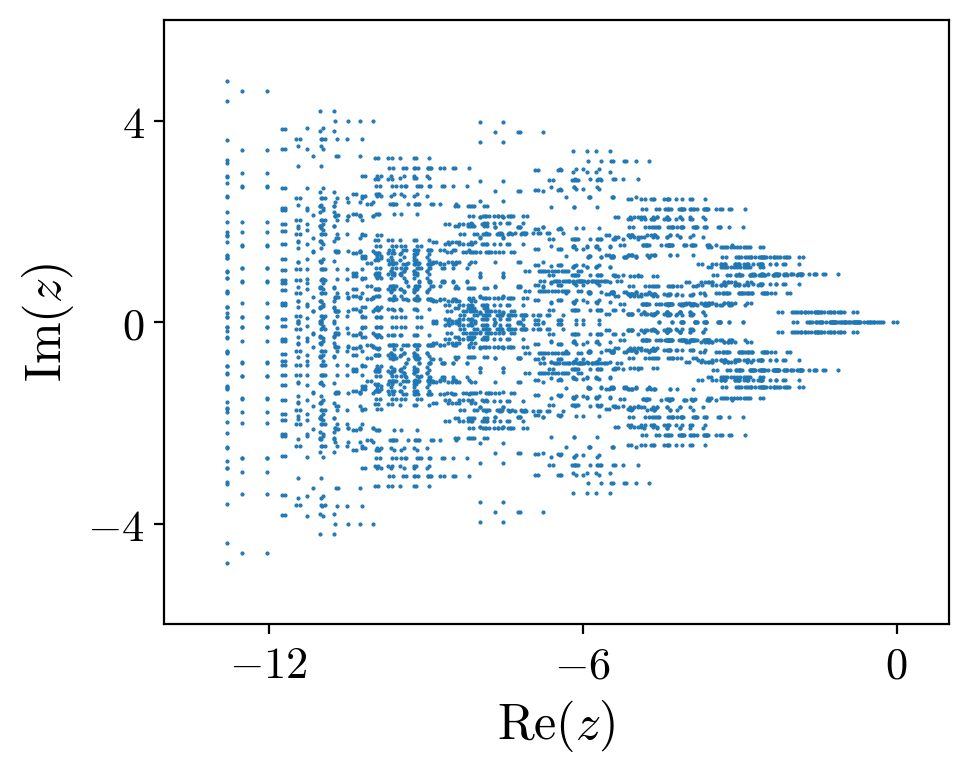}
\end{minipage}
\hspace*{\fill} 
\begin{minipage}[t]{0.3\textwidth}
\includegraphics[width=\linewidth,keepaspectratio=true]{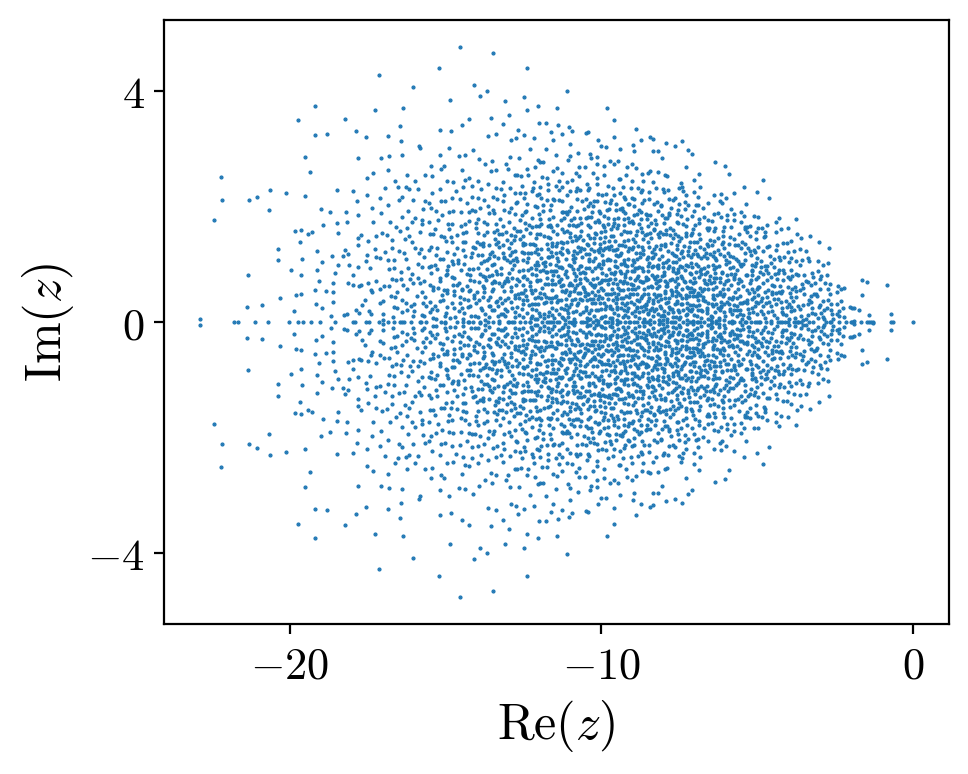}
\end{minipage}
\hspace*{\fill} 
\begin{minipage}[t]{0.3\textwidth}
\includegraphics[width=\linewidth,keepaspectratio=true]{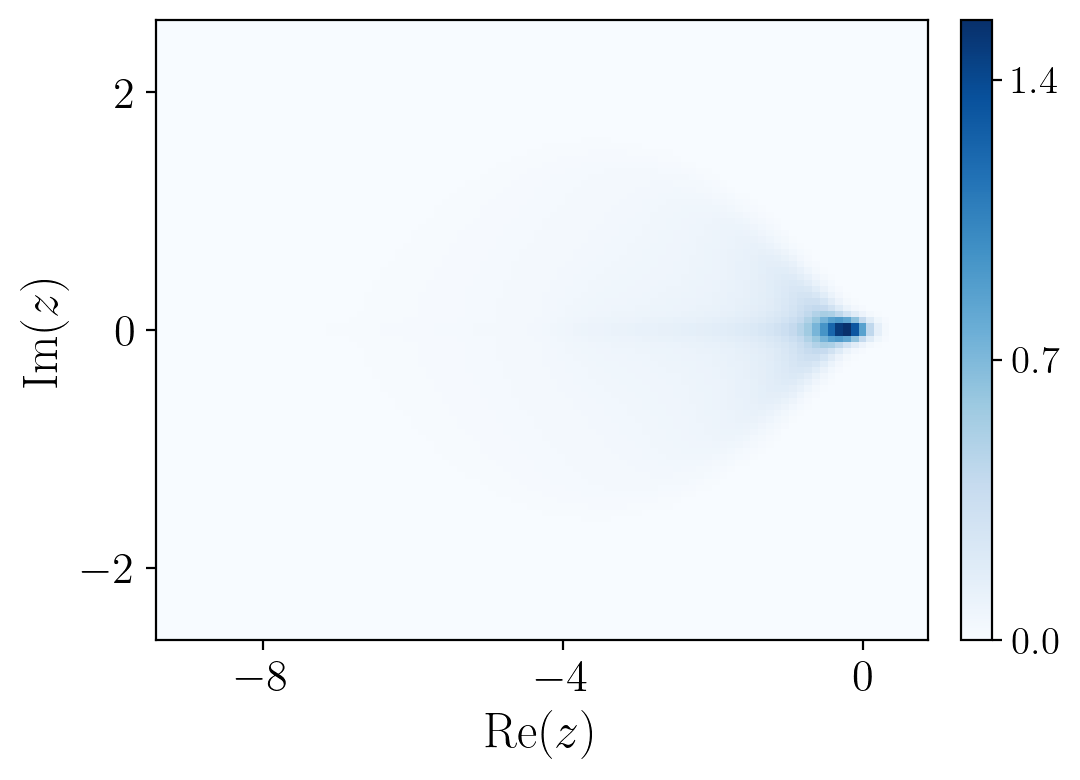}
\end{minipage}
\hspace*{\fill} 
\begin{minipage}[t]{0.3\textwidth}
\includegraphics[width=\linewidth,keepaspectratio=true]{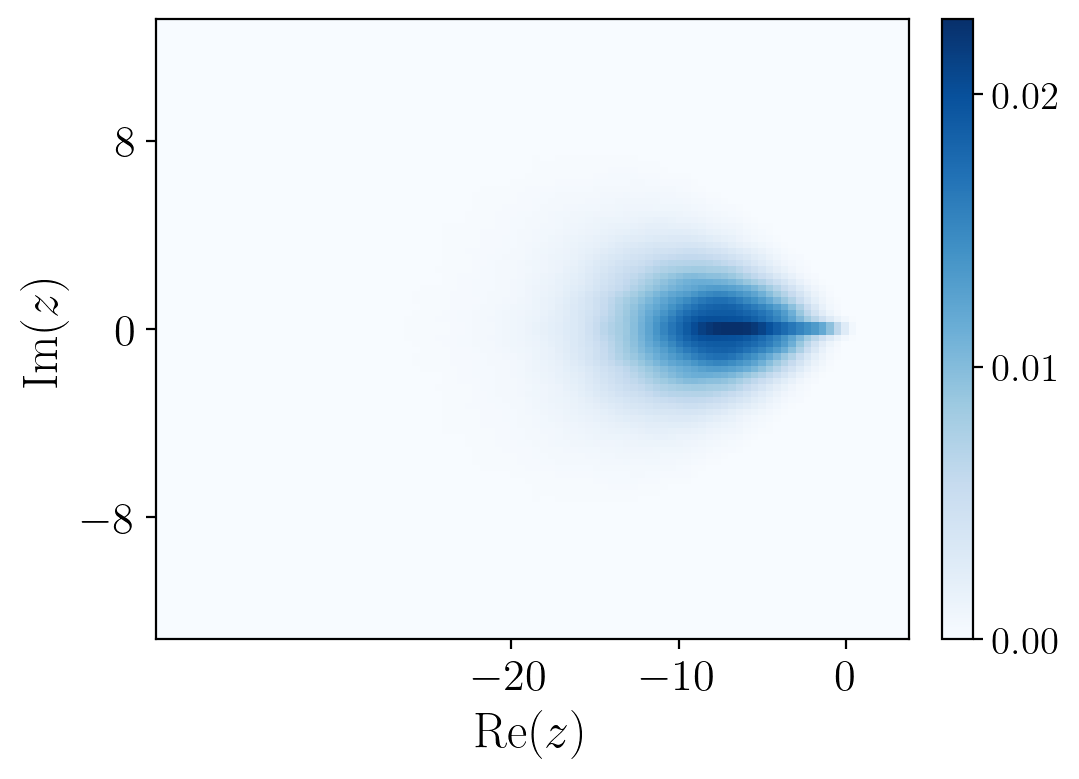}
\end{minipage}
\hspace*{\fill} 
\begin{minipage}[t]{0.3\textwidth}
\includegraphics[width=\linewidth,keepaspectratio=true]{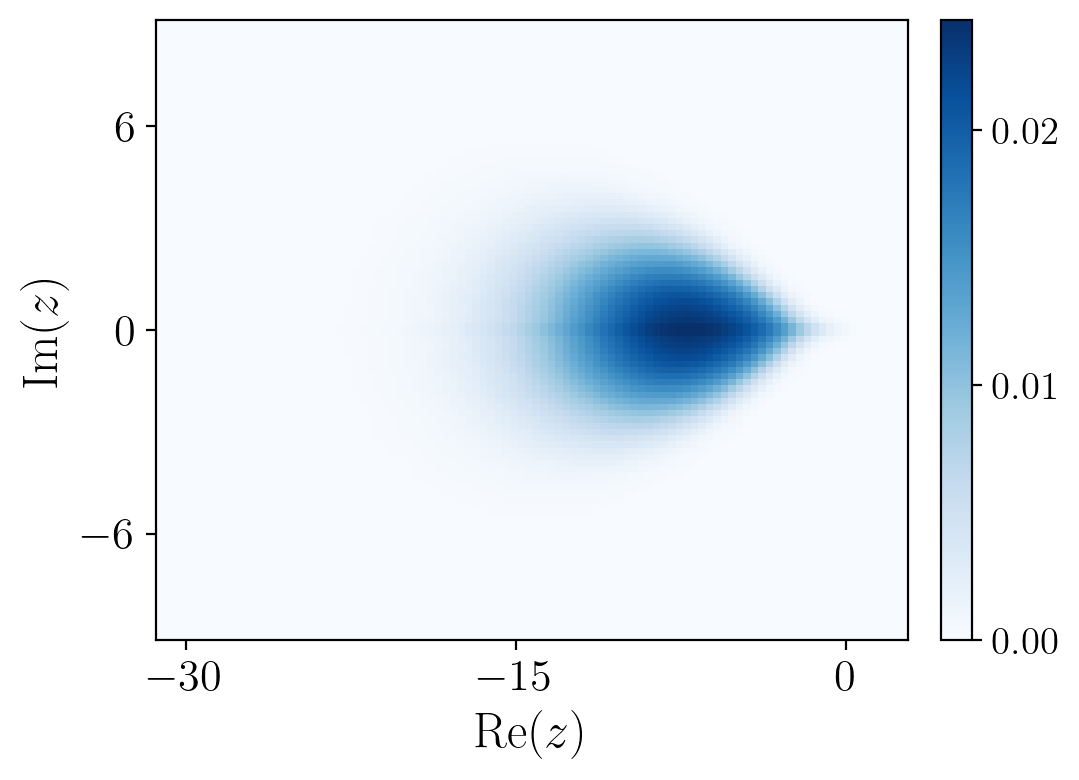}
\end{minipage}
    \caption{
    Spectrum of a single realization (first row) and  the heat map of the DOS (second row) of, from left to right, the 0D-RL-JO, the 1D-RL-1JO and the 1D-RL-2JO with $L=6$.
    }
    \label{fig:single_spectra_JO_models}
\end{figure}

For Lindbladians, we include the all the models studied, from (\autoref{fig:spectra_0DRL_1DRL}) to (\autoref{fig:single_spectra_JO_models}).

\section{Nearest neighbour spacing distribution}
For a given complex eigenvalue $z_n$ and its $m$-th nearest neighbour $z_n^{(m)}$, we define the $m$-th spacing distance of $z_n$ to be  
\be
s^{(m)}_{n} = \left|z^{(m)}_n -z_n \right| \, .
\ee
The nearest neighbour spacing distribution (NNSD) is the set $\left\{ s^{(1)} \right\}$, and is used in non-Hermitian systems~\cite{sommers1988} to
diagnose chaos in OQMBS, as in Hermitian systems~\cite{mehta,bohigas1984characterization}. 
%
%
%
%
Unfolding procedure has to be applied to the spectrum so that variation of spectral density in the complex plane is removed. 
To this end, we unfold the spectrum $\{z_n\}$ in three steps: (i) Apply a conformal transformation described in Appendix \ref{sec:unfolding} to obtain $\{ z'_n\}$, which in particular removes sharp peaks in the DOS;
(ii) Obtain and rescale the NNSD $\left\{s^{(1)}_n \right\}_R$ of $\{ z'_n\}_R$  with a rescaling factor dependent on local spectral density following~\cite{ueda2020univclass}:
\begin{equation}\label{eqn:nn_spacing_def}
    d^{(m)}_n  = s^{(1)}_n \sqrt{\frac{m}{\pi {s^{(m)}_n}^2}} \,;
\end{equation}
(iii) apply a filter $f(z_n; \alpha, \mu)$ as described in Appendix \ref{app:filtering} to keep only the transformed eigenvalues in some region $R$ defined by $\alpha, \mu$ which has approximately uniform spectral density.  Specifically, we average over many realizations by computing:

\begin{equation}\label{eqn:nn_spacing_filter_def}
    P(d^{(m)}) = \frac{\Big\langle 
  \sum_n  \delta(d^{(m)} -  d_n^{(m)} ) \times  f(z_n ; \alpha, \mu)
    \Big\rangle}{\Big\langle \sum_n 
    f(z_n ; \alpha, \mu)
    \Big\rangle}
    \,.
\end{equation}
%

%
%
%
\subsection{Random Kraus circuits}
\begin{figure}[H]
\begin{minipage}[t]{0.31\textwidth}
\includegraphics[width=\linewidth,keepaspectratio=true]{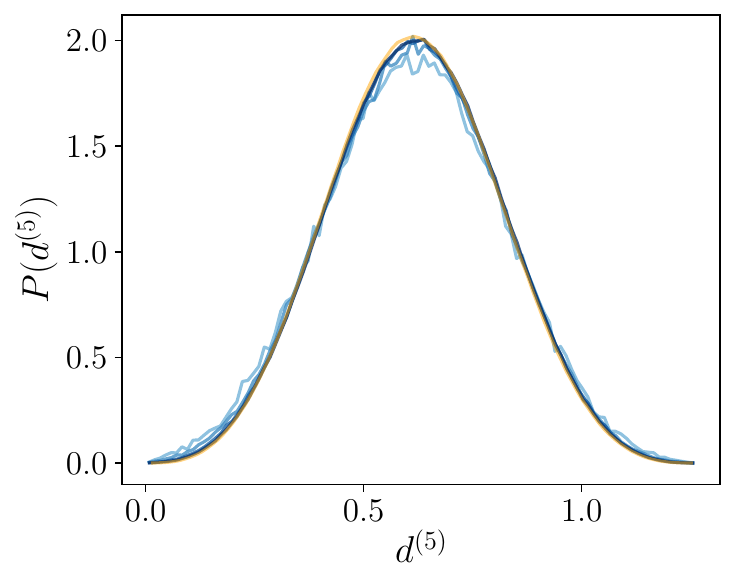}
\end{minipage}
\hspace*{\fill} 
\begin{minipage}[t]{0.31\textwidth}
\includegraphics[width=\linewidth,keepaspectratio=true]{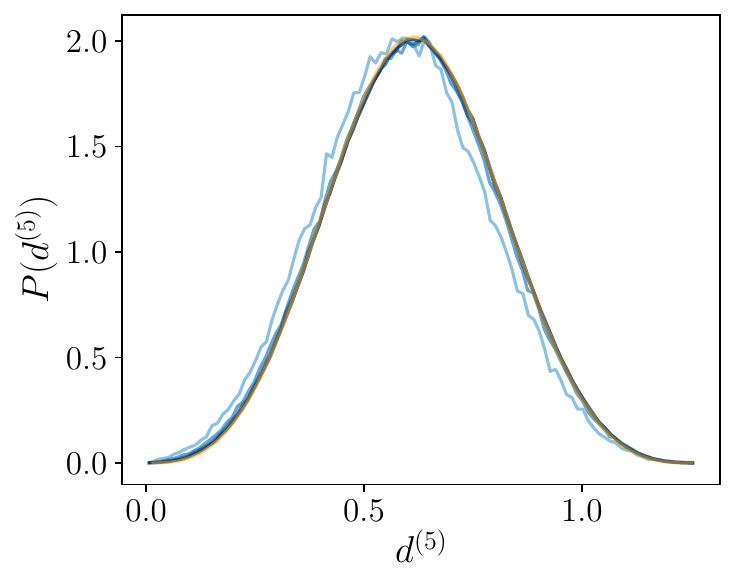}
\end{minipage}
\hspace*{\fill} 
\begin{minipage}[t]{0.31\textwidth}
\includegraphics[width=\linewidth,keepaspectratio=true]{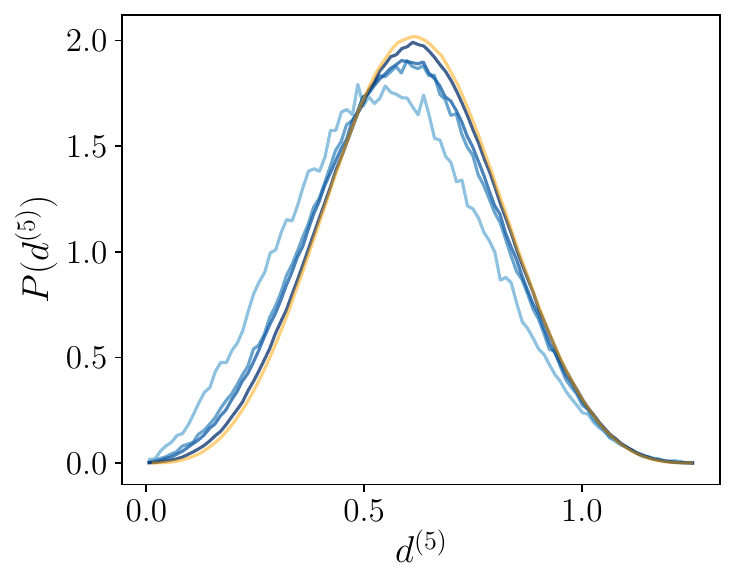}
\end{minipage}
    \caption{Left:
    The NNSD~\eqref{eqn:nn_spacing_def} with $m=5$ for the RKO, where we have implemented step (ii) and (iii), with increasing $N = 4^3, 4^5, 4^6$ from light to dark blue.  If we only implemented step (ii), we would see deviations from the RMT spacing at $d^{(5)} > 1$ due to
    a finite-size edge effect in the DOS.
    Middle: 
     The NNSD~\eqref{eqn:nn_spacing_def} with $m=5$ for the RKC, where we have implemented step (i)-(iii), with increasing $L \in [3,6]$ from light to dark blue.
     For both cases, the sample size is $5000$.
    Right:
    The NNSD~\eqref{eqn:nn_spacing_def} after steps (i)-(iii) with $m=5$ 
    for the $U(1)$-RKC with increasing $L \in [5,8]$ in darker blue. The
    sample size is $5000$.
    }
    \label{fig:ko_nn_spacing}
\end{figure}

Here we provide the NNSD for the RKO, RKC and $U(1)$-RKC.  
For RKO and RKC in Fig.~\ref{fig:ko_nn_spacing}, NNSD have converged in the accessible system sizes
and exhibit universal
eigenvalue correlations of the RMT GinUE NNSD. 
Consequently, we expect the DSFF to exhibit universal behavior at late complex time.  Crucially, we find that for the NNSD to match that of RMT, steps (i)-(iii) are 
necessary.  This in part justifies the need to use both filtering and unfolding.
For $U(1)$-RKC in Fig.~\ref{fig:ko_nn_spacing} right, NNSD  has not yet 
converged in the accessible system size, but the distribution is clearly approaching the RMT GinUE NNSD.
%
%

%

\subsection{Random Lindbladians}\label{app:C_rl}
\begin{figure}[H]
\begin{minipage}[t]{0.22\textwidth}
\includegraphics[width=\linewidth,keepaspectratio=true]{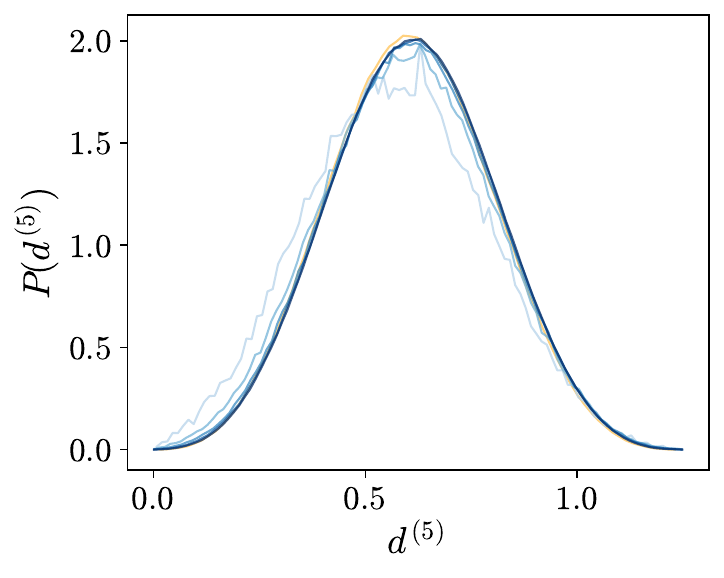}
\end{minipage}
\hspace*{\fill} 
\begin{minipage}[t]{0.22\textwidth}
\includegraphics[width=\linewidth,keepaspectratio=true]{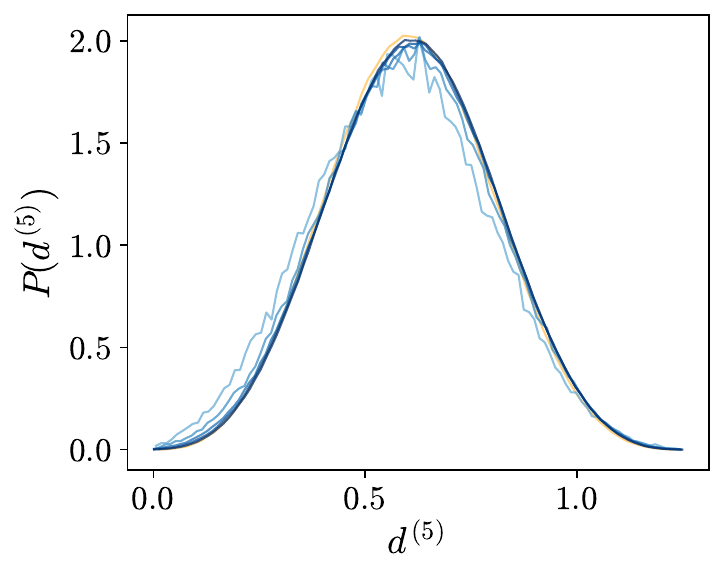}
\end{minipage}
\hspace*{\fill} 
\begin{minipage}[t]{0.22\textwidth}
\includegraphics[width=\linewidth,keepaspectratio=true]{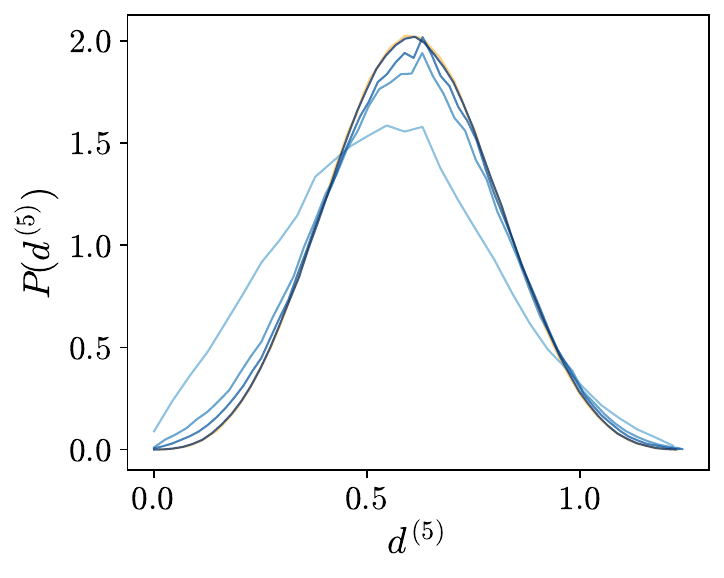}
\end{minipage}
\hspace*{\fill} 
\begin{minipage}[t]{0.22\textwidth}
\includegraphics[width=\linewidth,keepaspectratio=true]{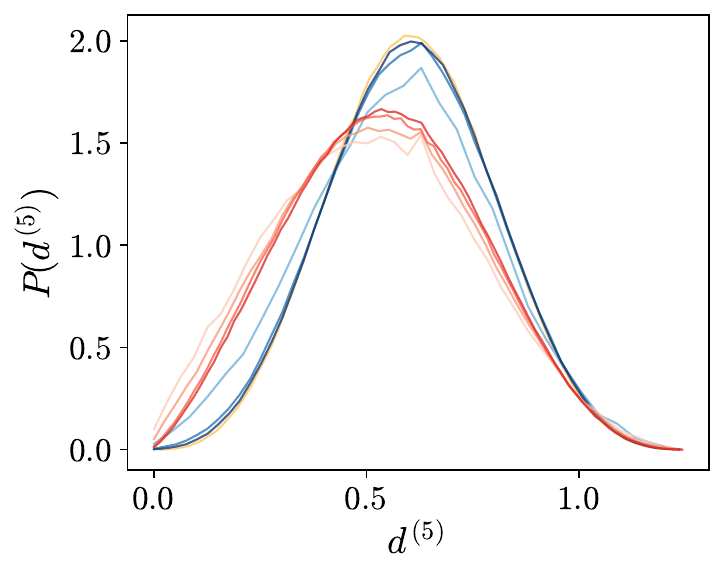}
\end{minipage}
\hspace*{\fill} 
\begin{minipage}[t]{0.19\textwidth}
\includegraphics[width=\linewidth,keepaspectratio=true]{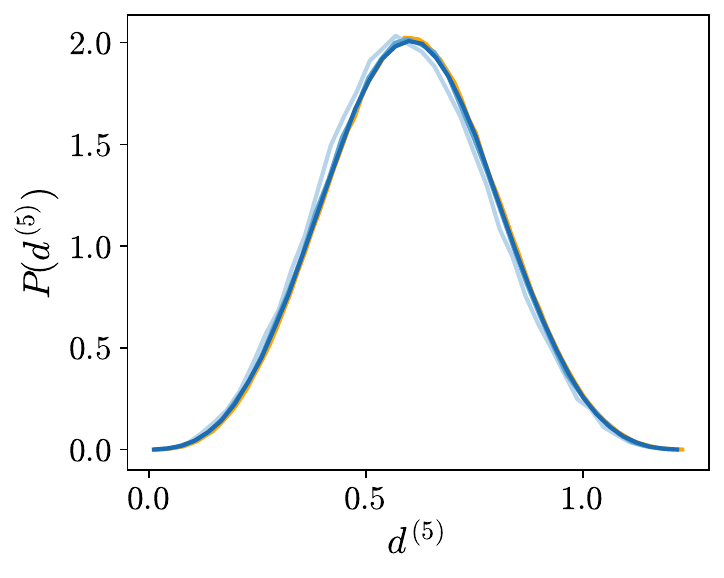}
\end{minipage}
\hspace*{\fill} 
\begin{minipage}[t]{0.19\textwidth}
\includegraphics[width=\linewidth,keepaspectratio=true]{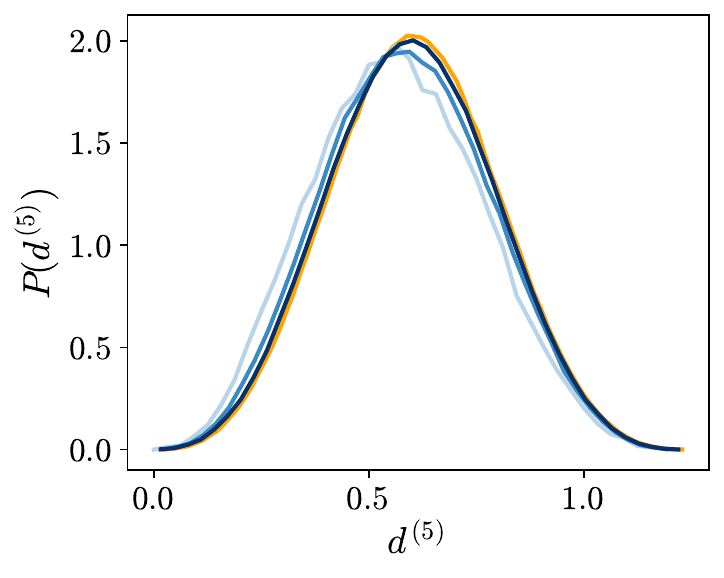}
\end{minipage}
\hspace*{\fill} 
\begin{minipage}[t]{0.19\textwidth}
\includegraphics[width=\linewidth,keepaspectratio=true]{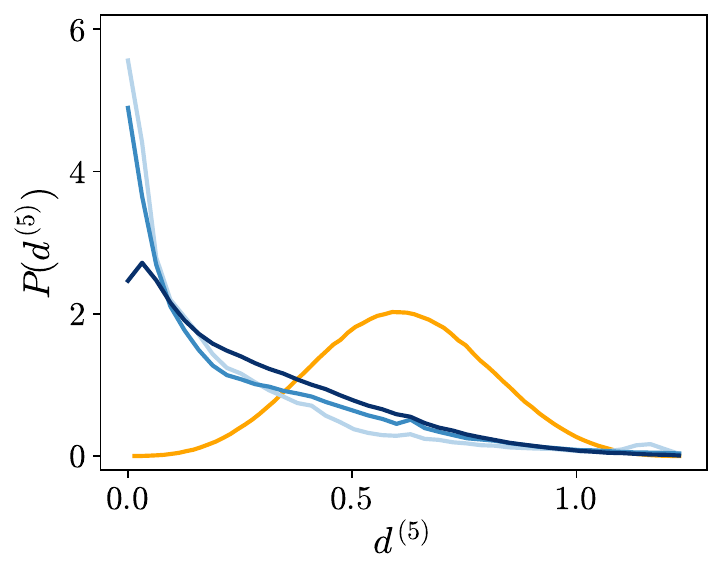}
\end{minipage}
\hspace*{\fill} 
\begin{minipage}[t]{0.19\textwidth}
\includegraphics[width=\linewidth,keepaspectratio=true]{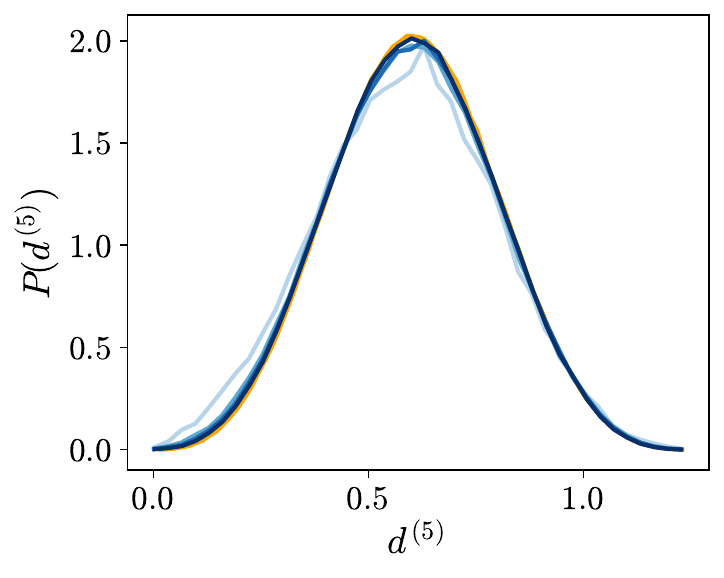}
\end{minipage}
\hspace*{\fill} 
\begin{minipage}[t]{0.19\textwidth}
\includegraphics[width=\linewidth,keepaspectratio=true]{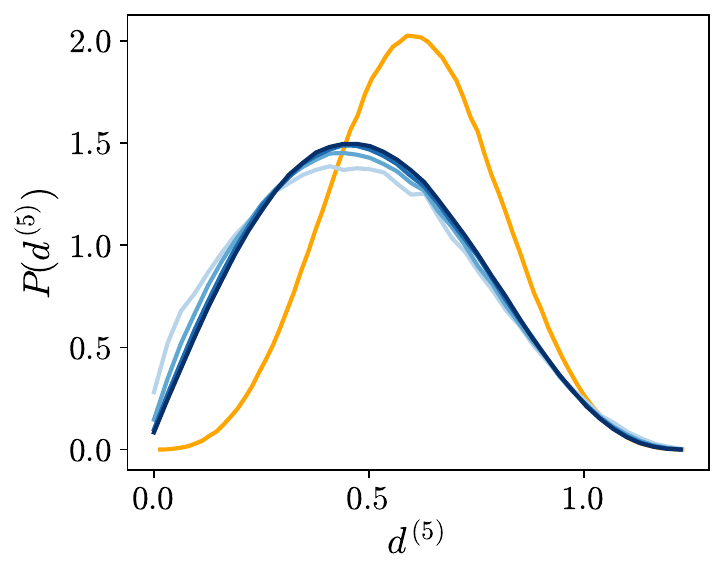}
\end{minipage}
    \caption{
    The NNSD~\eqref{eqn:nn_spacing_def} with $m=5$ for (a) the 0D-RL with $L \in [3,7]$,  (b) the 1D-RL with $L \in [3,7]$, (c) the WS-$U(1)$-RL with the symmetry sector $M=L-2$ and $L \in [5,8]$, (d) for the $q=3$ ($q=2$) SS-$U(1)$-RL with $L \in [4,6]$ ($L \in [6,9]$) in blue (red) and the symmetry section $S_z=0$,
    (e) SYK-L with $L/2 \in [4,6]$ and the symmetry section $p=1$
    (f) dXXZ with $L/2 \in [4,6]$ and the symmetry section
    (g) dXX
    (h) dIsing-Chaos
    and (l) dIsing-MBL. For all cases, the system size increases from light to dark blue, and the sample size is $2000$.
    }
    \label{fig:lindblad_nn_spacing}
\end{figure}

Here we provide the NNSD for the 0DRL, 1DRL, WS-$U(1)$-RL, and SS-$U(1)$-RL in Fig.~\ref{fig:lindblad_nn_spacing}, NNSD have converged in the accessible system sizes and exhibit universal eigenvalue correlations of the RMT GinUE NNSD. 

\begin{figure}[H]
\begin{minipage}[t]{0.3\textwidth}
\includegraphics[width=\linewidth,keepaspectratio=true]{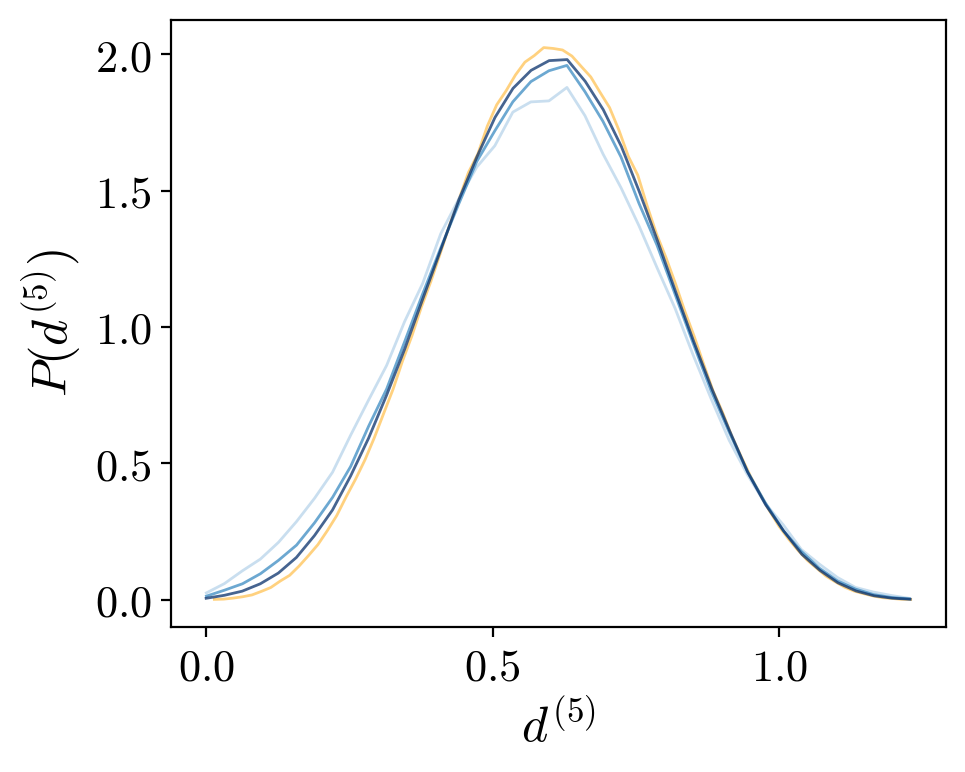}
\end{minipage}
\hspace*{\fill} 
\begin{minipage}[t]{0.3\textwidth}
\includegraphics[width=\linewidth,keepaspectratio=true]{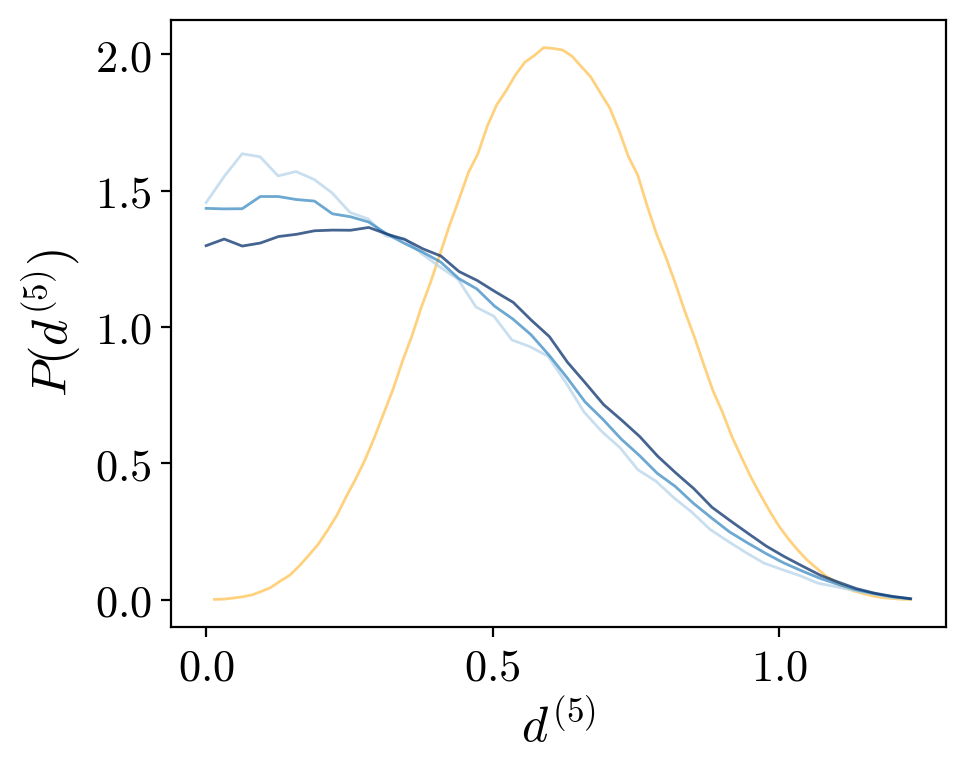}
\end{minipage}
\hspace*{\fill} 
\begin{minipage}[t]{0.3\textwidth}
\includegraphics[width=\linewidth,keepaspectratio=true]{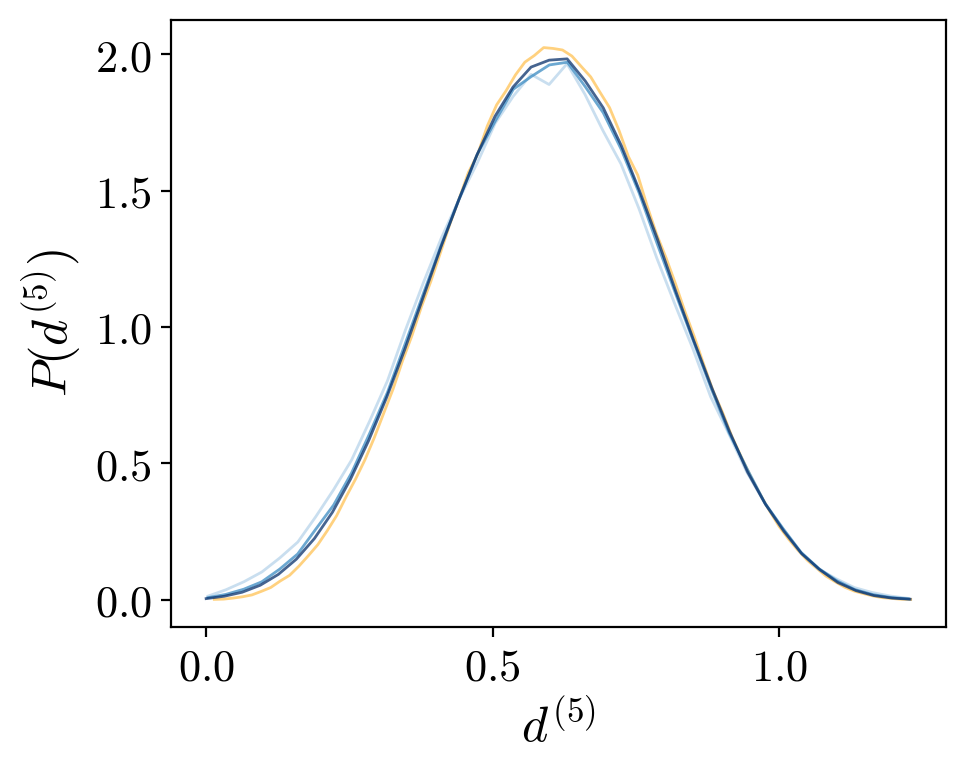}
\end{minipage}
    \caption{
    The NNSD~\eqref{eqn:nn_spacing_def} with $m=5$ for (a) the 0D-RL-JO, (b) the 1D-RL-1JO and (c) the 1D-RL-2JO where we have implemented step (i-ii), with increasing $L \in [4,6]$ from light to dark blue.
    }
    \label{fig:JO_nn_spacing}
\end{figure}
    
\section{Complex spacing ratio} 
We consider the complex spacing ratios (CSR) defined as~\cite{sa2019a}
\begin{equation}\label{eqn:complex_ratio_def}
     r_m = \frac{z_m^{(1)} - z_m}{z_m^{(2)} - z_m} \,,
\end{equation}
where $z_m^{(1)}$ and $z_m^{(2)}$ represent the nearest- and next-nearest-neighboring 
eigenvalues to $z_m$, respectively. 
The CSR has the advantage that no unfolding procedure is required. However, we find that the convergence of CSR (and related quantities) is slower. 

\begin{figure}[H]
%
\begin{minipage}[t]{0.24\textwidth}
\includegraphics[width=\linewidth,keepaspectratio=true]{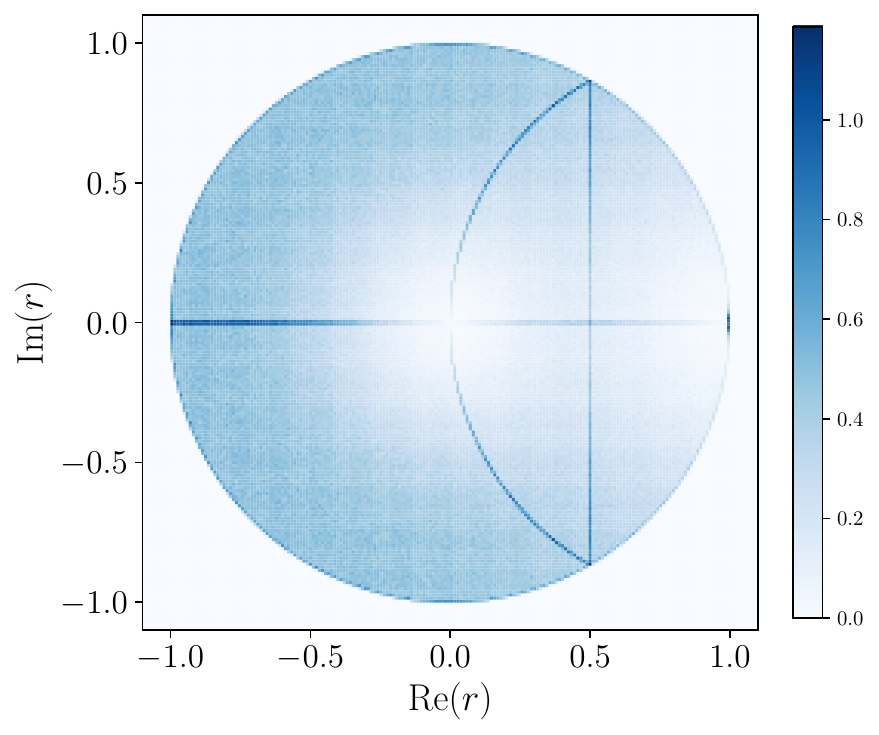}
\end{minipage}
%
%
\begin{minipage}[t]{0.24\textwidth}
\includegraphics[width=\linewidth,keepaspectratio=true]{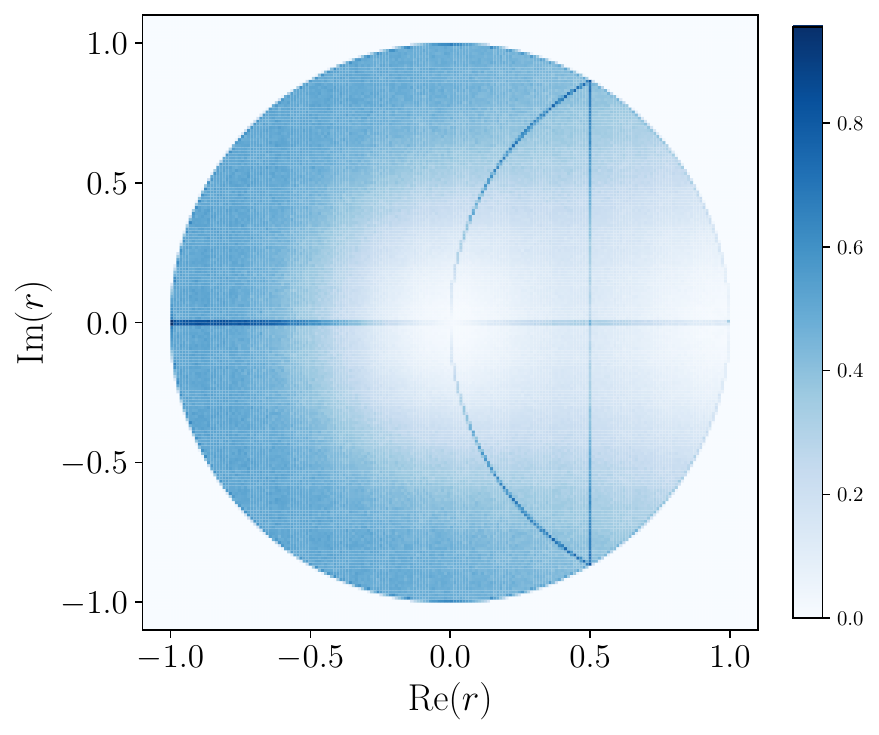}
\end{minipage}
%
%
%
\begin{minipage}[t]{0.24\textwidth}
\includegraphics[width=\linewidth,keepaspectratio=true]{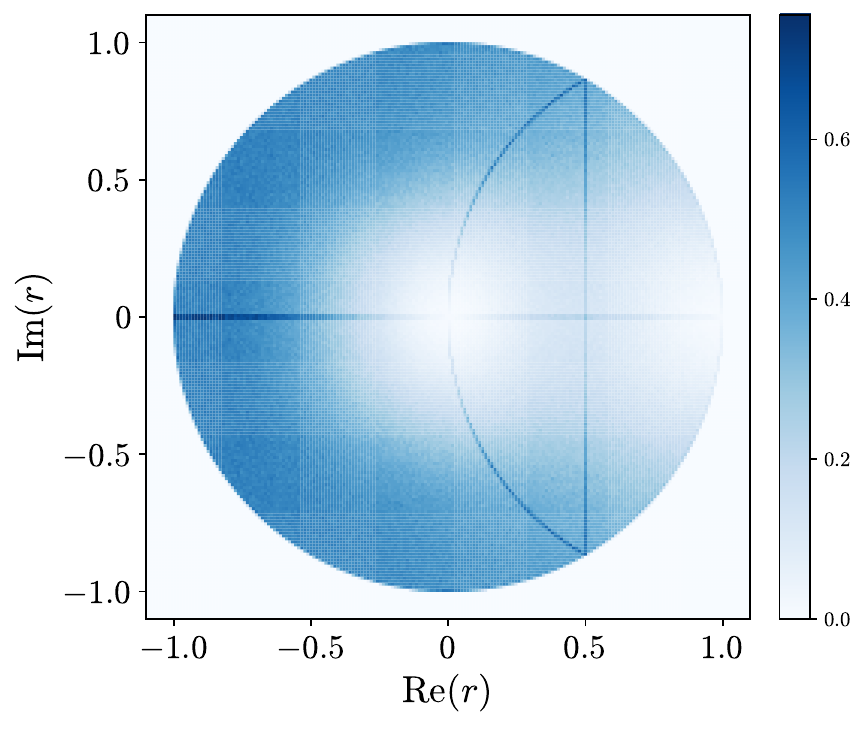}
\end{minipage}
%
%
\begin{minipage}[t]{0.24\textwidth}
\includegraphics[width=\linewidth,keepaspectratio=true]{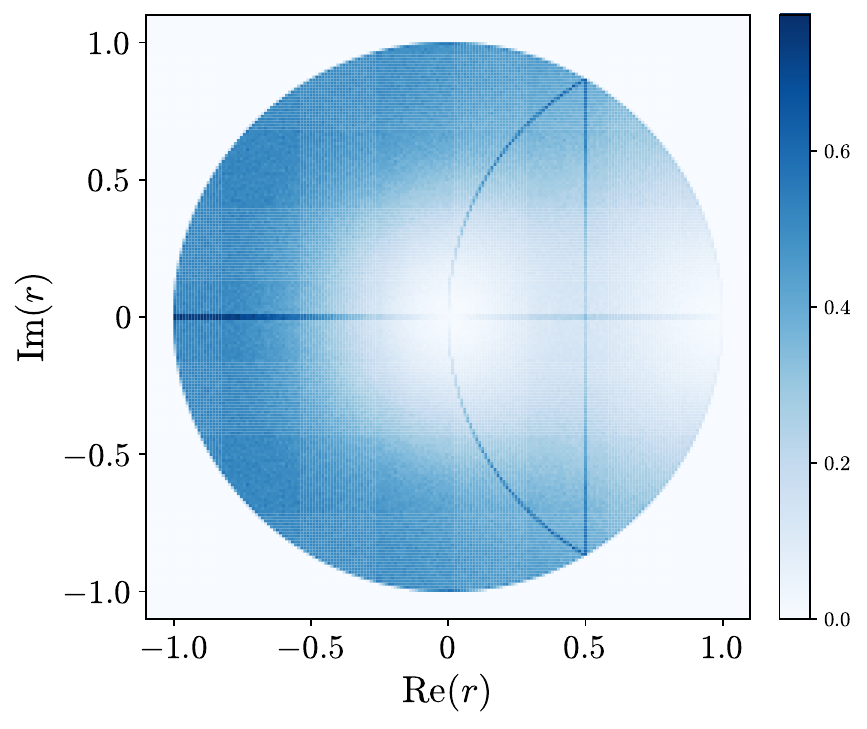}
\end{minipage}
    \caption{The density function $P(r)$ for the CSR~\eqref{eqn:complex_ratio_def} plotted for two example models of Kraus operators, the 
    RKO (first left) and the RKC (second left), and two example models of Lindbladians, the 
    0D-RL (third left) and the 1D-RL (fourth left).  For all cases, the sample size is $5000$.
    }
    \label{fig:ko_complex_ratio}
\end{figure}

\begin{figure}[H]
%
\begin{minipage}[t]{0.19\textwidth}
\includegraphics[width=\linewidth,keepaspectratio=true]{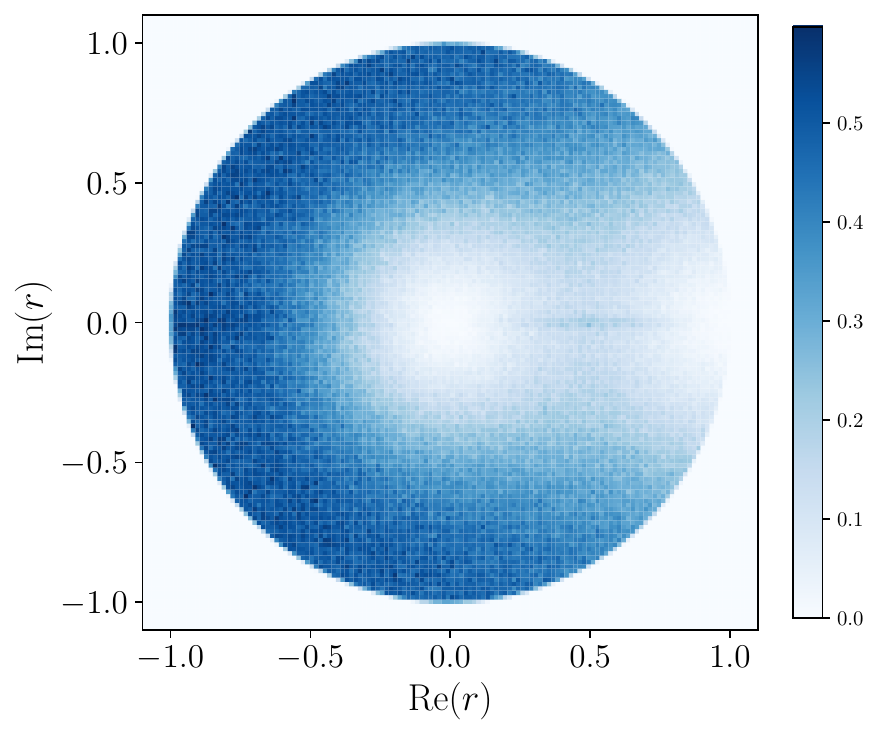}
\end{minipage}
%
%
\begin{minipage}[t]{0.19\textwidth}
\includegraphics[width=\linewidth,keepaspectratio=true]{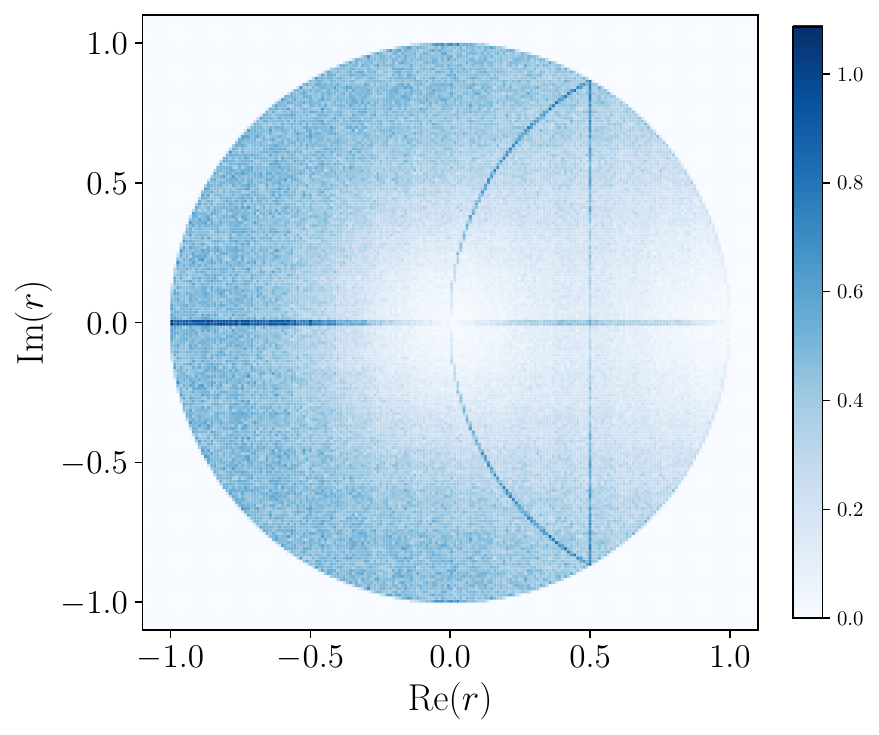}
\end{minipage}
%
%
%
\begin{minipage}[t]{0.19\textwidth}
\includegraphics[width=\linewidth,keepaspectratio=true]{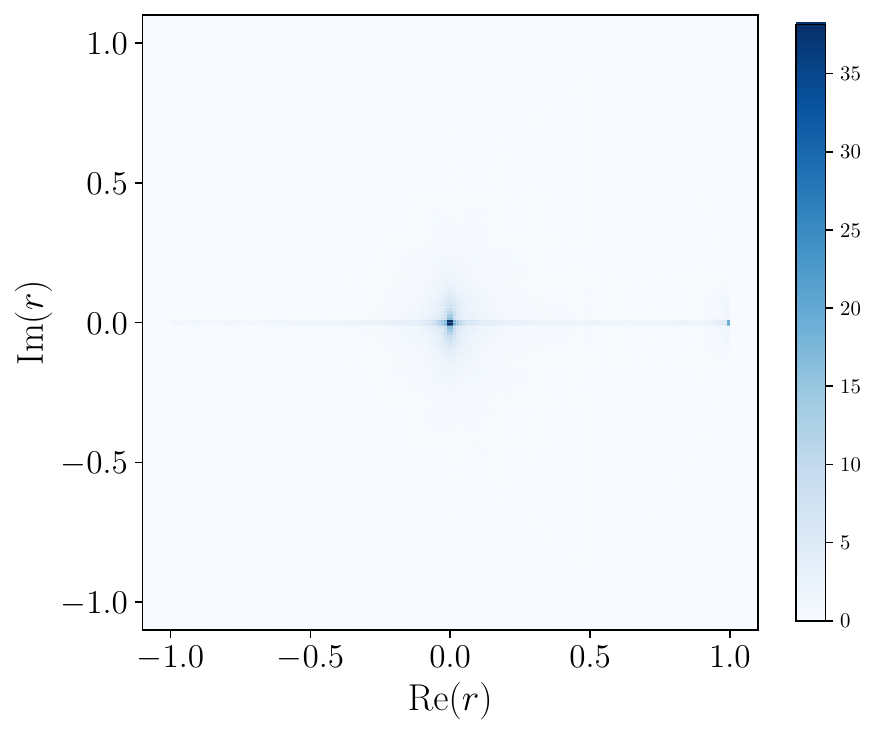}
\end{minipage}
%
%
\begin{minipage}[t]{0.19\textwidth}
\includegraphics[width=\linewidth,keepaspectratio=true]{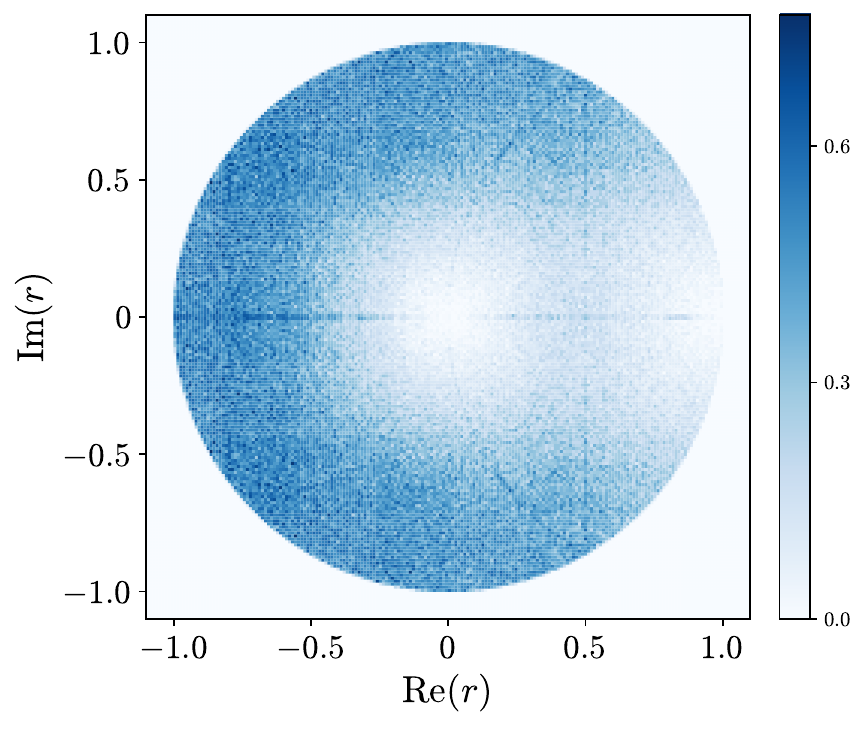}
\end{minipage}
\begin{minipage}[t]{0.19\textwidth}
\includegraphics[width=\linewidth,keepaspectratio=true]{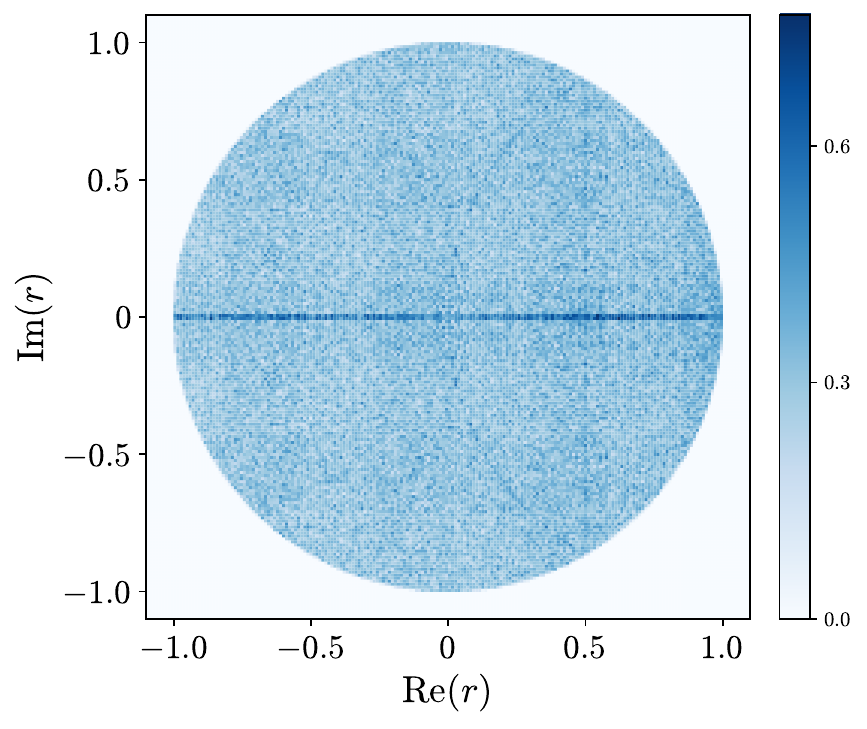}
\end{minipage}
    \caption{The density function $P(r)$ for the CSR~\eqref{eqn:complex_ratio_def} plotted for SYK-L, dXXZ, dXX, dIsing-Chaos ($W=0.6$) and dIsing-MBL ($W=5$) from the left to the right. For dXX and dIsing-MBL, the signature of level repulsion around $z=0$ and $z=1$ is absent. For all cases, the sample size is $5000$.
    }
    \label{fig:csr_lindbladtwo}
\end{figure}

The density of the CSR $P(r)$ is plotted 
for nine example models in~\autoref{fig:ko_complex_ratio} and~\autoref{fig:csr_lindbladtwo}. For chaotic open systems, $P(r)$ has lower value around the origin due to spectral rigidity,
and lower value around $r=1$ since $z_m^{(1)}$ and $z_m^{(2)}$ tend to lie on the opposite side of $z_m$ due to level repulsion.
For dIsing-MBL, we see that $P(r)$ is approximately flat, as consistent with the  Poissonian distributed complex spectrum.
%
%
%

Furthermore, there are a few other salient features of the CSR. First, the CSR exhibits reflection symmetry across the real axis in the complex plane.  
This is due to the fact that the eigenvalues of the superoperator occur in conjugate pairs.
Indeed, within the bulk, the nearest- and next-nearest neighbors for $\overline{z}_n$ are given by
$\overline{z}_n^{(1)}$ and $\overline{z}_n^{(2)}$ such that the CSR
is given by $\overline{r}_n$, i.e. for each $r_n$ we also have $\overline{r}_n$. 
As a sanity check, we find that if we compute the CSR for only eigenvalues in the upper half plane with $\imm(z) > 0$, the CSR is no longer reflection symmetric.

Second, the CSR exhibits a concentration along the real axis.
This is due to the fact that the spectrum of has a concentration along the real axis, and that the ratio of real numbers is real.
%
As a sanity check, if we remove eigenvalues lying on the real axis before computing the CSR, we eliminate the concentration along the real axis.

Finally, the density of the spacing ratio exhibits an ``arc"-like feature, as well as a vertical
line located at $\ree(r) = \frac{1}{2}$. 
These are caused by the case where an eigenvalue's two
nearest neighbors are its complex conjugate and another eigenvalue lying on the real axis.  
Let 
the distance between the eigenvalue and its complex conjugate be $2y$, and the distance to the 
real eigenvalue be $a$.  The angle between the lines connecting the two pairs of eigenvalues, 
$\theta$, is also the argument of the complex spacing $r$, and is given by $\cos \theta = y/a$ and
$\sin \theta  = \sqrt{1-y^2/a^2}$.  Now there are two cases.

In the first case, $2y > a$, and so we have that:
\begin{equation}
    \ree (r) = |r| \cos\theta = \frac{a}{2y} \frac{y}{a} = \frac{1}{2} \,,
\end{equation}
which explains the concentration along the vertical line $\ree (r) = \frac{1}{2}$.

In the second case, $2y < a$, and so we have that $\ree (r) = (2y/a) \cos \theta = 2 y^2 / a^2$.  Since $a > 2y$, this 
implies that $\ree (r) < 1/2$.  On the other hand, we have that:
\begin{equation}
    \imm (r) = \frac{2d}{a} \sin \theta = \frac{2d}{a} \sqrt{1-\frac{y^2}{a^2}}
    = \pm \sqrt{\ree(r) \left( 2 - \ree(r) \right)} \,.
\end{equation}

This explains the concentration along the ``arc," as one finds that it is well-fit by the 
curve defined by the above relation for $0 < \ree (r) < \frac{1}{2}$.

\section{Unfolding}\label{sec:unfolding}
To compute long-range spectral correlation including the DSFF, we need to ``unfold'' the complex spectrum, i.e. to remove the variation of spectral density in the complex plane.
The unfolding procedure typically involves a transformation of the spectra which makes the associated DOS flatter.
In fact, unfolding for complex spectra is considerably more challenging than the analogous procedure for real spectra. 
One particular obstacle is that the complex spectral transformation must be a conformal transformation, which restrict the possible choices of transformations.
As a result, we often cannot perfectly unfold the spectrum, and a further ``filtering'' procedure needs to be applied to  favour DSFF contributions from flatter regions in the spectra (see Appendix~\ref{app:filtering}).
%
%
%

Here we comment on the necessity of unfolding (and filtering) in the context of DSFF. 
Consider an ensemble of a chaotic open system with a highly non-uniform DOS.
%
Each sufficiently small section of the DOS will yield a contribution to DSFF that is well-described by the GinUE ensemble, but with an effective $\tauhei$ that depends locally on the DOS. 
Because the ``ramp" of the DSFF for the GinUE ensemble grows quadratically~\cite{li2021spectral} as opposed to linearly in the Gaussian unitary ensemble for Hermitian systems, 
DSFF for ensembles with non-unform DOS will deviate significantly from the GinUE DSFF behavior, and hence unfolding is required.

%
%


\subsection{Random Kraus circuits}
%

 First we consider the zero-dimensional RKO model.
As pictured in~\autoref{fig:single_spectra_0d}, for sufficiently large $d$, the number of jump operators, the DOS of a RKO is characterized by a roughly flat DOS which has support over a circular region in the complex plane.  
The radius of the circle does not depend on the Hilbert space dimension, and 
is qualitatively similar to the DOS of a RMT ensemble.
Therefore, unfolding is not needed.

\begin{figure}[H]
\begin{minipage}[t]{0.28\textwidth}
\includegraphics[width=\linewidth,keepaspectratio=true]{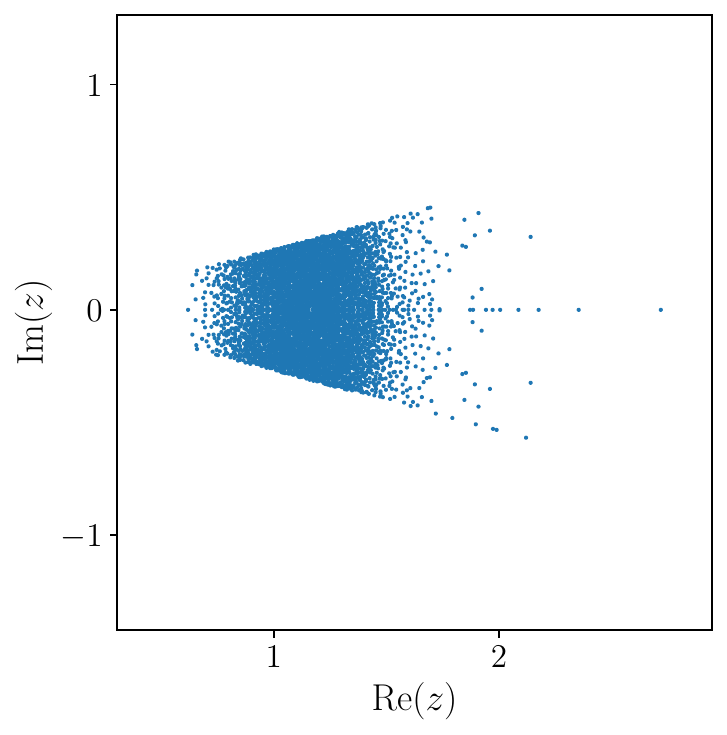}
\end{minipage}
\hspace*{\fill} 
\begin{minipage}[t]{0.33\textwidth}
\includegraphics[width=\linewidth,keepaspectratio=true]{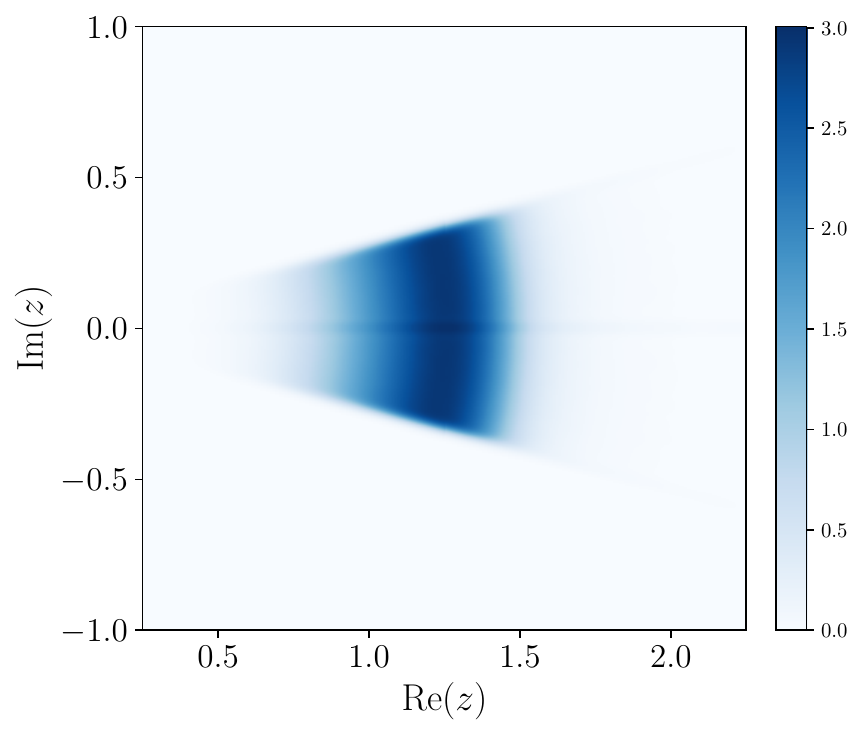}
\end{minipage}
\hspace*{\fill} 
\begin{minipage}[t]{0.35\textwidth}
\includegraphics[width=\linewidth,keepaspectratio=true]{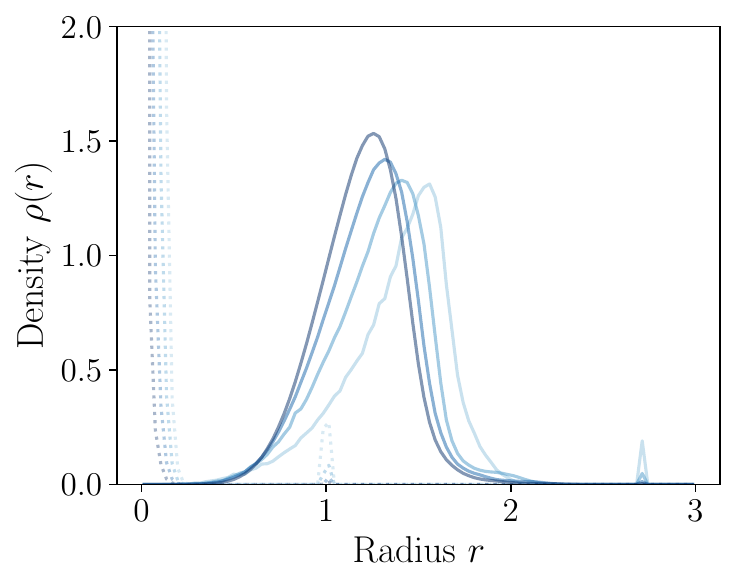}
\end{minipage}
    \caption{Left and middle: Spectrum of a single realization (left) and the density of state (right) of the RKC model with physical dimension $L=6$ after unfolding.
    The spectra is spread out over a sector of an annulus centered at the origin.
    Right: Radial density function $\rho(r)$ for the RKC
    before (dashed lines) and after (solid lines) unfolding,  given for different system
    sizes $L \in [3,6]$ from light to dark blue.
    }
    \label{fig:unfold_spectra_2x2}
\end{figure}

\begin{figure}[H]
\begin{minipage}[t]{0.39\textwidth}
\includegraphics[width=\linewidth,keepaspectratio=true]{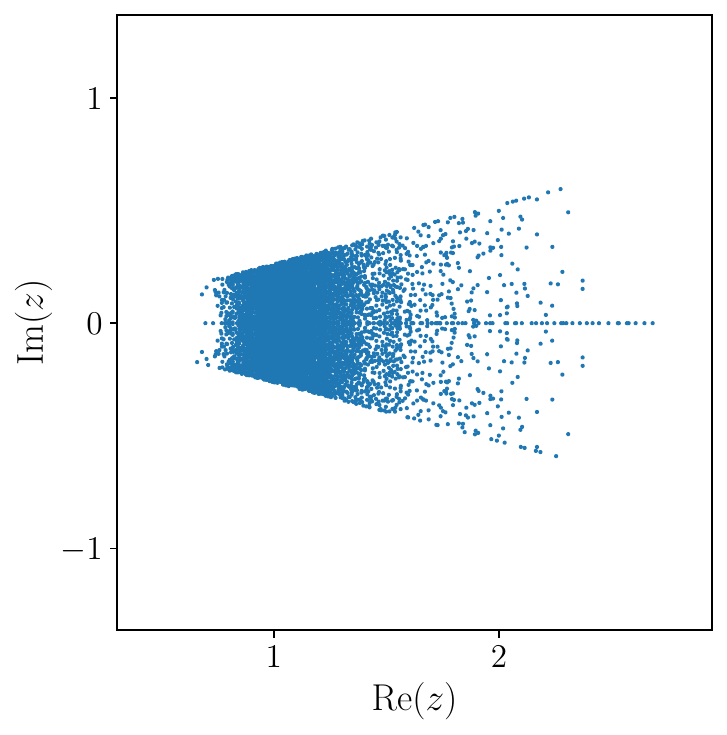}
\end{minipage}
\hspace*{\fill} 
\begin{minipage}[t]{0.475\textwidth}
\includegraphics[width=\linewidth,keepaspectratio=true]{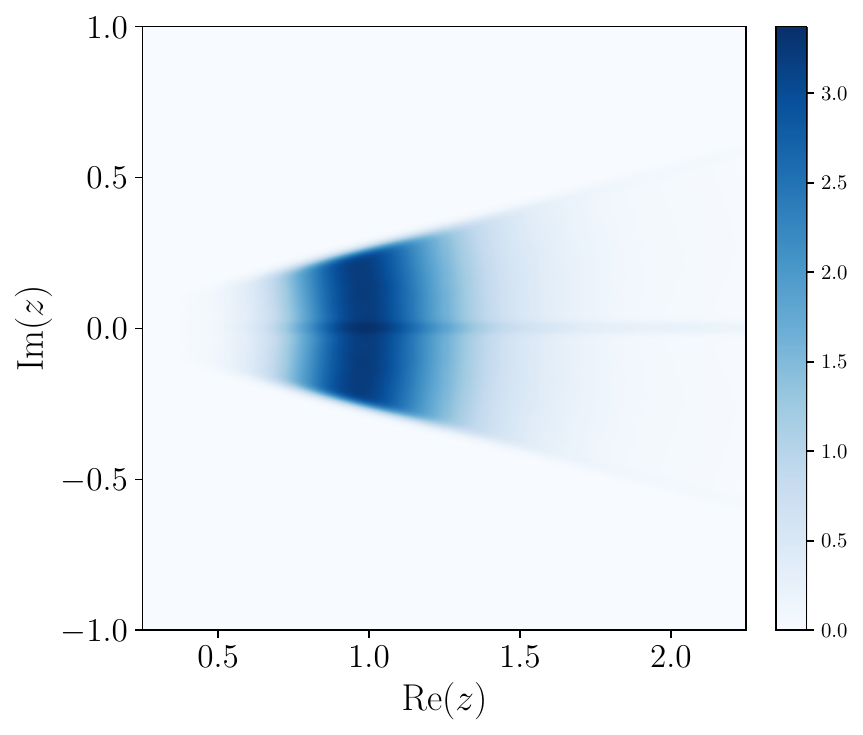}
\end{minipage}
    \caption{Spectrum of a single realization (left) and the DOS (right) of the $U(1)$-RKC model with physical dimension $L=6$ after unfolding (left).
    The spectra is spread out over a sector of an annulus centered at the origin.
    }
    \label{fig:unfold_spectra_u1}
\end{figure}

This is to be contrasted with the situation for the one-dimensional model, the RKC.
In this case, the DOS is sharply peaked about the origin in the complex plane, with the strength of the peak growing as the physical dimension is increased, i.e. the DOS becomes less uniform. 
To unfold the spectra,  we choose a similar 
unfolding function to that used in~\cite{akemann2012}, which for our case is given by:
\begin{equation}\label{eqn:unfolding_function}
    z \mapsto A z^{1/\log{N}} \,,
\end{equation}
where $A$ is some arbitrary scaling constant, which we chose to be $A=e$.
The branch cuts here and in all Kraus models are chosen to be $(-\infty, 0]$. 
The results of this unfolding scheme for the RKC and $U(1)$-RKC are presented in~\autoref{fig:unfold_spectra_2x2} and~\autoref{fig:unfold_spectra_u1}.
By comparing with the unfolded spectra in \autoref{fig:single_spectra_1DRKC}, we
see that unfolding simultaneously increases the support and uniformity of the spectrum.
Furthermore, suppose we have a density of state $\rho(r, \theta)$. If we assume that the DOS before unfolding is rotationally symmetric, we may define a local radial density function $\rhor(r)$ via
\begin{equation}\label{eq:radial_dos}
   \iint dr \, d\theta \, r  \rho(r,\theta) = \int dr \, 2 \pi  r \, \rhor(r) \,,
\end{equation}
such that unfolding is successful if our unfolding function~\eqref{eqn:unfolding_function}
transforms $\rho(r)$ into a flat function of $r$ (Note that we have a factor of $r$ in \eqref{eq:radial_dos}). 
The radial density function $\rho(r)$ before and after unfolding for the RKC is provided in~\autoref{fig:unfold_spectra_2x2} right.  
We see that unfolding creates a local maxima in the radial
density $\rho(r)$.  At this maxima, the radial density does not change to the first order in $r$, and so the
DOS can be taken to be approximately uniform in a small radius window.

Generally, we consider unfolding functions which are multi-valued, and could potentially introduce artificial edges due to the branch cut. Ideally, we would like to isolate the bulk of the spectrum by avoiding these edges. 
However, in~\autoref{app:sanity} we consider the sanity check of unfolding the GinUE spectrum, and find that such edges do not qualitatively distort the DSFF even in the absence of a filtering scheme which removes the edge.  We expect this to be true for other models as long as $\rho_\mathrm{av}$ is sufficiently uniform. 

\subsection{Random Lindbladians}
\begin{figure}[H]
\begin{minipage}[t]{0.3\textwidth}
\includegraphics[width=\linewidth,keepaspectratio=true]{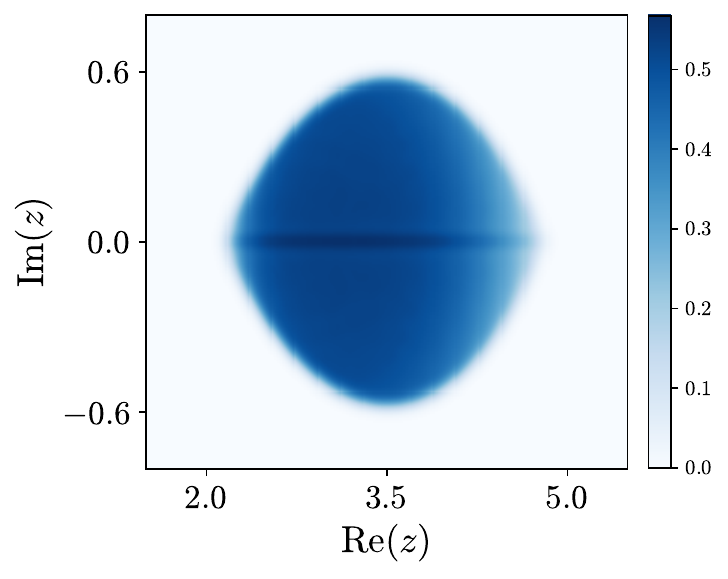}
\end{minipage}
\hspace*{\fill} 
\begin{minipage}[t]{0.3\textwidth}
\includegraphics[width=\linewidth,keepaspectratio=true]{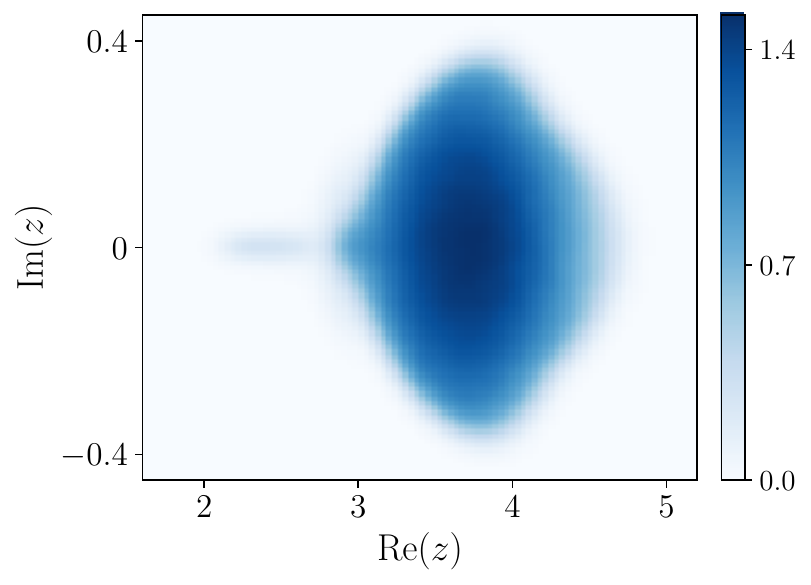}
\end{minipage}
\hspace*{\fill} 
\begin{minipage}[t]{0.3\textwidth}
\includegraphics[width=\linewidth,keepaspectratio=true]{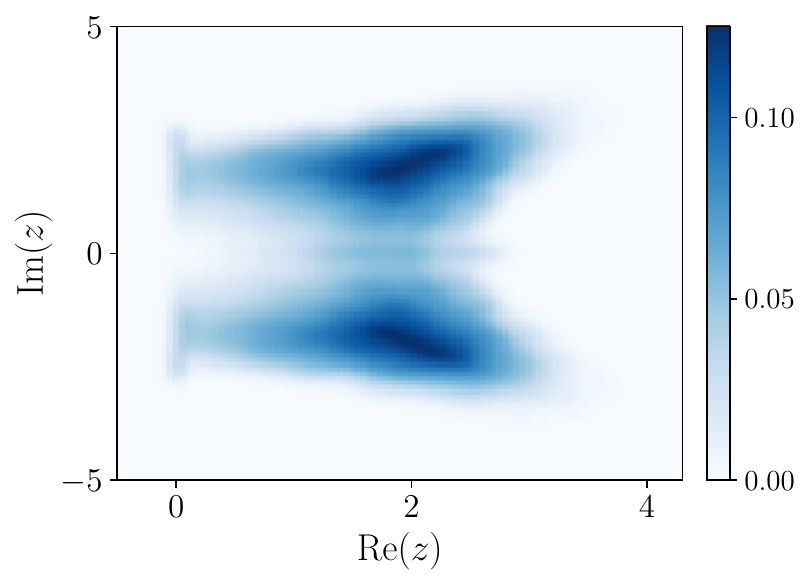}
\end{minipage}
    \caption{The density of state of the 0D-
        RL model with dimension $N=4096$, SYK-L with $L=6$ and dXXZ with $L=6$ after unfolding.
    %
    %
    }
    \label{fig:unfold_spectra_0DRL}
\end{figure}

As shown in Appendix~\ref{app:dos_lindblad}, the spectrum of 1D-RL and dIsing-MBL are relatively flat and gives nice DSFF behavior after proper filtering. We find that an transformation similar to \eqref{eqn:unfolding_function} is sufficient to unfold the spectrum of 0D-RL, SYK-L,  dXXZ, 0D-RL-JO, 1D-RL-1JO, and 1D-RL-2JO: 
\begin{equation}\label{eqn:lind_unfolding_function}
    z \mapsto A (z-z_0)^{\beta} \,.
\end{equation}
where $\beta=\frac{1}{2}$ for all cases except for SYK-L and 0D-RL-JO, where $\beta= \frac{1}{3}$.  $A=-i$ and $z_0$ is chosen at the maximum DOS, which is model-dependent. The branch cut for each model is chosen to be $[z_0, \infty)$. 
The DOS after unfolding is shown in \ref{fig:unfold_spectra_0DRL}. 

\section{Filtering}\label{app:filtering}
\subsection{Definitions}
In Appendix \ref{sec:unfolding}, we described an unfolding procedure that creates a region in the complex plane with approximately uniform density.
%
%
In the context of DSFF, 
to further isolate the spectral correlation contributed from regions with uniform DOS, we apply an additional ``filtering'' procedure, analogous to the filtering procedure used in studying SFF for Hermitian systems \cite{Gharibyan_2018},
%
%
\be\label{eq:dsff_filter}
\fkc(t, s) = \left\langle 
        \left|
        \sum_n e^{i x_n t + i y_n s } \ff 
        \left(x_n,y_n ; \{ \valpha\} \right)
        \right|^2
        \right\rangle - \left| \left\langle
        \sum_n e^{i x_n t + i y_n s }
        \ff \left(x_n,y_n; \{ \valpha\} \right)
        \right\rangle \right|^2
        \,,
\ee
where $\ff$ is the filter function whose strength is controlled by a set of parameters $\valpha$. Note that we use $\tilde{K}$ (as supposed to $K$) to denote the DSFF for spectra after filtering. Two  possible choices of  filtering function we consider are
\begin{enumerate}
    \item Sharp Cutoff: 
    We choose the filter as an indicator function over some region $R$, i.e.
    \be
    \ff \left(x_n,y_n; R \right)
    =
    \begin{cases}
    0, &  \text{ if } z_n \not\in R
    \\
    1, &  \text{ if } z_n \in R
    \end{cases}
    \,.
    \ee
    where $R$ is the region whose DSFF contribution we want to include.
    \item Gaussian Filter: The filter function is taken to be  Gaussian function(s) with certain center labelled by $\vec{\mu}$ and strength parametrised by $\vec{\alpha}$. For a general coordinate ${x_m}$, we define 
\begin{equation}\label{eqn:gauss_filter_def_coord}
        \ff \left(\vec{x}; \vec{\alpha}, \vec{\mu}  \right) =
        e^{- \sum_m {\alpha_m}({x}_m - {\mu}_m)^2 }
        \,,
    \end{equation}
More concretely, for Lindbladians, the coordinate is taken to be the Cartesian coordinate in the complex plane, we define \begin{equation}\label{eqn:gauss_filter_def2}
        \ff \left(x_n,y_n; \alpha_x, \alpha_y, \mu_x,\mu_y  \right) =
        e^{- \alpha_x( x_n- \mu_x )^2 - \alpha_y(y_n - \mu_y)^2}
        \,,
    \end{equation}
    where $(\mu_x,\mu_y)$ is the center of the Gaussian filter, and $(\alpha_x, \alpha_y)$ parametrize the strength in the real and imaginary directions respectively.
    For the Kraus circuits, we use  Gaussian filters along the radial axis,
\begin{equation}\label{eqn:gauss_filter_radial}
        \ff \left(z_n ;  \alpha, \mu_r\right) =
        e^{- \alpha (|z_n| - \mu )^2}
        \,,
    \end{equation}
 \end{enumerate}   
    
%
    %
The center of the Gaussian filter $\vec{\mu}$ is chosen to be the peak of the averaged DOS $\rho(z)$, and $\vec{\alpha}$ is chosen so that we focused on data around the peak where $\rho(z)$ is relatively flat. We introduce a single tunable parameter, the dimensionless filtering strength $\tilde{\alpha}$, which is chosen to be the same for all direction. In turn, $\alpha_m$ for the $m$-th coordinate is fixed by the relation $\tilde{\alpha} =\alpha_{m}\Delta_{\mathrm{D},m}^2$. $\Delta_{\mathrm{D},m}$ is the half width at half maximum of the peak of $\rho(z)$ along coordinate $m$, i.e. $\Delta_{\mathrm{D},m} = \frac{x_{m+}-x_{m-}}{2}$ where $\rho(x_{m+})$ and $\rho(x_{m-})$ reach half of its peak value. 
For Kraus, we have $\alpha_\theta = 0$, $\alpha_r = \tilde{\alpha} /  \Delta_{\mathrm{D},r}^2$ and $\Delta_{D,r} = \frac{r_+ - r_-}{2}$. For Lindbladians, we have $\alpha_{m} = \tilde{\alpha} /  \Delta_{\mathrm{D},m}^2(L)$ with $m=x,y$, where the horizontal and vertical `peak width' are given by $\Delta_{\mathrm{D},x} = \frac{x_{+} - x_{-}}{2}$ and $\Delta_{\mathrm{D},y} = \frac{y_{+} - y_{-}}{2}$ respectively.

\subsection{Early time effect of filtering}\label{app:early_filter}
\begin{figure}[H]
\begin{minipage}[t]{0.45\textwidth}
\includegraphics[width=\linewidth,keepaspectratio=true]{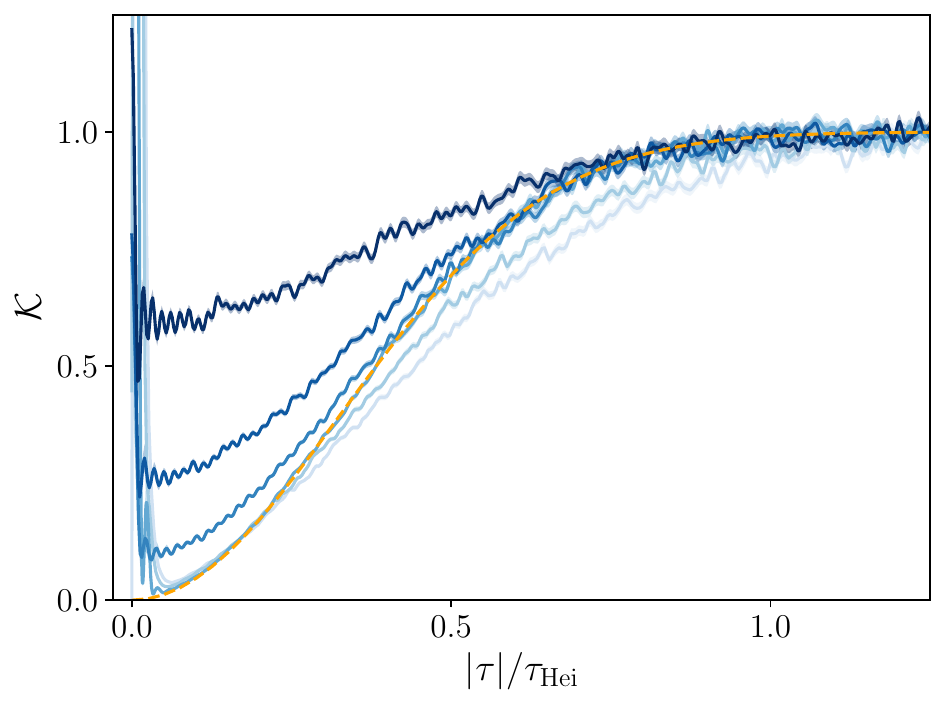}
\end{minipage}
\hspace*{\fill} 
\begin{minipage}[t]{0.45\textwidth}
\includegraphics[width=\linewidth,keepaspectratio=true]{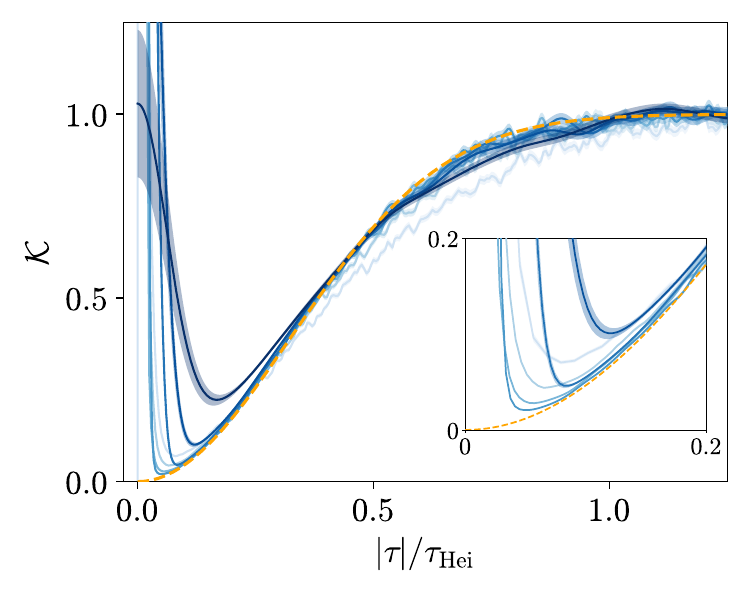}
\end{minipage}
    \caption{Left(Right) plot is respectively DSFF for RKC (1DRL) with $L=6$ at $\theta=\pi/4$
 with 5000 samples. From light blue to dark blue, the Gaussian filter strength is varied from
    $\tilde \alpha \in [1/20, 1/5, 1, 20, 40, 100]$ ($[1/5, 2/3, 2, 20, 60, 200]$).  In orange, theoretical DSFF curves for the GinUE random matrix ensemble are
    fit to the late-time DSFF for selected values of $\alpha$. 
    }\label{fig:ko_2x2_1drl_vary_alpha}
\end{figure}

The introduction of filtering leads to artifacts in the early-time DSFF due to edge-effects in the 
filtering procedure (\autoref{fig:ko_2x2_1drl_vary_alpha}). 
For the sharp cutoff filter, the early-time DSFF is roughly given by
\begin{equation}
    \tilde{K}_c(t,s) \big|_{t,s \ll 1} \approx 
    \left\langle
    \left|
        \sum_{z_n \in R} 1
        \,\,
    \right|^2
    \right\rangle
    -
    \left|
    \left\langle
        \sum_{z_n \in R} 1
        \,
    \right\rangle
    \right|^2
    \equiv
    \left\langle \delta \mathcal{N}^2( R)\right\rangle  \, .
\end{equation}
That is, the early-time value of the connected DSFF is non-zero, and measures the variance of the number of eigenvalues $\left\langle \delta \mathcal{N}^2( R)\right\rangle $ within region $R$. 
Since our unfolding procedure is constructed such that the uniform part of the DOS has roughly fixed region size as we increase $L$, 
%
%
the number of eigenvalues in $R$ grows with the Hilbert space dimension.
Thus, the early-time artifacts of the filtering procedure are a finite-size effect, that disappears with
increasing Hilbert space dimension.

For the Gaussian filter,  the early-time value of DSFF is determined by the variance of $\sum_n e^{ - \alpha (|z_n| - \mu)^2}$ given by
\begin{equation}
    \tilde{K}_c(t,s) \big|_{t,s \ll 1} \approx 
    \left\langle
    \left|
         \sum_n e^{- \alpha(|z_n| - \mu)^2}
        \,\,
    \right|^2
    \right\rangle
    -
    \left|
    \left\langle
        \sum_n e^{- \alpha(|z_n| - \mu)^2}
        \,
    \right\rangle
    \right|^2 \, .
\end{equation}
We observe empirically that the 
early-time artifacts caused by the Gaussian filter are smaller in magnitude for the system sizes of interest (see examples in Appendices \ref{app:sanity} and \ref{app:dsff}).  
The difference in relative strength
of the early-time artifacts can be understood as a difference in the sharpness of the two filters.  
The Gaussian filter
can be thought of as a cutoff with smoothened edges which is less sensitive to fluctuations in the number of eigenvalues.  
Since we are interested in the early-time features of the DSFF, namely the ``bump" and the Thouless time, we opt to use the Gaussian filter to minimize such filtering artifacts.

\subsection{Late time effect of filtering}\label{app:late_filter}

The unfolding and the (Gaussian) filtering procedure alter both the late-time plateau value as well as the scaling 
of $\tauhei$.  
The late-time plateau value can be determined by taking $|\tau| \to \infty$ in $\fk_c$ given by~\eqref{eq:dsff_filter}, 
which averages the phase differences in the first term to a delta function $\delta_{nm}$ and in the second term to zero. 
Thus, the late-time plateau value is
\begin{equation}\label{eqn:plateau_value}
  \lim_{|\tau| \to \infty} \fk_c(t,s) =  \left\langle
    \sum_n e^{- 2{\alpha (|z_n| - \mu)^2}}
    \right\rangle
    \,.
\end{equation}
The Heisenberg time $\tauhei$ is determined empirically
by computing the ``weighted" nearest-neighbor spacing.  
For a given filter strength $\alpha$, we 
estimate the ``weighted" nearest-neighbor spacing as
\begin{equation}\label{eqn:weight_nn_spacing_def}
    \tilde{s}_n^{(1)} = |z_n^{(1)} - z_n| 
    \times
    e^{-\alpha(|z_n| - \mu)^2 - \alpha (|z_n^{(1)}| - \mu)^2} \,.
\end{equation}
Averaging over all eigenvalues $z_n$ across many 
realizations, we obtain the averaged weighted nearest-neighbor
spacing $\tilde{s}^{(1)}(\mu(L), \alpha(L))$.
\be\label{eqn:weighted_mean_level_spacing}
\tilde{s}^{(1)}(\mu(L), \alpha(L)) =  
\langle
\mathbb{E}_n
[
    \tilde{s}_n^{(1)} 
]
\rangle
\ee
where $\mathbb{E}$ averages over all eigenvalues within the spectrum of a single realization.

\subsection{Definition of Heisenberg time after filtering}
For the GinUE, the Heisenberg time in the DSFF is proportional to the mean level spacing by a constant we call $\chi_{\mathrm{GinUE}}$, 
\begin{equation}
     \tau_{\mathrm{Hei, GinUE}} = \frac{\chi_{\mathrm{GinUE}}}{s^{(1)}_{\mathrm{GinUE}}} \;. 
\end{equation}
We suppose the corresponding ``weighted'' Heisenberg time in the filtered DSFF is related to the weighted mean level spacing by the same $\chi_{\mathrm{GinUE}}$, i.e.
\begin{equation}\label{eqn:weighted_tauhei_def}
{\tau}_{\mathrm{Hei}} := \frac{\chi_{\mathrm{GinUE}}}{\tilde{s}^{(1)}} \;. 
\end{equation}
Indeed, this is good definition of the weighted Heisenberg time since it allows us to collapse DSFF of many-body open quantum systems with different system sizes into a single curve, see \autoref{fig:many_model_same_L}. Note that $\tauhei$ refers to a time scale of the filtered DSFF, but we have suppressed the tilde notation on $\tauhei$. 



\subsection{Choice of filtering strength \texorpdfstring{$\alpha$}{alpha}}
%
%
The introduction of the Gaussian filter means that $\tauhei$ and the general shape of the DSFF depend on the 
filter strength $\alpha$ in a non-trivial way.
%
%
As we reasoned in Appendix~\ref{sec:unfolding}, due to the non-uniformity of the DOS, 
our DSFF does not exhibit universal RMT behavior even
at late-times, and thus filtering is required.  However, 
when filtering is too strong, we might hide physically 
relevant features.  Specifically, there are three energy scales $\DeltaD \geq  \Deltath \geq \Deltahei$ 
in our problem, 
 and we seek for a value of $\alpha$ such that
\be \label{app_eq:scales}
\frac{1}{\DeltaD^2} \ll \alpha \ll \frac{1}{\Deltath^2} \leq \frac{1}{\Deltahei^2}\; .
\ee   
Here $\DeltaD$ represents the energy scale 
over which the DOS varies. 
We describe below an approach to identify $\alpha$ that lies in this regime described by \autoref{app_eq:scales}.

When the filter is too weak, i.e. $\tilde{\alpha}\equiv \alpha \DeltaD^2 \ll 1$,  we observe $\taudev \gtrsim \tauhei$. This is because
the DSFF is distorted by a non-uniform DOS, which is expected to be dependent on the microscopic details of the open system (light blue in~\autoref{fig:ko_2x2_1drl_vary_alpha}).
Thus, the weakly filtered DSFF largely deviates from that of the Ginibre ensemble, even at large times. 
When the filter is too strong, i.e. $ \alpha \Deltahei^2 \gg 1$,  we also observe $\taudev \gtrsim \tauhei$.
This is because the filter will wash out the signatures of level repulsion, and therefore the filtered DSFF will not display the ramp-plateau behavior of GinUE (dark blue in~\autoref{fig:ko_2x2_1drl_vary_alpha}).

\begin{figure}[H]
\begin{minipage}[t]{0.45\textwidth}
\includegraphics[width=\linewidth,keepaspectratio=true]{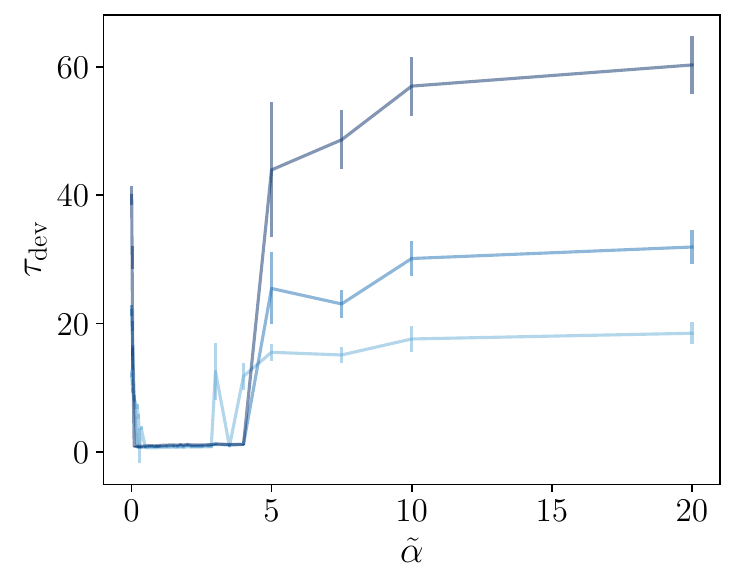}
\end{minipage}
\hspace*{\fill} 
\begin{minipage}[t]{0.45\textwidth}
\includegraphics[width=\linewidth,keepaspectratio=true]{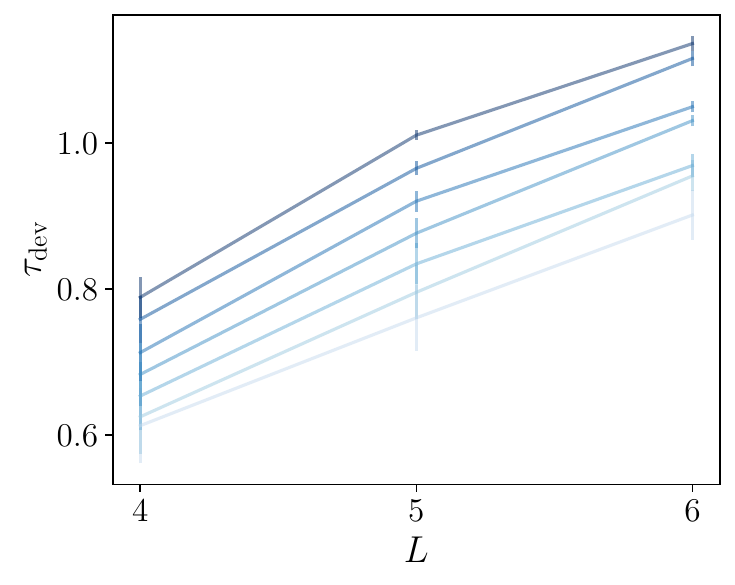}
\end{minipage}
    \caption{The effect of changing filter strength $\tilde \alpha$ and system size $L$ on $\taudev$ for the RKC model.  Left: Trend of $\taudev$ for different $\tilde \alpha$.  Plotted for different system sizes $L \in [4, 6]$ in
    light to dark blue.  
    Right: Trend of $\taudev$ versus
    system size for multiple $\tilde \alpha \in [1/2, 2]$ in light to dark blue.
    %
   }
    \label{fig:t_star_trends_rkc}
\end{figure}

\begin{figure}[H]
\begin{minipage}[t]{0.23\textwidth}
\includegraphics[width=\linewidth,keepaspectratio=true]{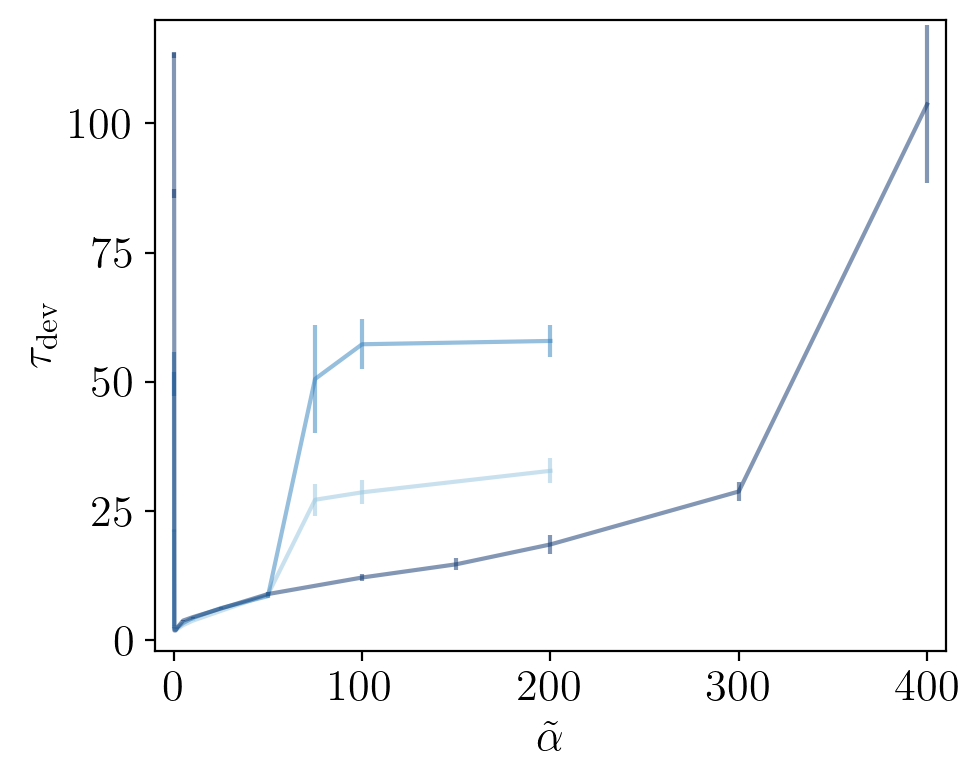}
\end{minipage}
\hspace*{\fill} 
\begin{minipage}[t]{0.23\textwidth}
\includegraphics[width=\linewidth,keepaspectratio=true]{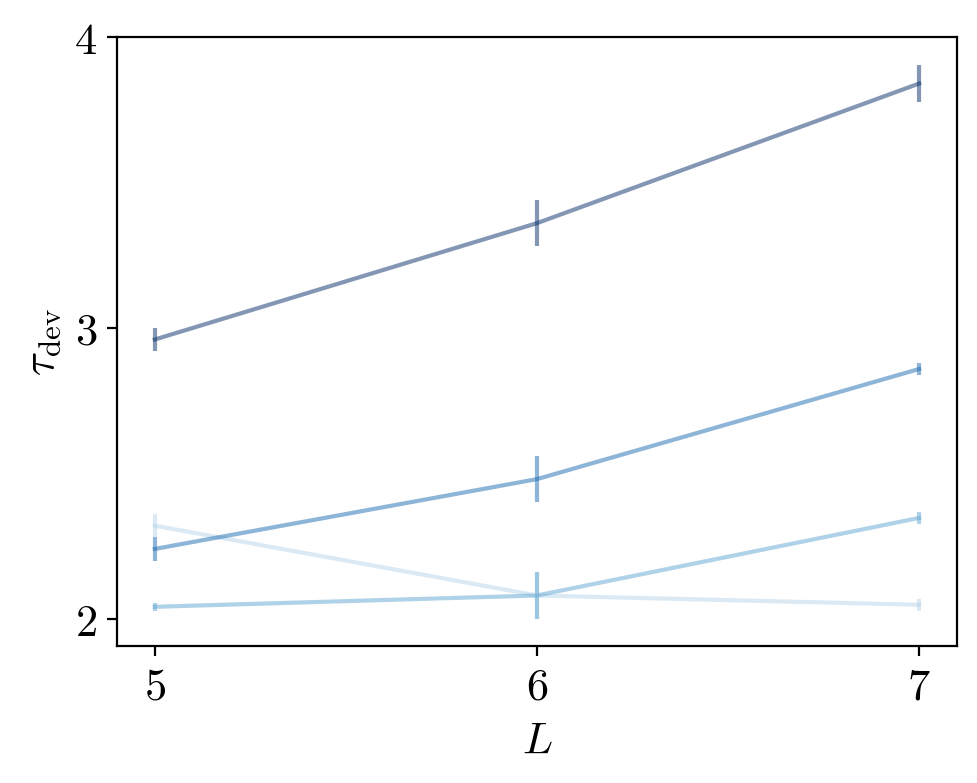}
\end{minipage}
\hspace*{\fill} 
\begin{minipage}[t]{0.23\textwidth}
\includegraphics[width=\linewidth,keepaspectratio=true]{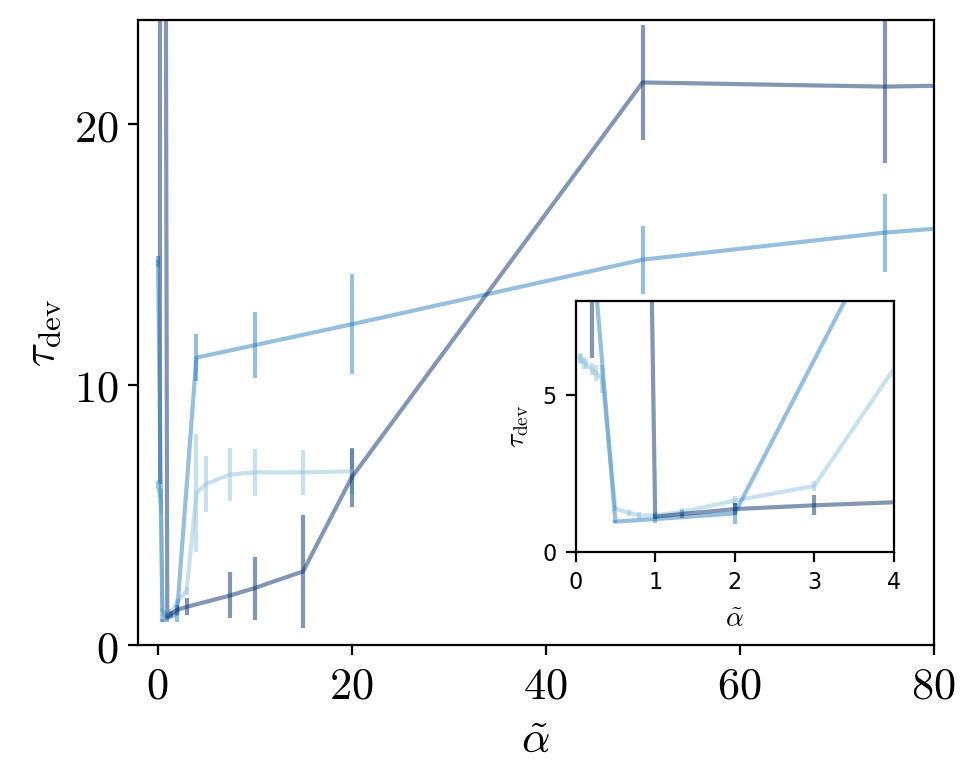}
\end{minipage}
\hspace*{\fill} 
\begin{minipage}[t]{0.23\textwidth}
\includegraphics[width=\linewidth,keepaspectratio=true]{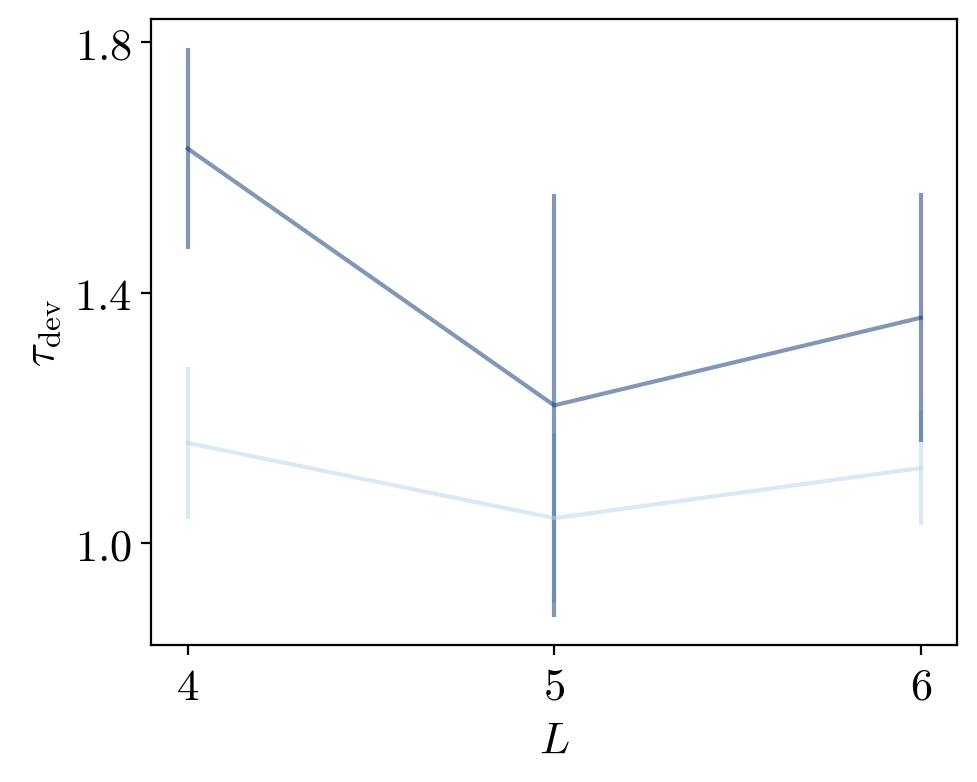}
\end{minipage}
    \caption{The effect of changing filter strength $\tilde \alpha$ and system size $L$ on $\taudev$ for the 1D-RL and SYK-L models.  First from the left:  $\taudev$ against $\tilde \alpha$ for 1D-RL for system size $L \in [5, 7]$ from
    light to dark blue. Second: $\taudev$ against $L$ for 
    1D-RL for multiple $\tilde \alpha = 0.5, 1, 2, 5$ from light to dark blue.
    Third:  $\taudev$ against $\tilde \alpha$ for SYK-L  for  $L \in [4, 6]$ from light to dark blue. Fourth: $\taudev$  against $L$ for 
    SYK-L for  $\tilde \alpha = 1, 2$ from light to dark blue.
    }
    \label{fig:t_dev_trends_rl}
\end{figure}

\begin{figure}[H]
\begin{minipage}[t]{0.27\textwidth}
\includegraphics[width=\linewidth,keepaspectratio=true]{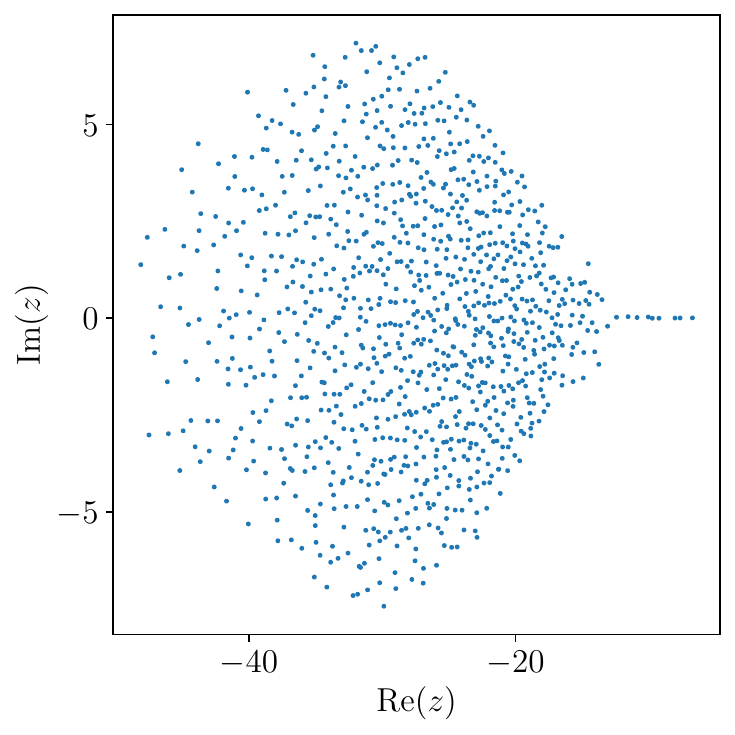}
\end{minipage}
\hspace*{\fill} 
\begin{minipage}[t]{0.27\textwidth}
\includegraphics[width=\linewidth,keepaspectratio=true]{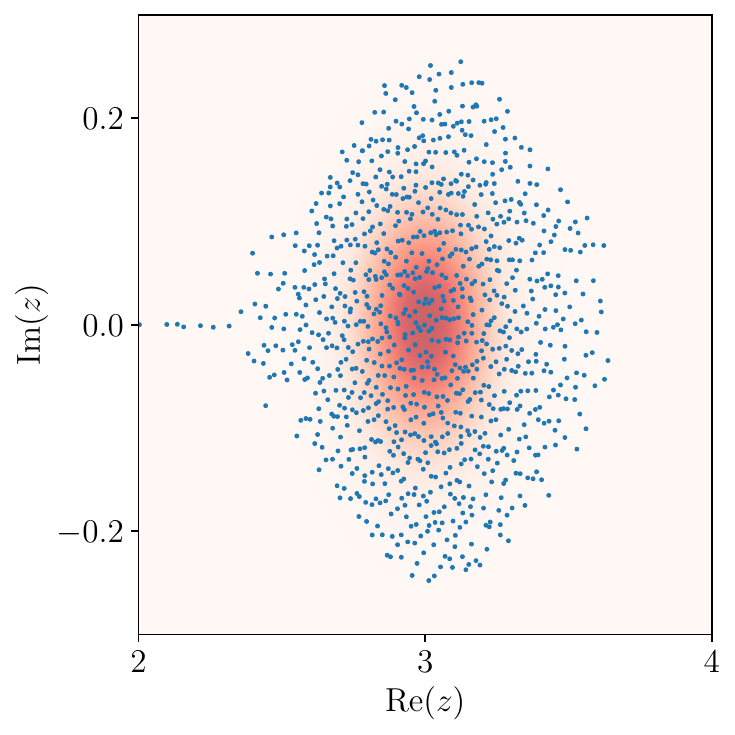}
\end{minipage}
\hspace*{\fill} 
\begin{minipage}[t]{0.34\textwidth}
\includegraphics[width=\linewidth,keepaspectratio=true]{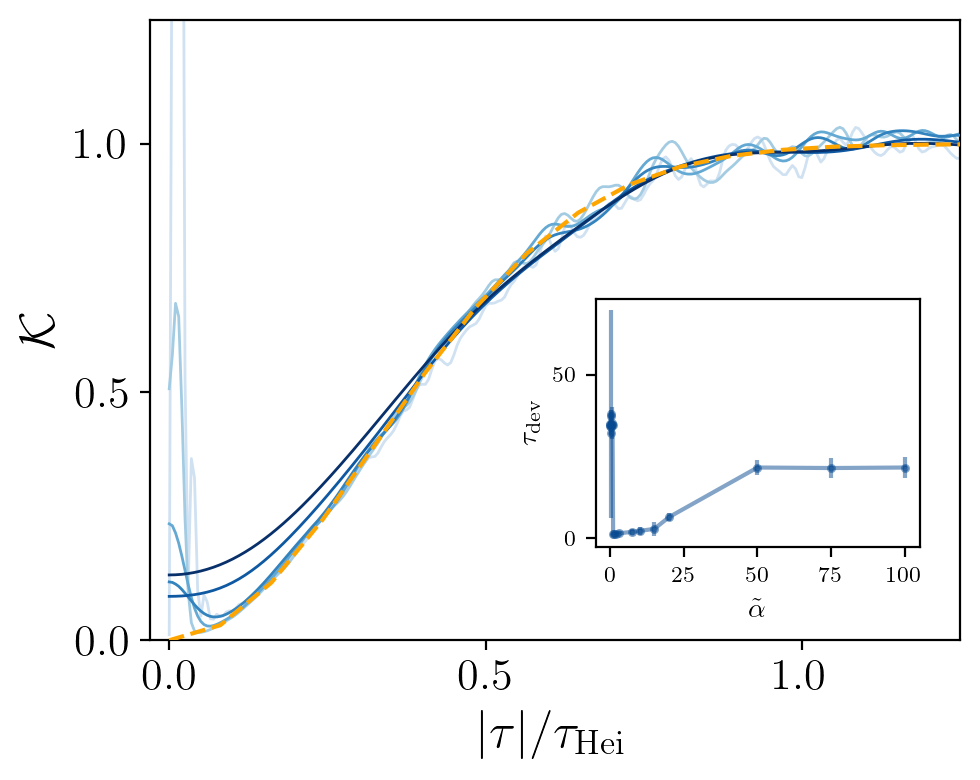}
\end{minipage}
    \caption{Left: A single-realization spectrum $\{z\}$ of SYK-L model for $L=14$. Middle: The unfolded spectrum $\{ z^{1/3}\}$ (of the single-realization spectrum) is subsequently filtered  with the filtering function $f(z)= e^{-(\mathrm{Re}(z)-3)^2/0.09-\mathrm{Im}(z)^2/0.02}$. Right: DSFF of the SYK-L model for varying filtering strength $\tilde{\alpha} \in [1/100, 100]$ from light to dark blue. Inset: $\taudev$ against $\tilde{\alpha}$ shows a trough in which the suitable filtering strength lies. 
    }
    \label{fig:t_dev_trends_sykl}
\end{figure}

\begin{figure}[H]
\begin{minipage}[t]{0.27\textwidth}
\includegraphics[width=\linewidth,keepaspectratio=true]{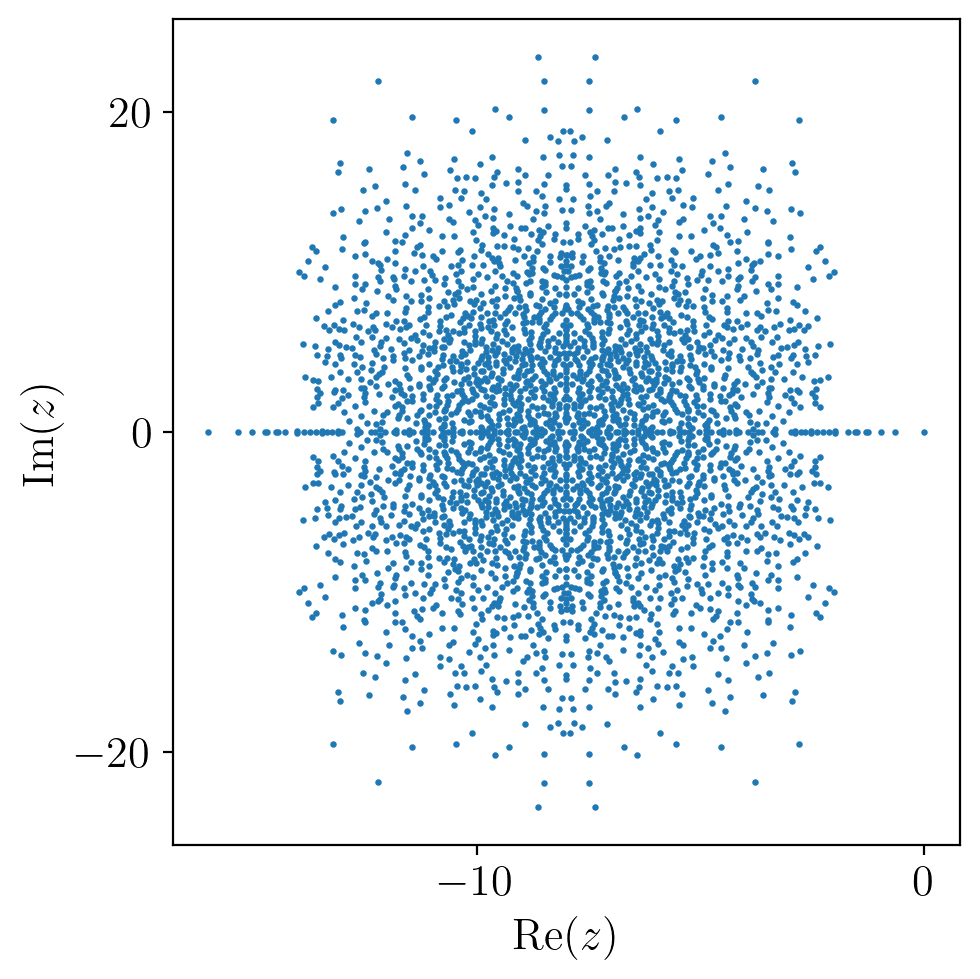}
\end{minipage}
\hspace*{\fill} 
\begin{minipage}[t]{0.27\textwidth}
\includegraphics[width=\linewidth,keepaspectratio=true]{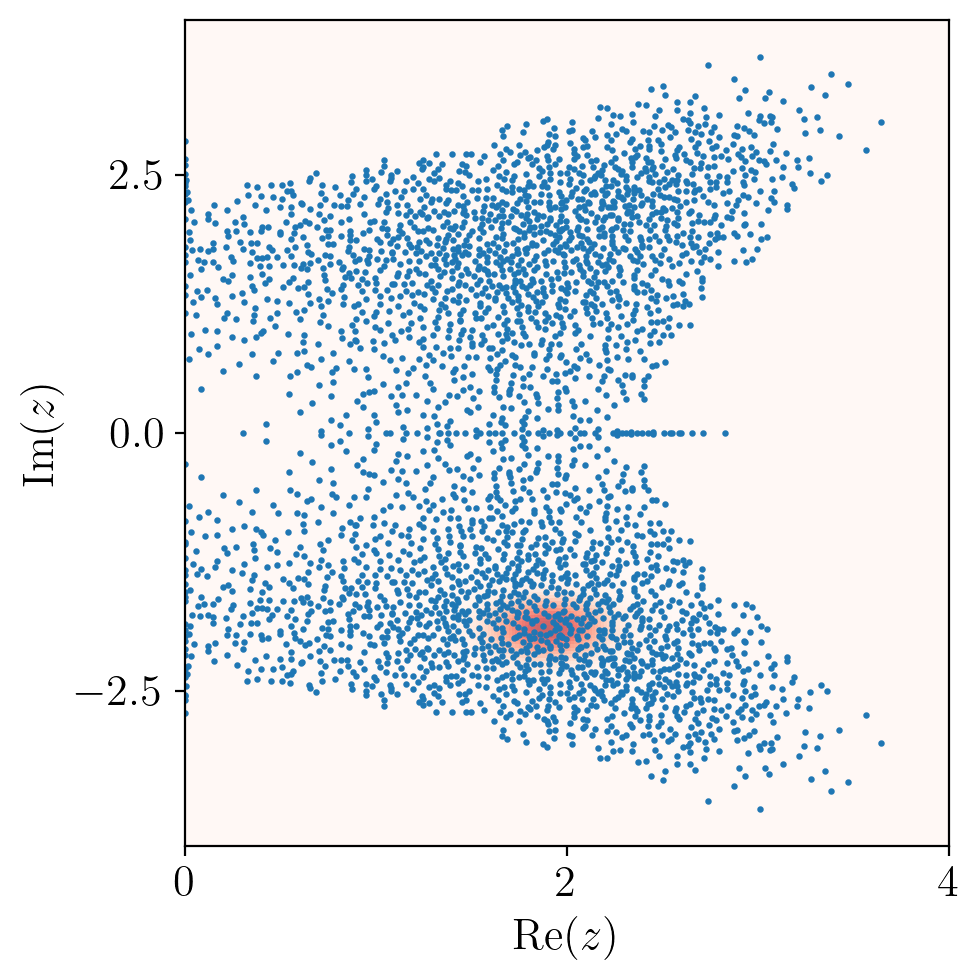}
\end{minipage}
\hspace*{\fill} 
\begin{minipage}[t]{0.34\textwidth}
\includegraphics[width=\linewidth,keepaspectratio=true]{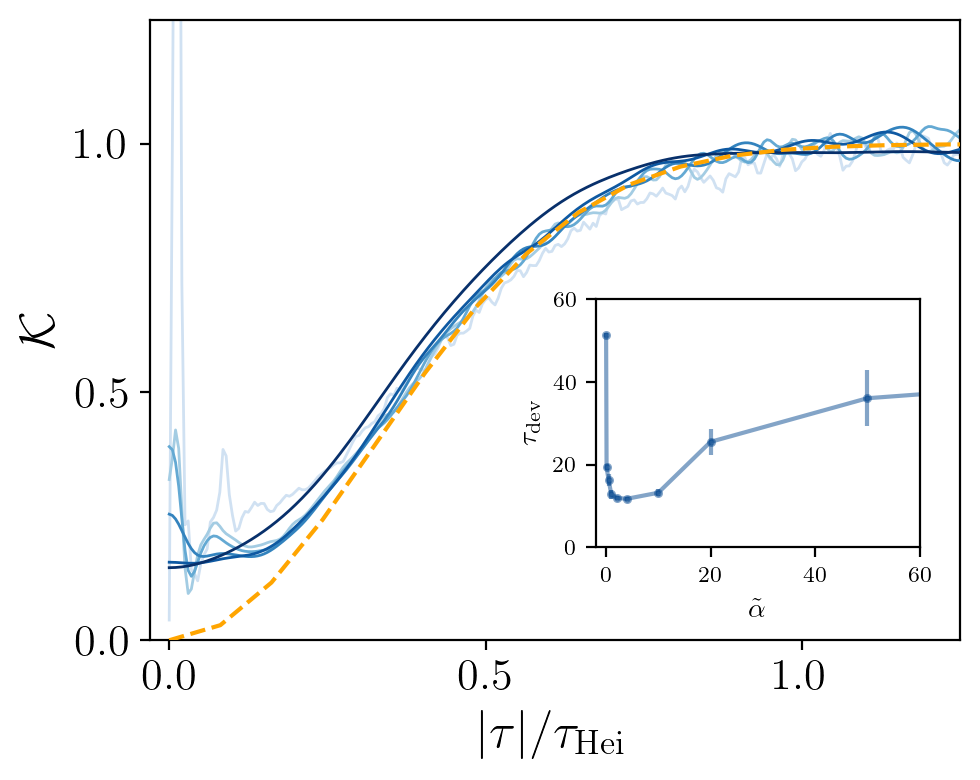}
\end{minipage}
    \caption{Left: A single-realization spectrum $\{z\}$ of dXXZ model for $L=7$. Middle: The unfolded spectrum $\{ (z+8)^{1/2}\}$ (of the single-realization spectrum) is subsequently filtered  with the filtering function $f(z)= e^{-(\mathrm{Re}(z)-1.9)^2/0.16-(\mathrm{Im}(z)+1.9)^2/0.14}$. Right: DSFF of the dXXZ model for varying filtering strength $\tilde{\alpha} \in [1/100, 75]$ from light to dark blue. $\taudev$ against $\tilde{\alpha}$ shows a trough in which the suitable filtering strength lies.
    }
    \label{fig:t_dev_trends_dxxz}
\end{figure}

As $L$ increases, we expect there to be a wider range of $\alpha$ which satisfies $\frac{1}{\DeltaD^2} \ll \alpha \ll \frac{1}{\Deltahei^2}\;$, since the number of eigenvalues grow exponentially in $L$.
Further, if $\Deltath$ behaves like its Thouless energy analogue in closed generic quantum many-body systems, which scales polynomially in $L$~\cite{cdc2, Gharibyan_2018}, the window, $\frac{1}{\DeltaD^2} \ll \alpha \ll \frac{1}{\Deltath^2}\;$, also increases as $L$ increases.
%
This is consistent with \autoref{fig:t_star_trends_rkc}, for example.
The above arguments and empirical observation mean that $\taudev$ against $\alpha$ displays a trough shape, and the appropriate $\alpha$ should be located at the bottom of the trough. 
%
%
For certain models, we observe empirically that the bottom of the trough is becoming flatter and increasing in size (as argued above), which suggest that the ratio $\DeltaD^2/\Deltath^2$ is increasing in system size.
The flatness of the trough in $\taudev$ versus $\alpha$ implies that $\taudev$ becomes less and less dependent on the choice of $\alpha$ (within the trough) as $L$ increases.
%
%
For example, consider the $\taudev$ against $\tilde{\alpha}$  plot for RKC in~\autoref{fig:t_star_trends_rkc}, and for SYK-L in~\autoref{fig:t_dev_trends_rl}. The valley shape is increasingly apparent as $L$ increases, which allows us to identify the regime $\frac{1}{\DeltaD^2} < \alpha < \frac{1}{\Deltahei^2}$.  Furthermore, we see that
in this regime, for most models, neither the value of $\taudev$ at fixed $L$, 
nor the behavior of $\taudev$ as a function of $L$ 
vary greatly with $\tilde \alpha$. 
Note that however, for 1D-RL and SYK-L, the dependence of $\taudev$ on $\tilde \alpha$ is not stable except for the largest two system sizes. This sensitivity is present especially since $\taudev$ are small, i.e. these models become RMT-like quickly for small $L$. 
Therefore, for sufficiently large system size, we may identify
$\taudev$ with the Thouless time $\tauth$, independent from details of the filtering protocol like the precise value of $\tilde \alpha$ as long as $\tilde{\alpha}$ is within the trough. While all models display a trough structure in $\taudev$ against $\tilde{\alpha}$, some do not display the trough-widening behaviour in larger $L$, e.g. 1D-RL \autoref{fig:t_dev_trends_rl}, and the suitable choice of $\tilde{\alpha}$ may depend on $L$ in general.
In practice, 
we choose $\tilde{\alpha}= 1$, which lies roughly in the trough of the $\taudev$ versus $\tilde\alpha$ curve.
%
%
Similar to Fig.~\ref{fig:unfold} in the main text, in \autoref{fig:t_dev_trends_sykl}
and \autoref{fig:t_dev_trends_dxxz}, we provide plots for the single-realization spectrum, unfolded spectrum, DSFF for varying filtering strength, and $\taudev$ against $\tilde{ \alpha}$, which indeed display the trough structure.
%
%

\section{Sanity checks}\label{app:sanity}
As sanity checks for our filtering procedure, we compute the DSFF of GinUE after deforming the DOS with a variety of conformal transformations.  We also compare with the DSFF of the RKO, which has a flat spectrum and is expected to be described by RMT.

\begin{figure}[H]
\begin{minipage}[t]{0.31\textwidth}
\includegraphics[width=\linewidth,keepaspectratio=true]{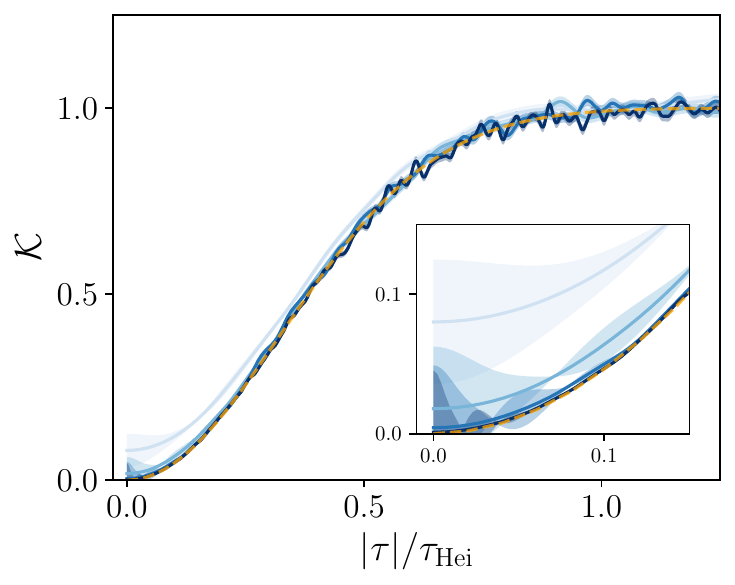}
\end{minipage}
\hspace*{\fill} 
\begin{minipage}[t]{0.31\textwidth}
\includegraphics[width=\linewidth,keepaspectratio=true]{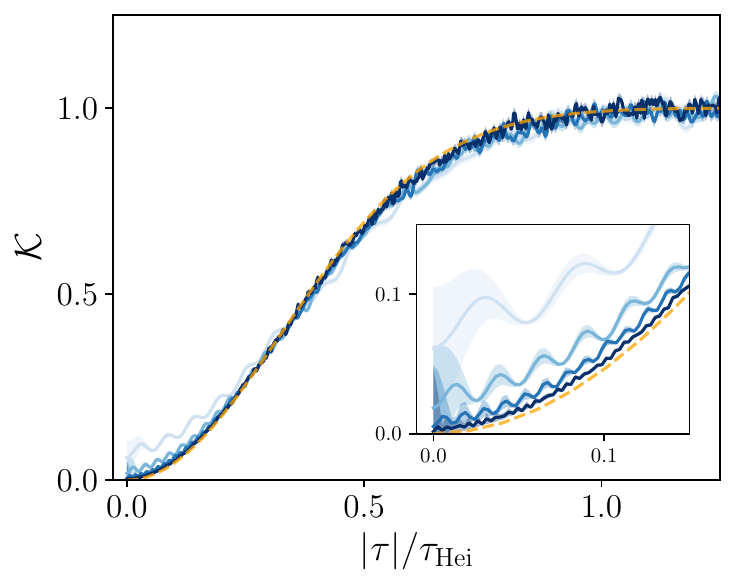}
\end{minipage}
\hspace*{\fill} 
\begin{minipage}[t]{0.31\textwidth}
\includegraphics[width=\linewidth,keepaspectratio=true]{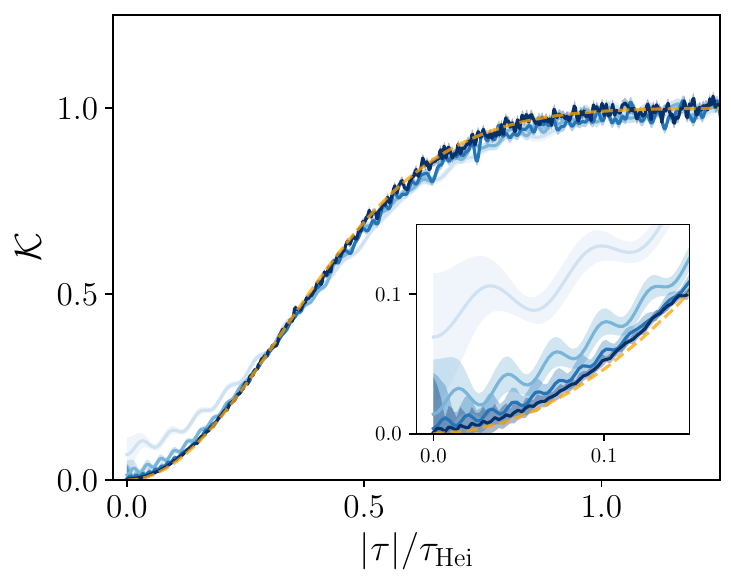}
\end{minipage}
    \caption{DSFF for $\{z\}$ of GinUE (left), $\{\log z\}$ of GinUE (middle), and $\{z^p\}$ of GinUE (right) at 
    $\theta = \pi/4$ with 5000 samples. System size is varied from $N = 4^3,4^4,4^5,4^6$ from light blue to
    dark blue.
    All presented curves are computed with filter strength $\tilde \alpha = 1$.
    The DSFF analytical solution for  GinUE  are plotted in orange. 
The DSFF behaviour for small $|\tau|/\tauhei$ are plotted in the insets.
       }
    \label{fig:san_check_vary_L_fixed_alpha}
\end{figure}

\begin{figure}[H]
\begin{minipage}[t]{0.31\textwidth}
\includegraphics[width=\linewidth,keepaspectratio=true]{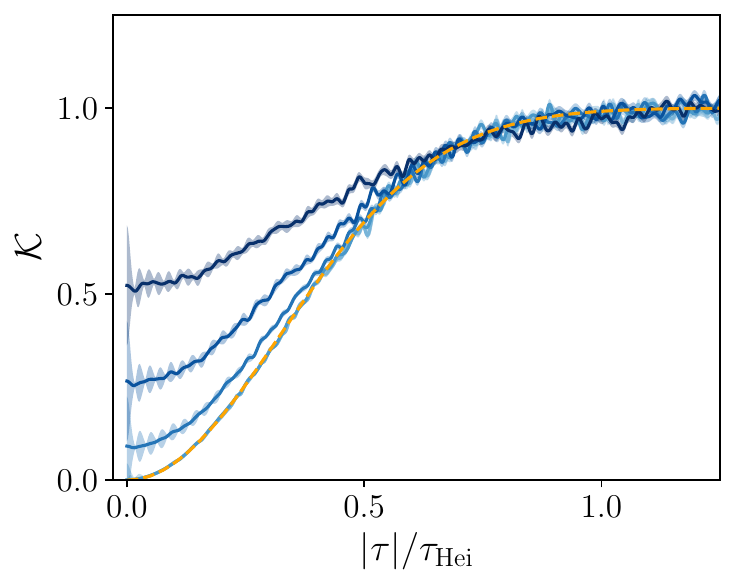}
\end{minipage}
\hspace*{\fill} 
\begin{minipage}[t]{0.31\textwidth}
\includegraphics[width=\linewidth,keepaspectratio=true]{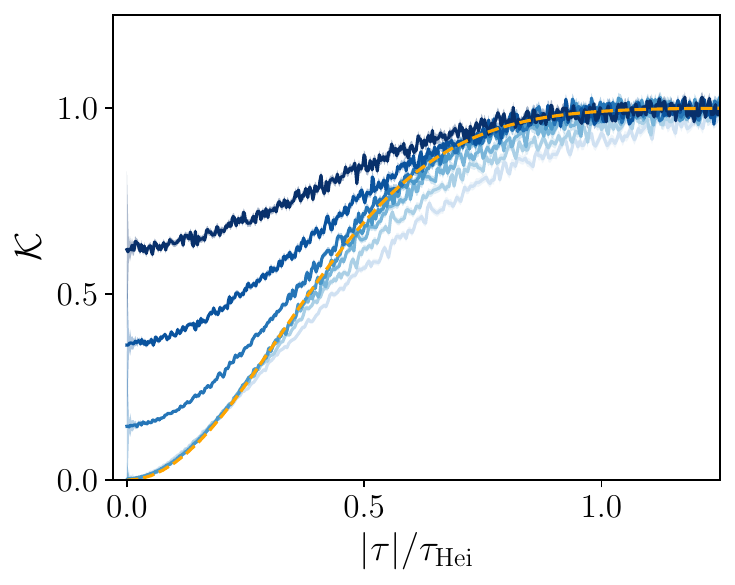}
\end{minipage}
\hspace*{\fill} 
\begin{minipage}[t]{0.31\textwidth}
\includegraphics[width=\linewidth,keepaspectratio=true]{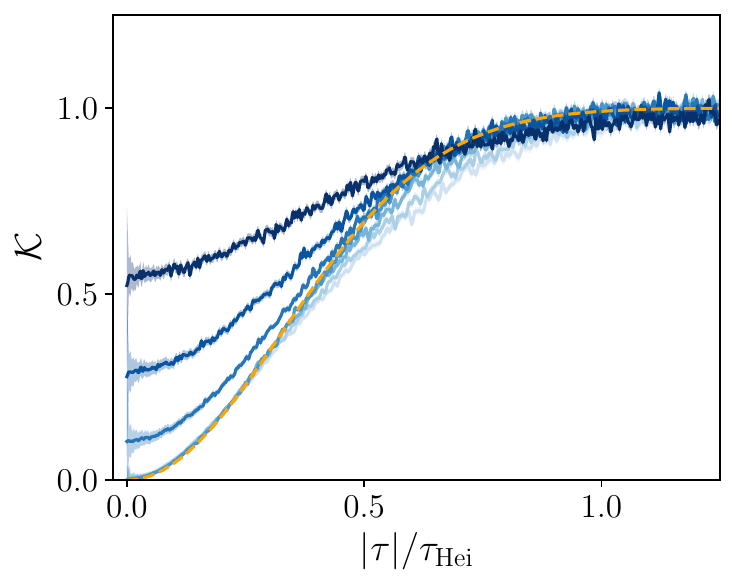}
\end{minipage}
    \caption{ 
    DSFF for $\{z\}$ of GinUE (left), $\{\log z\}$ of GinUE (middle), and $\{z^p\}$ of GinUE (right) at 
    $\theta = \pi/4$ with {5000 samples}.  
    The dimension is chosen with $N = 4^6$.
    The filter strength is increased from $\tilde \alpha \in [1/20, 1/10, 1/5, 1, 10, 20, 40]$  from light blue
    to dark. 
    For the weakest filters (lightest blue), $\tilde \alpha = 1/20, 1/10, 1/5, 1$,
    the DSFF collapses well onto the theoretical curve (orange).
    }
    \label{fig:san_check_fixed_L_vary_alpha}
\end{figure}

Our sanity checks are:
\begin{enumerate}
    \item  DSFF of $\{z \} $ of GinUE with filtering, and
    \item DSFF of $\{\log z \} $ of GinUE with filtering, and
    \item  DSFF of $\{z^{p} \} $ of GinUE with filtering and $p=\frac{1}{4}$. 
\end{enumerate}
These checks are particularly instructive because the form of DSFF for GinUE without filtering is exactly known~\eqref{eq:dsff_ginue_largeN}. As we show in \autoref{fig:san_check_vary_L_fixed_alpha}, for fixed $\tilde{\alpha}=1$, the DSFF approaches the GlnUE solution~\eqref{eq:dsff_ginue_largeN} as $N$ increases. In \autoref{fig:san_check_fixed_L_vary_alpha}, for fixed $N$, the DSFF fits the GinUE solution for sufficiently large  $\talpha$, until an early-time plateau forms for very large $\talpha$.

\subsection*{Example: DSFF of RKO with filtering}

\begin{figure}[H]
\begin{minipage}[t]{0.475\textwidth}
\includegraphics[width=\linewidth,keepaspectratio=true]{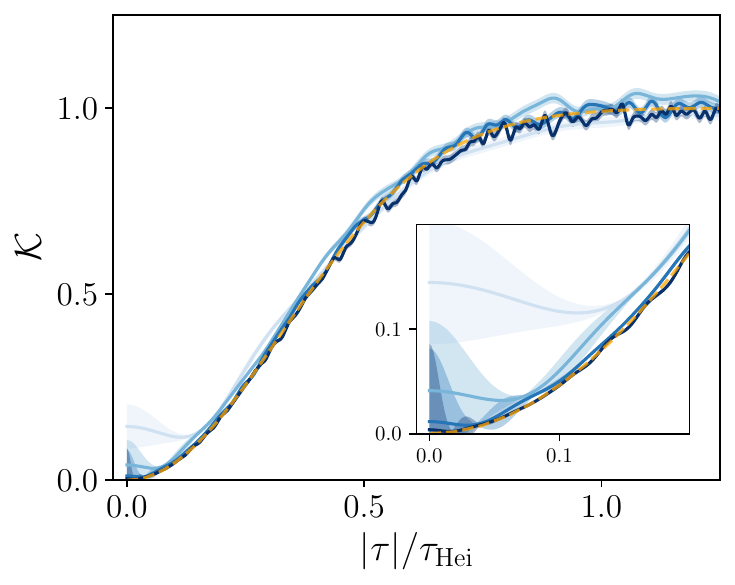}
\end{minipage}
\hspace*{\fill} 
\begin{minipage}[t]{0.475\textwidth}
\includegraphics[width=\linewidth,keepaspectratio=true]{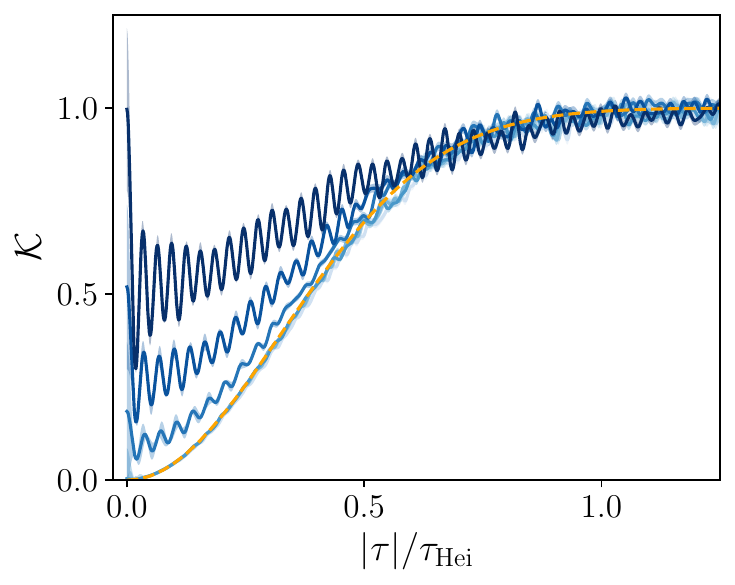}
\end{minipage}
    \caption{Left: DSFF for RKO at 
    $\theta = \pi/4$ with {5000 samples}.  System size is varied from $N = 4^3,4^4,4^5,4^6$ from light blue to
    dark blue.
    All presented curves are computed with filter strength $\tilde \alpha = 1$.
The DSFF behaviour for small $|\tau|/\tauhei$ are plotted in the insets.
    Right:
DSFF for RKO at 
    $\theta = \pi/4$ with {5000 samples}.  
    The dimension is chosen with $N = 4^6$.
    The filter strength is increased from light blue
    to dark, with $\tilde \alpha \in [1/20, 1/10, 1/5, 1, 10, 20, 40]$.
    The DSFF analytical solution for  GinUE  are plotted in orange. 
    %
    %
    }
    \label{fig:sanity_dsff_0d} \label{fig:sanity_dsff_vary_alpha_0D}
\end{figure}


\section{Summary table and additional numerics for dissipative spectral form factor (DSFF) }\label{app:dsff}

\begin{figure}[H]
\centering
\includegraphics[scale = 0.6]{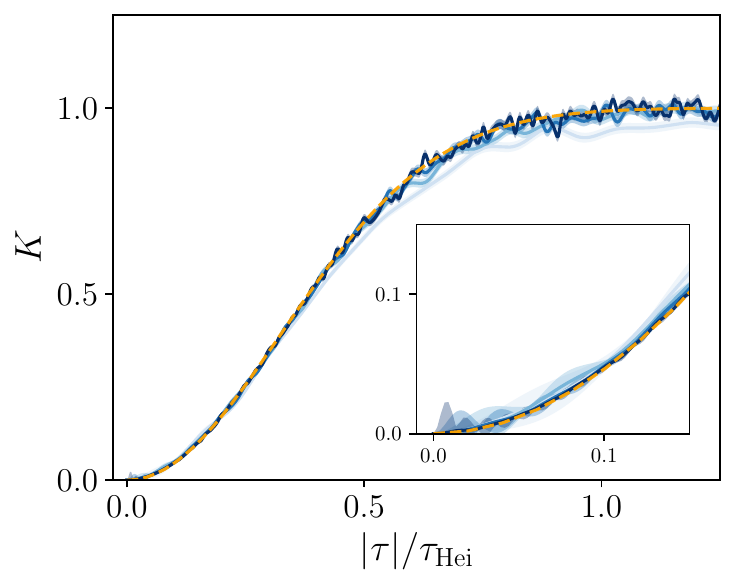}
    \caption{DSFF for RKO without filtering at $\theta=\pi/4$
    and with 5000 samples.  
    The dimension is increased from $N = 4^3,4^4,4^5,4^6$, from light blue to dark blue.
    The DSFF behaviour for small $|\tau|/\tauhei$ are plotted in the insets.
    The DSFF analytical solution for  GinUE  are plotted in orange. 
    } \label{fig:ko_dsff_0d_pure}
\end{figure}

\begin{figure}[H]
\begin{minipage}[t]{0.45\textwidth}
\includegraphics[width=\linewidth,keepaspectratio=true]{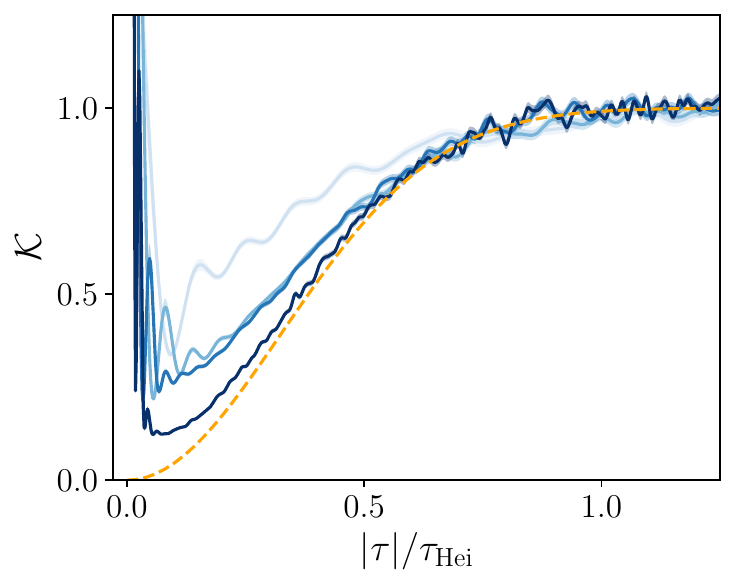}
\end{minipage}
\hspace*{\fill} 
\begin{minipage}[t]{0.48\textwidth}
\includegraphics[width=\linewidth,keepaspectratio=true]{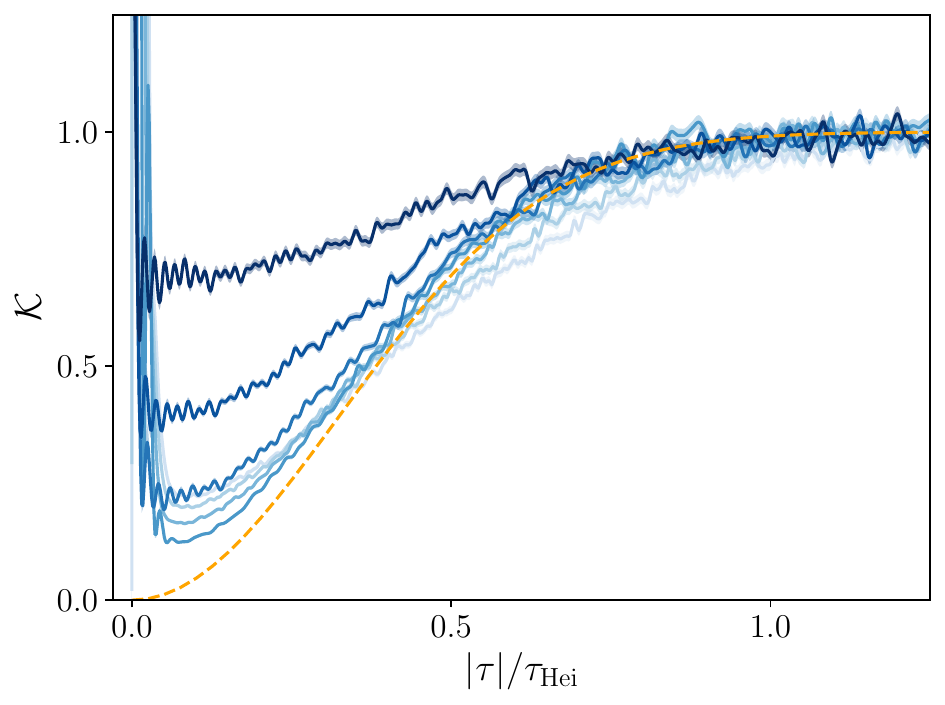}
\end{minipage}
    \caption{ 
    The DSFF for the $U(1)$-RKC model after unfolding and filtering 
    along $\theta = \pi/4$ and {5000 samples.}  
    On the left, we fix $\tilde \alpha = 1$ and vary $L = 5,6,7,8$ from light to dark blue respectively.
    On the right,
    we vary, from light to dark blue, $\tilde{\alpha} \in [1/20, 1/10, 1/5, 1, 10, 20, 40, 100]$ for fixed system size $L=8$.
    }
    \label{fig:ko_u1_dsff}
\end{figure}

\begin{figure}[H]
\begin{minipage}[t]{0.24\textwidth}
\includegraphics[width=\linewidth,keepaspectratio=true]{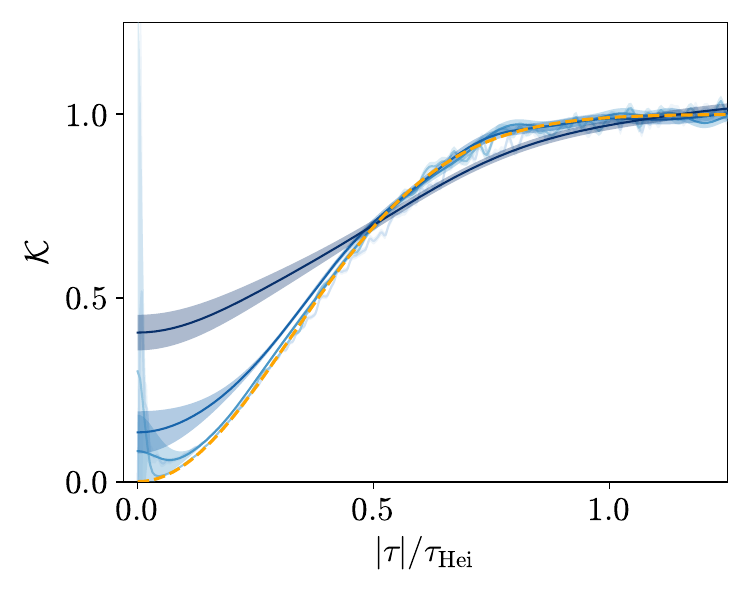}
\end{minipage}
\hspace*{\fill} 
\begin{minipage}[t]{0.24\textwidth}
\includegraphics[width=\linewidth,keepaspectratio=true]{Figures/1D_RMT_L6_gaussian_diff_sig_combined.pdf}
\end{minipage}
\hspace*{\fill} 
\begin{minipage}[t]{0.24\textwidth}
\includegraphics[width=\linewidth,keepaspectratio=true]{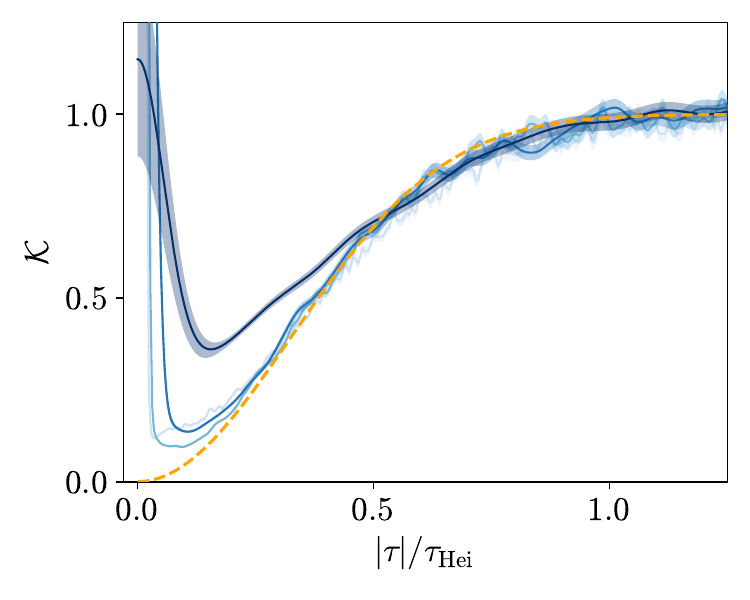}
\end{minipage}
\hspace*{\fill} 
\begin{minipage}[t]{0.24\textwidth}
\includegraphics[width=\linewidth,keepaspectratio=true]{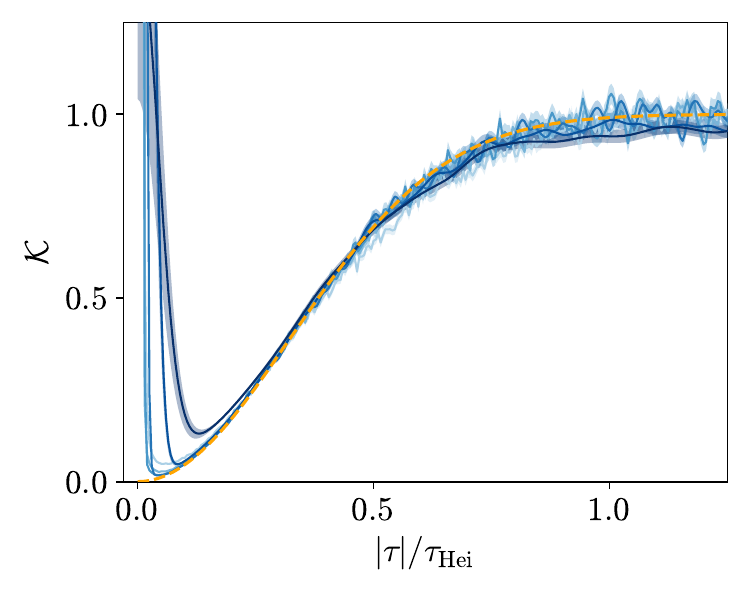}
\end{minipage}
    \caption{ The DSFF for the 0D-RL, 1D-RL, WS-$U(1)$-RL, SS-$U(1)$-RL after unfolding and filtering 
    along $\theta = \pi/4$ and with {5000 samples.}    We vary, from light to dark blue, $\tilde{\alpha} \in [0,\frac{1}{5},\frac{2}{3},2,20, 60,200]$ for the  fixed system size $L=6$ or $L_{\mathrm{eff}} = 6$.
    }
    \label{fig:lindblad_fixed_L_vary_alpha}
\end{figure}

In this section, we provide additional numerics of the DSFF. Il \autoref{fig:ko_dsff_0d_pure}, we provide the DSFF of RKO with different system sizes $L$ without filtering. In \autoref{fig:ko_u1_dsff} and \autoref{fig:lindblad_fixed_L_vary_alpha}, we provide the DSFF of different Kraus operators and Lindbladians with fixed $L$ and varying filtering strength. Lastly, we provide a table of DSFF of all models with the corresponding unfolding function, filtering function, DSFF against system sizes $L$, and deviation time $\taudev$ against system sizes $L$.

\begin{table}
\begin{center}
\caption{DSFF with fits from the GinUE, Unfolding and filtering methods, and $\taudev$ and $\tauplateau$ for all models. Note that for certain models (e.g. $U(1)$-RKC, dXXZ), DSFF has not converged to GinUE RMT behaviour for the accessible system sizes yet, and consequently the trend of $\taudev$ in system size is not representative.}
\begin{tabular}{c|c|c|c|c}
     & Unfolding & Filtering &  DSFF \& GinUE fits & $\tau_
     \mathrm{dev}$ \\
     \hline
     RKO & N/A & N/A 
     & 
     \begin{minipage}[h]{0.20\textwidth}
    \includegraphics[width=\linewidth,keepaspectratio=true]{Figures/000_DSFF_Sample_0d_d20.pdf}
    \end{minipage}
     & 
     \begin{minipage}[h]{0.20\textwidth}
    \includegraphics[width=\linewidth,keepaspectratio=true]{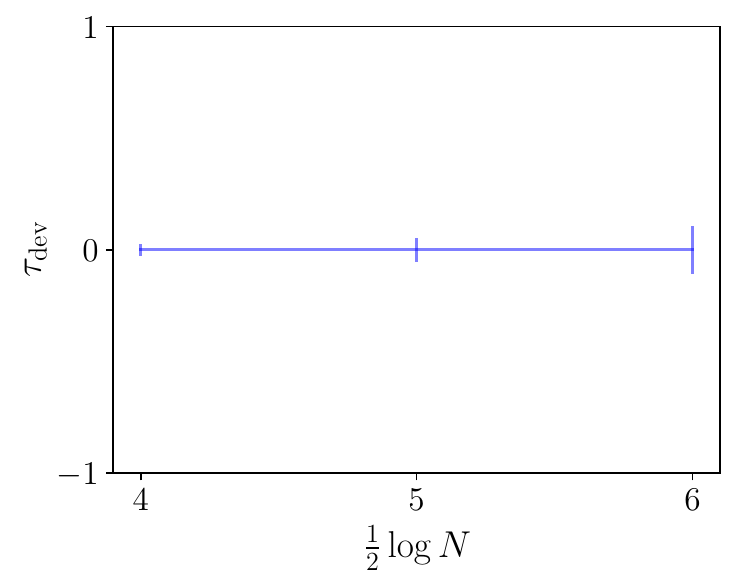}
    \end{minipage}
\\
\hline
     RKC & $z \to z^{1/\log_2 N}$ & Radial Gaussian 
     &
     \begin{minipage}[h]{0.2\textwidth}
    \includegraphics[width=\linewidth,keepaspectratio=true]{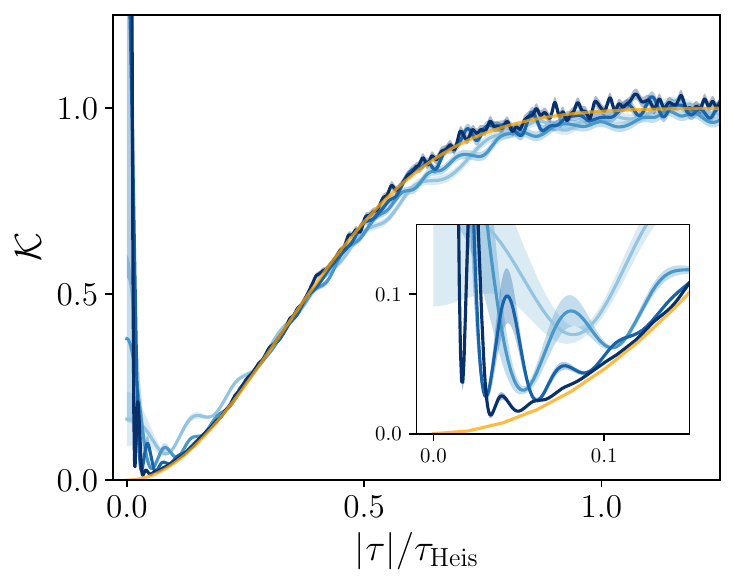}
    \end{minipage}
     &
     \begin{minipage}[h]{0.2\textwidth}
    \includegraphics[width=\linewidth,keepaspectratio=true]{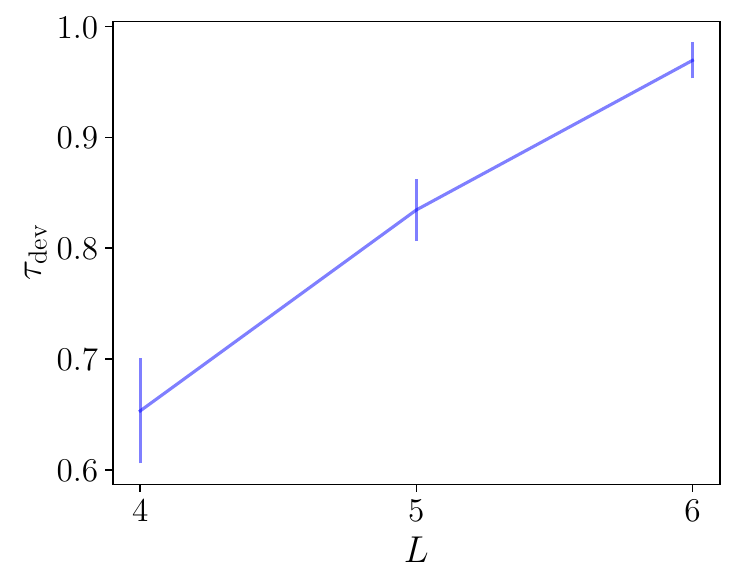}
    \end{minipage}
      \\
\hline
     $U(1)$-RKC & $z \to z^{1/\log_2 N}$ & Radial Gaussian 
     &
     \begin{minipage}[h]{0.2\textwidth}
\includegraphics[width=\linewidth,keepaspectratio=true]{Figures/000_DSFF_Sample_u1_d20.pdf}
    \end{minipage}
     &
     \begin{minipage}[h]{0.2\textwidth}
\includegraphics[width=\linewidth,keepaspectratio=true]{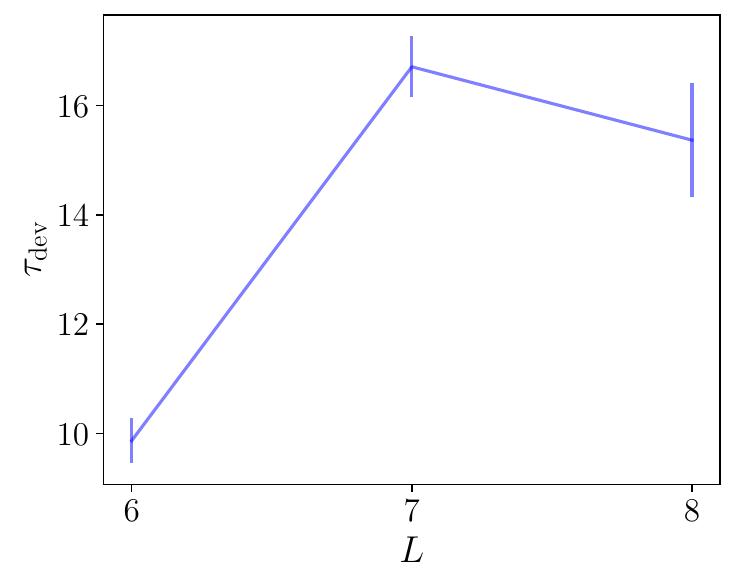}
    \end{minipage}
     \\
\hline
     $\mathbb{Z}_2$-RKC & $z \to z^{1/\log_2 N}$ & Radial Gaussian 
     &
     \begin{minipage}[h]{0.2\textwidth}
    \includegraphics[width=\linewidth,keepaspectratio=true]{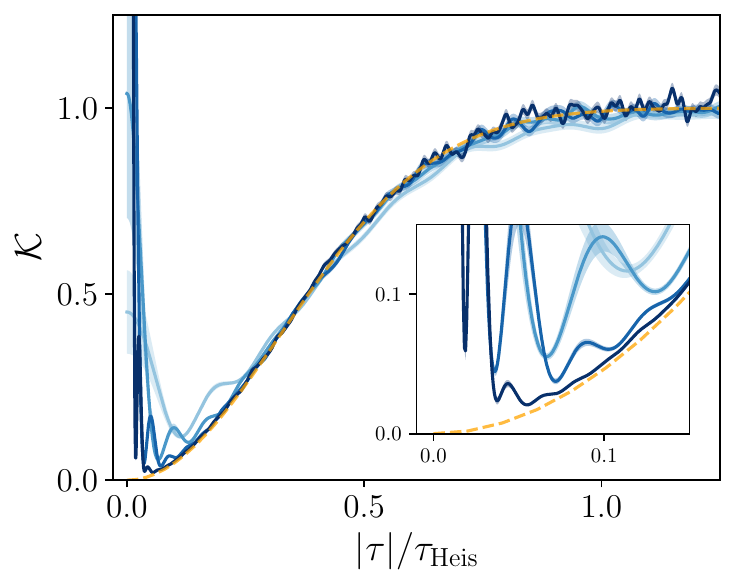}
    \end{minipage}
     &
     \begin{minipage}[h]{0.2\textwidth}
    \includegraphics[width=\linewidth,keepaspectratio=true]{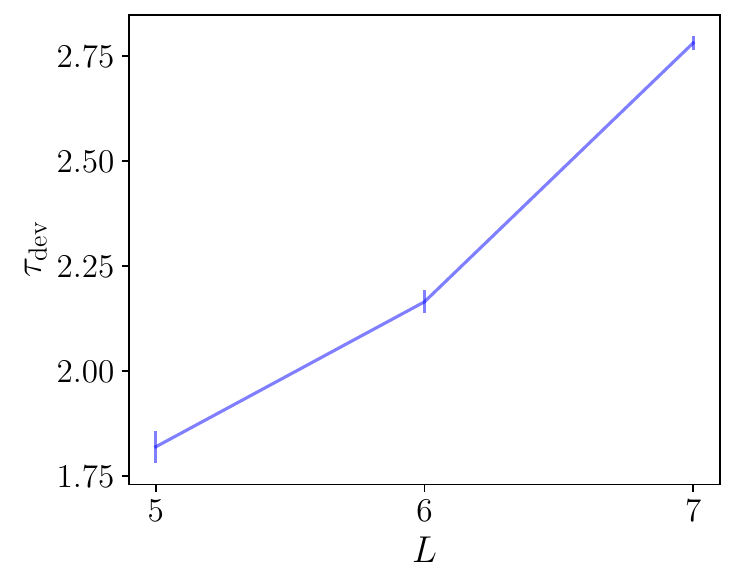}
    \end{minipage}
     \\
\hline
     0D-RL & $z \to z^{1/2}$ & 2D Gaussian 
     &
     \begin{minipage}[h]{0.2\textwidth}
    \includegraphics[width=\linewidth,keepaspectratio=true]{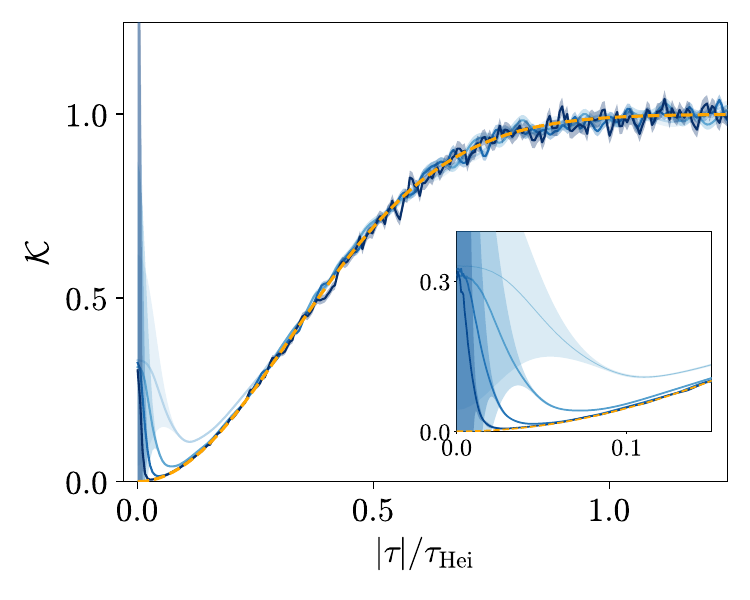}
    \end{minipage}
     &
     \begin{minipage}[h]{0.2\textwidth}
    \includegraphics[width=\linewidth,keepaspectratio=true]{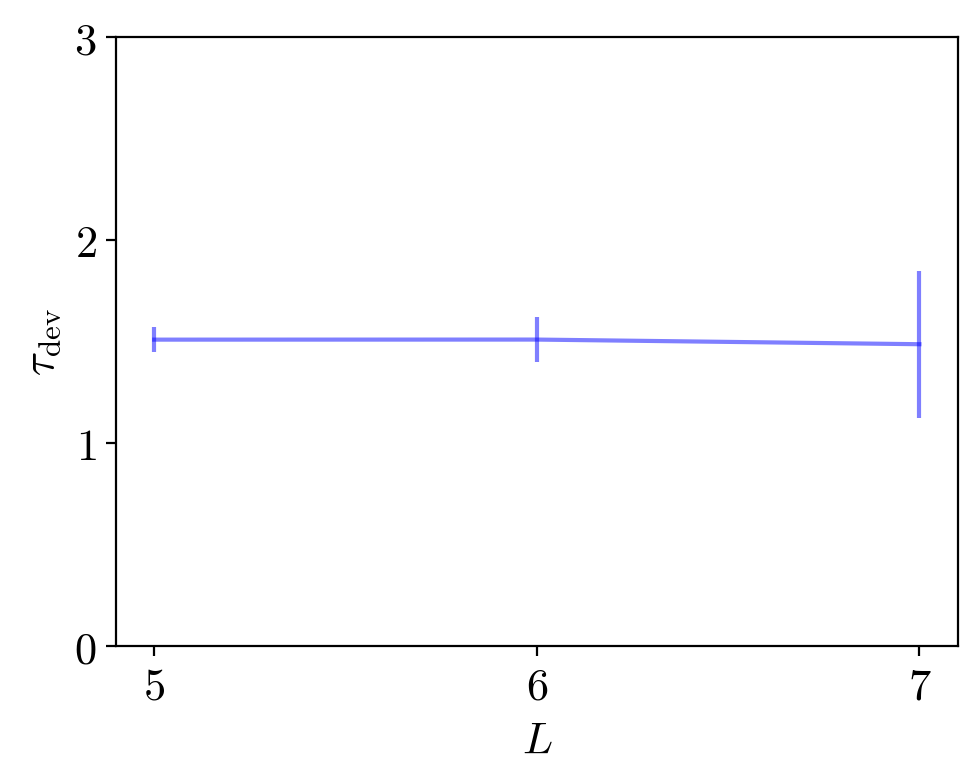}
    \end{minipage}
     \\
\hline
     1D-RL & N/A & 2D Gaussian 
     &
     \begin{minipage}[h]{0.2\textwidth}
    \includegraphics[width=\linewidth,keepaspectratio=true]{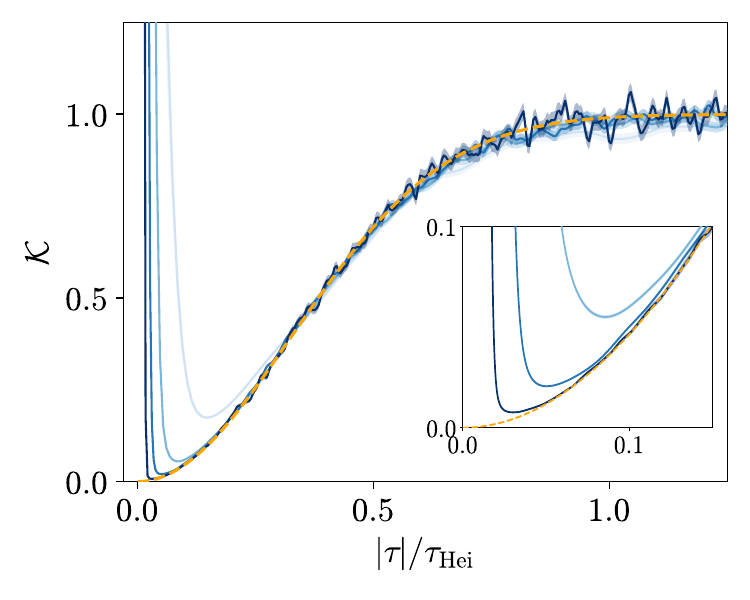}
    \end{minipage}
     &
     \begin{minipage}[h]{0.2\textwidth}
    \includegraphics[width=\linewidth,keepaspectratio=true]{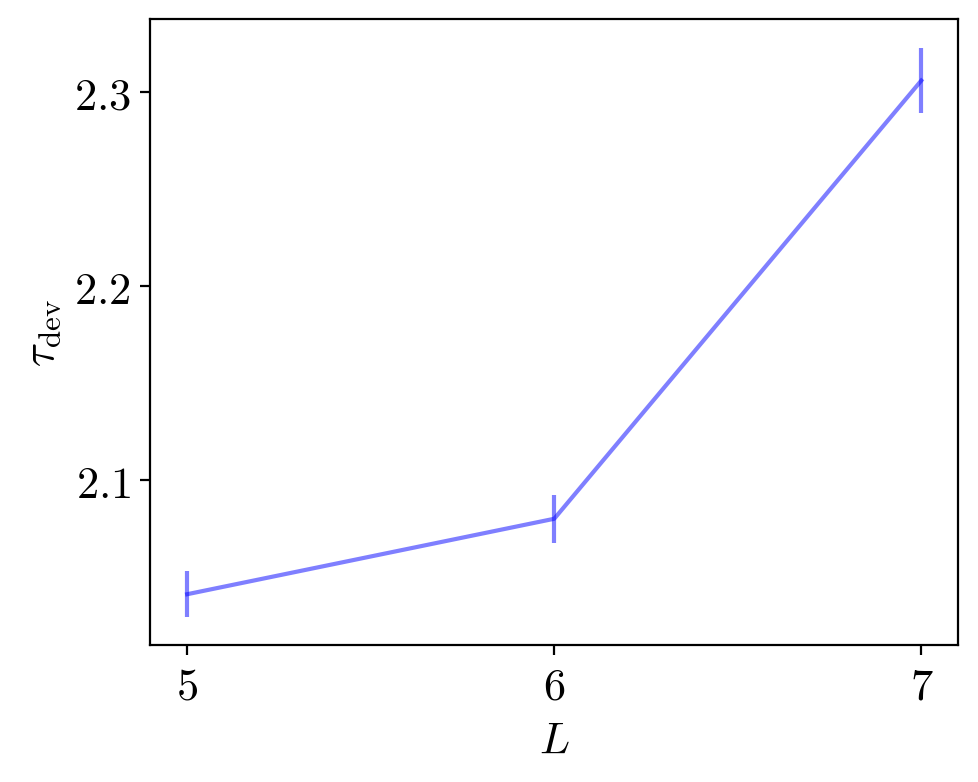}
    \end{minipage}
     \\
\hline
     WS-$U(1)$-RL & N/A & 2D Gaussian 
     &
     \begin{minipage}[h]{0.2\textwidth}
    \includegraphics[width=\linewidth,keepaspectratio=true]{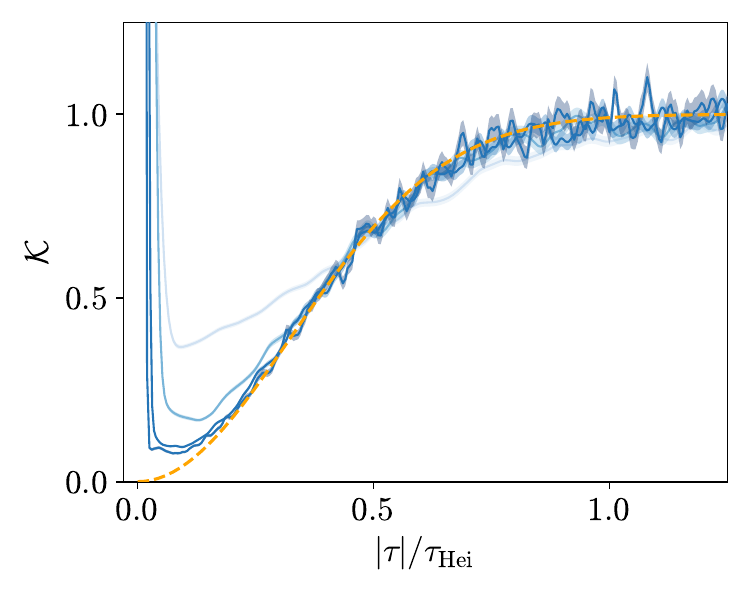}
    \end{minipage}
     &
     \begin{minipage}[h]{0.2\textwidth}
    \includegraphics[width=\linewidth,keepaspectratio=true]{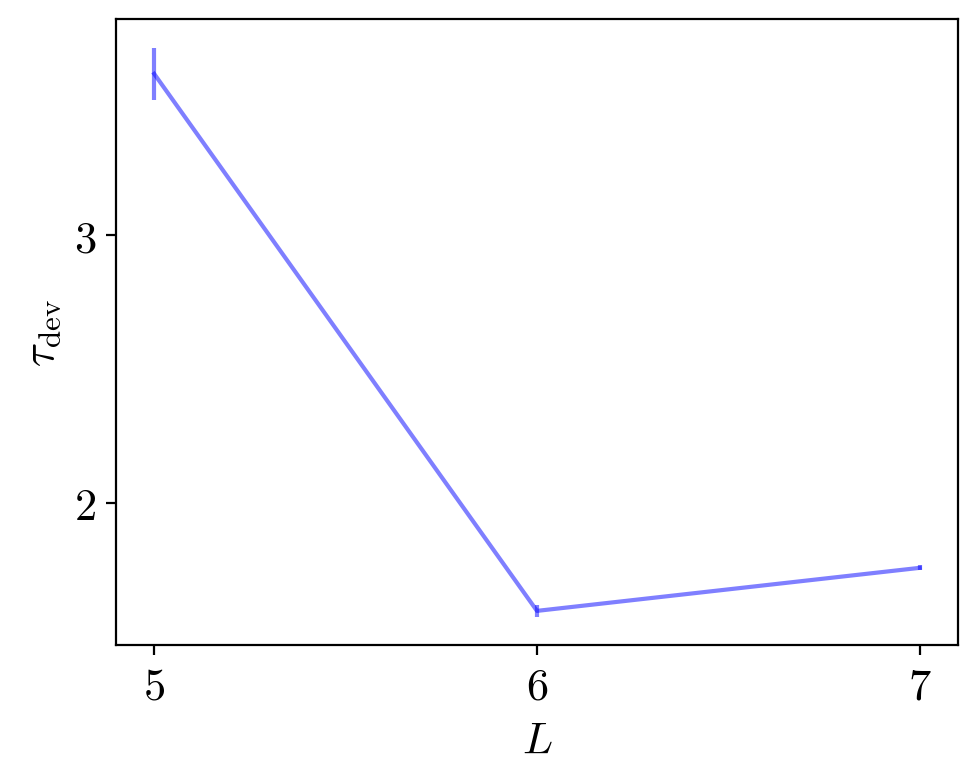}
    \end{minipage}
     \\
\hline
     SS-$U(1)$-RL & N/A & 2D Gaussian 
     &
     \begin{minipage}[h]{0.2\textwidth}
    \includegraphics[width=\linewidth,keepaspectratio=true]{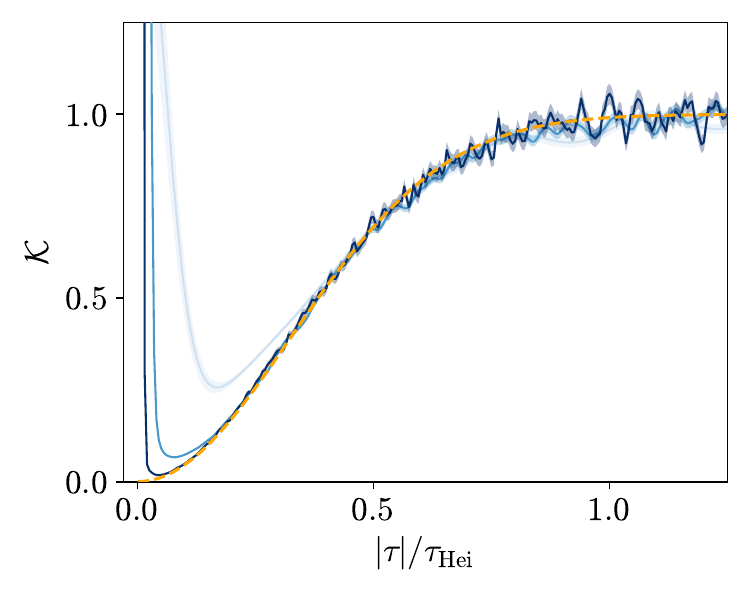}
    \end{minipage}
     &
     \begin{minipage}[h]{0.2\textwidth}
    \includegraphics[width=\linewidth,keepaspectratio=true]{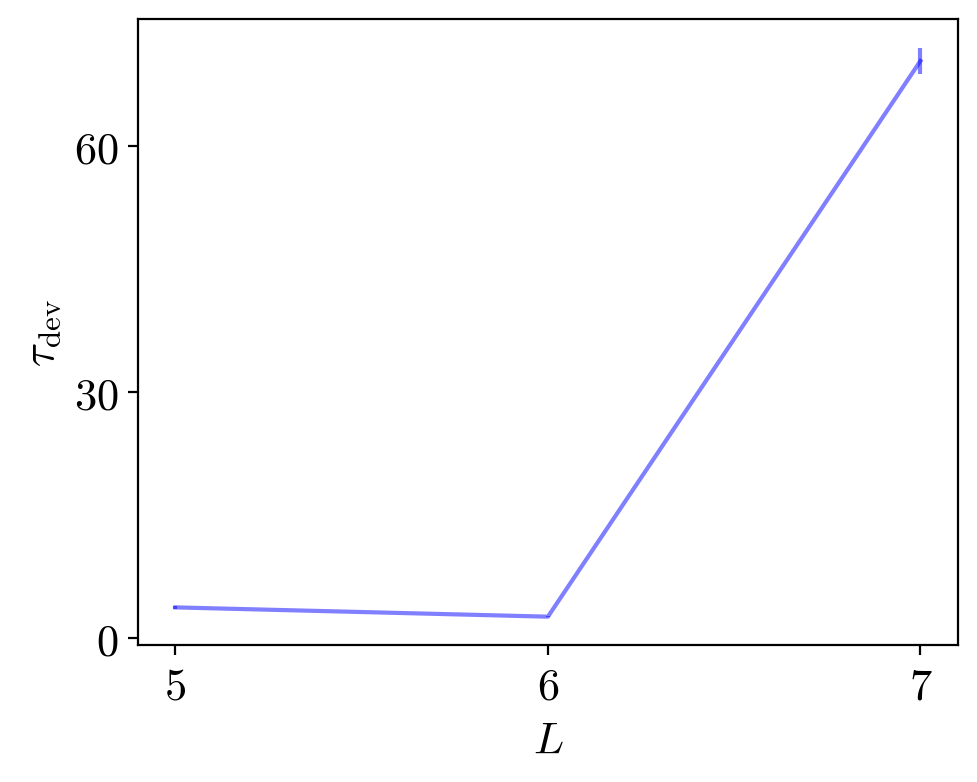}
    \end{minipage}
     \\
\hline
\end{tabular}
\end{center}
\end{table}

\begin{center}
\begin{tabular}{c|c|c|c|c}
     & Unfolding & Filtering & Fits GinUE & $\tau_\mathrm{dev}$ or $\tau_\mathrm{plat}$ \\
\hline
     SYK-L & $z \to z^{1/3}$ & 2D Gaussian 
     &
     \begin{minipage}[h]{0.2\textwidth}
    \includegraphics[width=\linewidth,keepaspectratio=true]{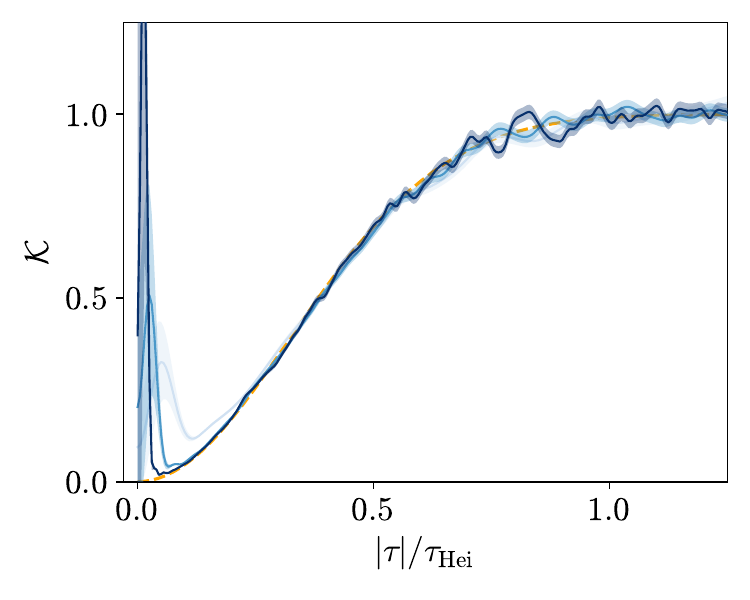}
    \end{minipage}
     &
     \begin{minipage}[h]{0.2\textwidth}
    \includegraphics[width=\linewidth,keepaspectratio=true]{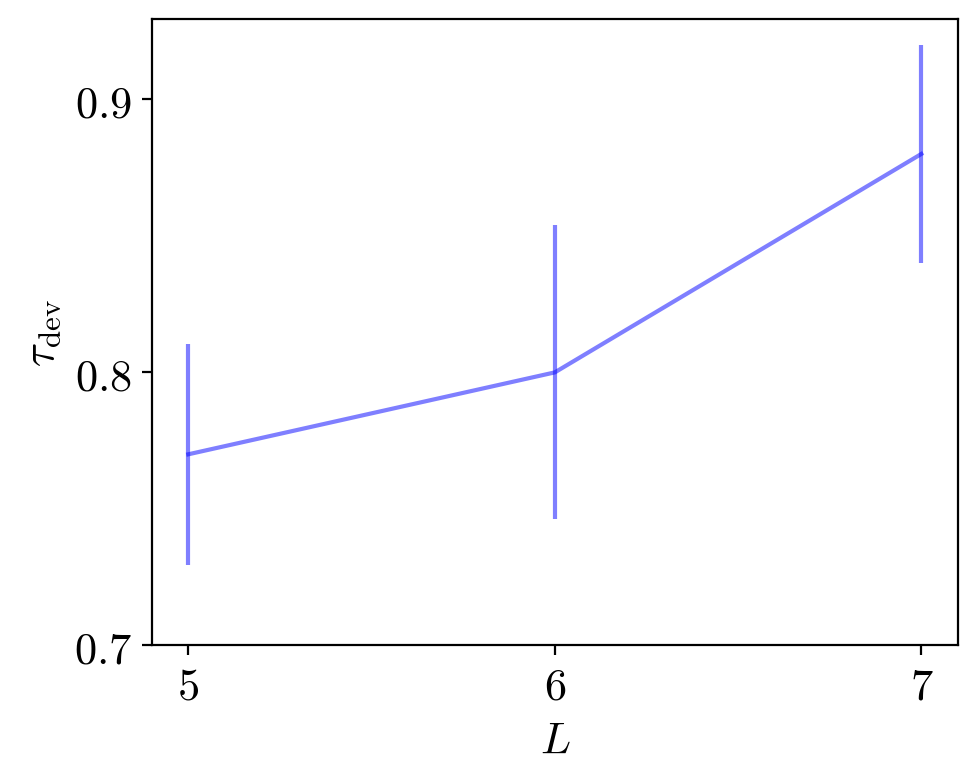}
    \end{minipage}
     \\
\hline
     dXXZ & $z \to (z-z_0)^{1/2}$ & 2D Gaussian 
     &
     \begin{minipage}[h]{0.2\textwidth}
    \includegraphics[width=\linewidth,keepaspectratio=true]{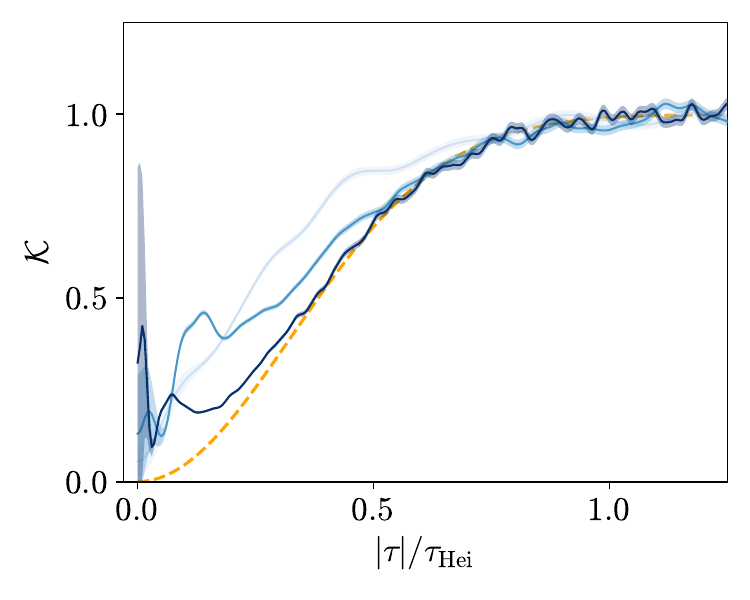}
    \end{minipage}
     &
     \begin{minipage}[h]{0.2\textwidth}
    \includegraphics[width=\linewidth,keepaspectratio=true]{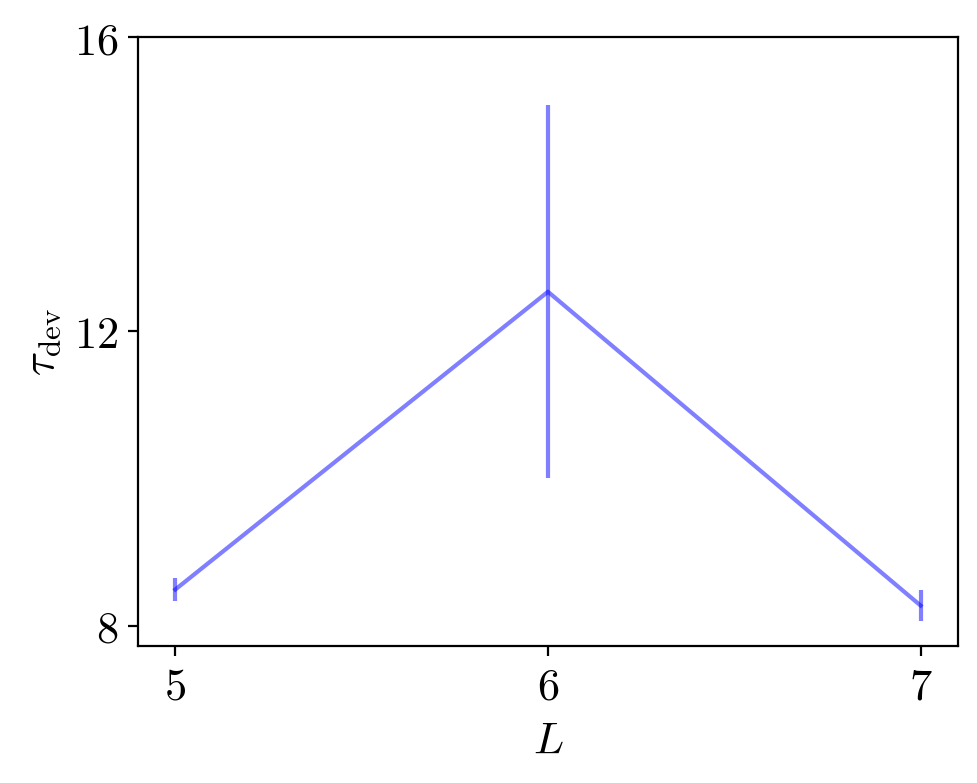}
    \end{minipage}
     \\
\hline
     dXX & N/A &  2D Gaussian
     &
     \begin{minipage}[h]{0.2\textwidth}
    \includegraphics[width=\linewidth,keepaspectratio=true]{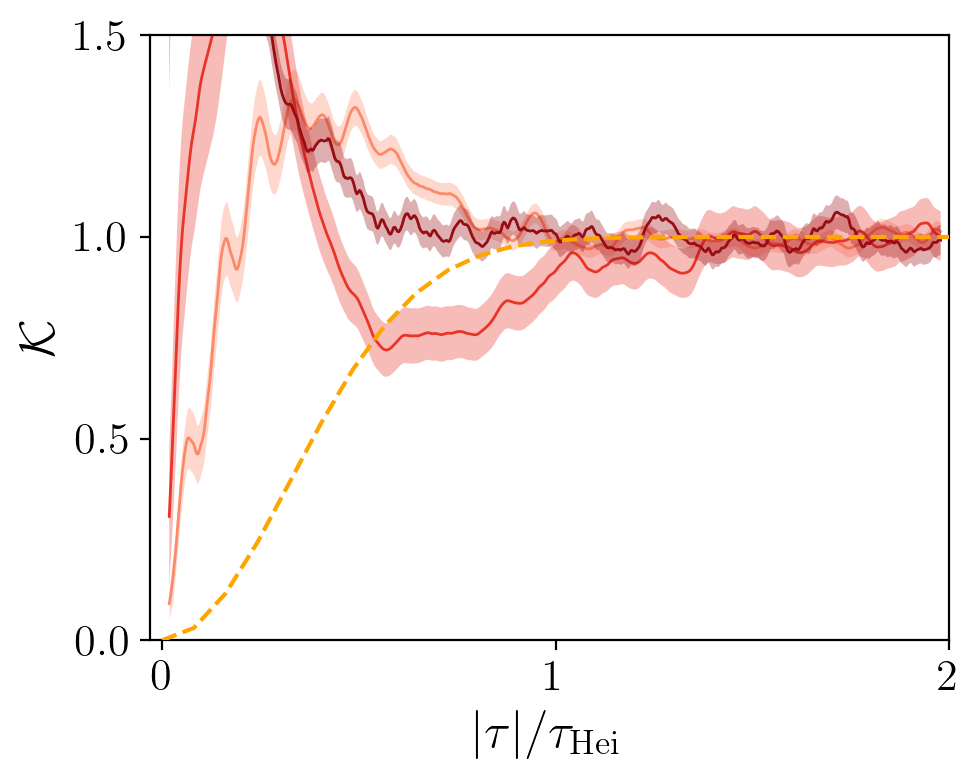}
    \end{minipage}
     &
     \begin{minipage}[h]{0.2\textwidth}
\includegraphics[width=\linewidth,keepaspectratio=true]{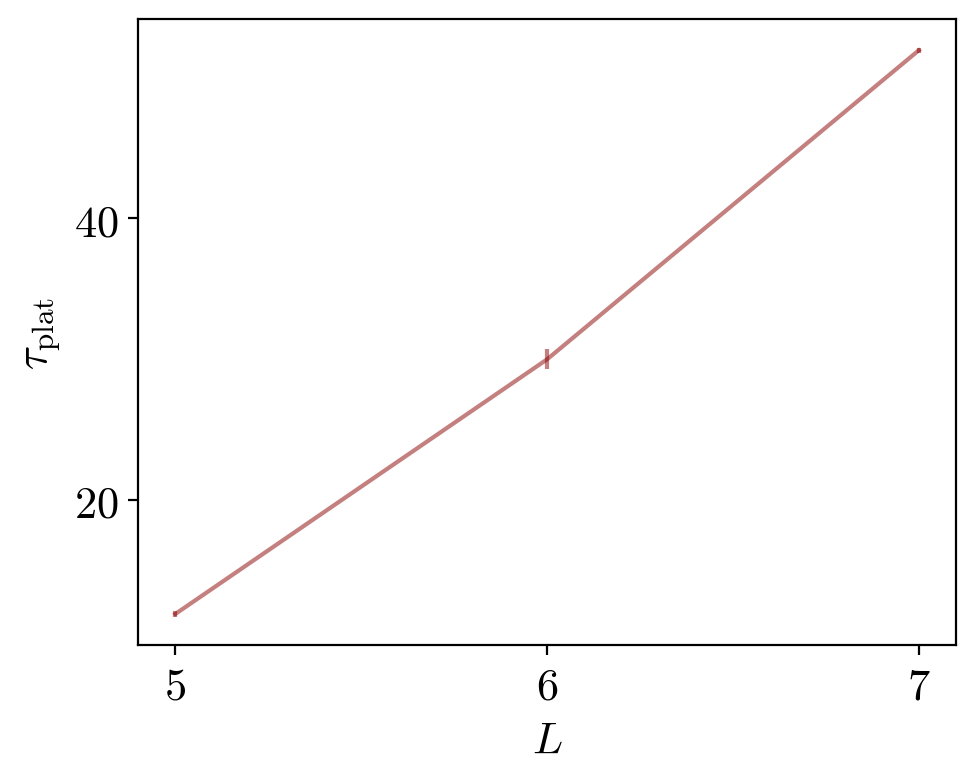}
    \end{minipage}
     \\
\hline
     dIsing-Chaos & N/A & 2D Gaussian 
     &
     \begin{minipage}[h]{0.2\textwidth}
    \includegraphics[width=\linewidth,keepaspectratio=true]{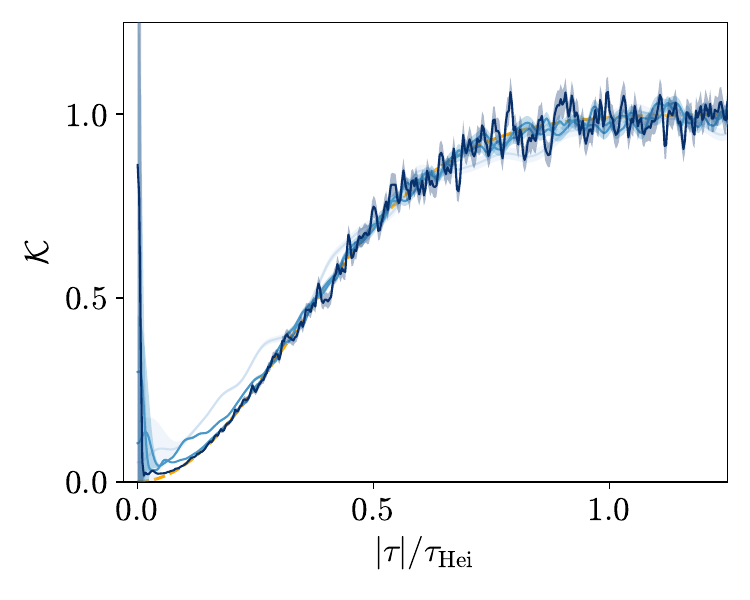}
    \end{minipage}
     &
     \begin{minipage}[h]{0.2\textwidth}
    \includegraphics[width=\linewidth,keepaspectratio=true]{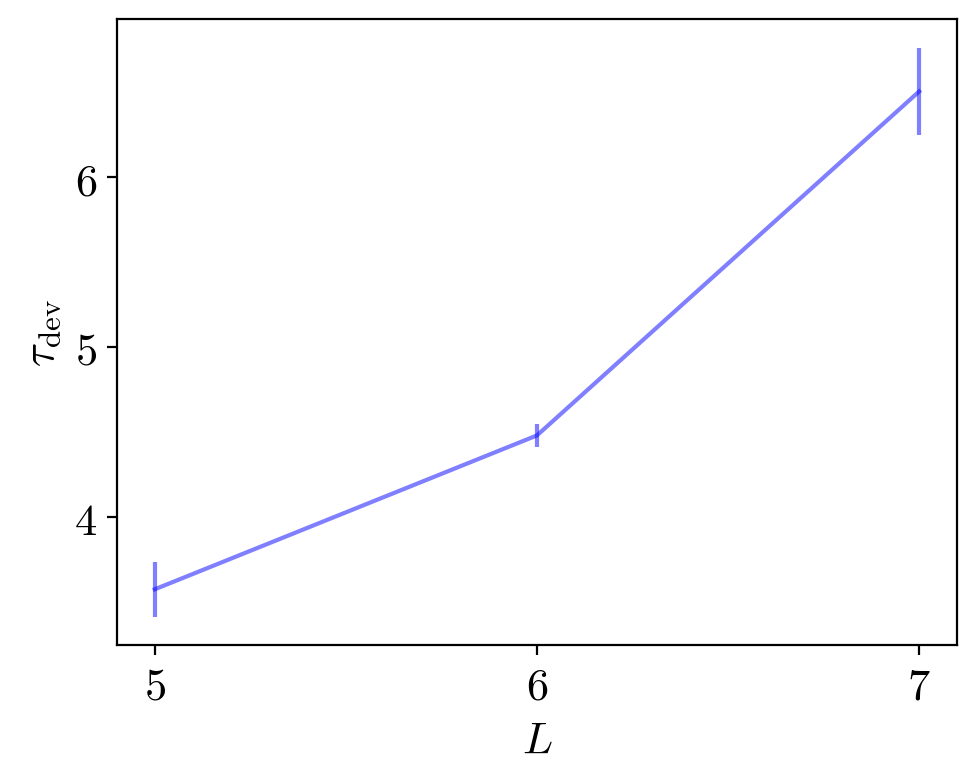}
    \end{minipage}
     \\
\hline
     dIsing-MBL & N/A & 2D Gaussian
     &
     \begin{minipage}[h]{0.2\textwidth}
    \includegraphics[width=\linewidth,keepaspectratio=true]{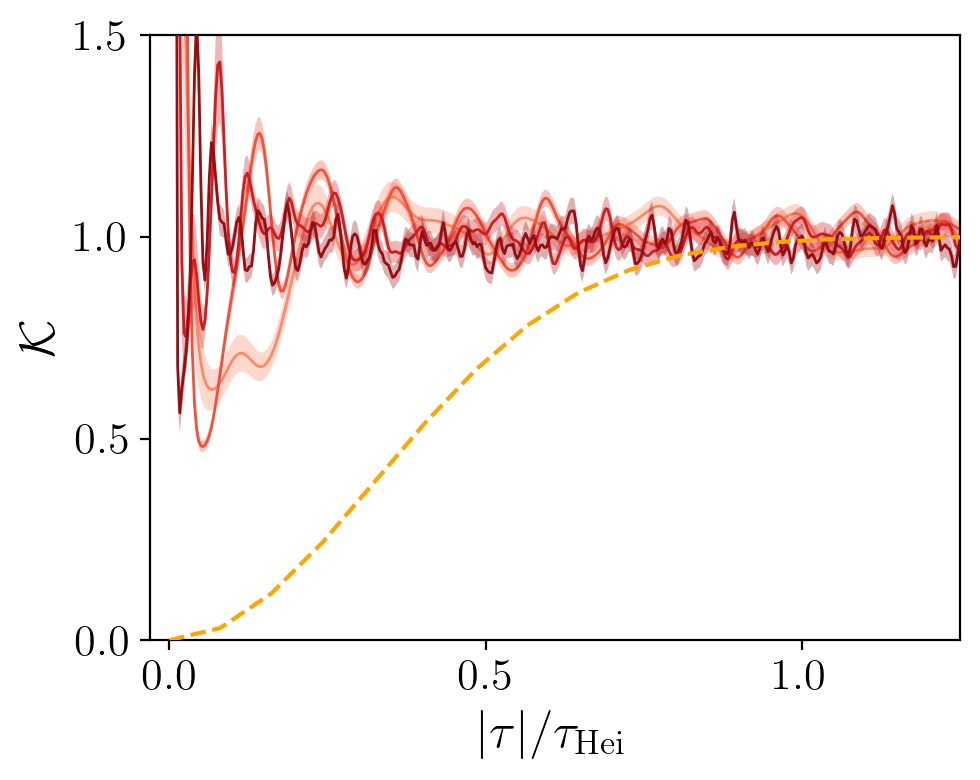}
    \end{minipage}
     &
      \begin{minipage}[h]{0.2\textwidth}
\includegraphics[width=\linewidth,keepaspectratio=true]{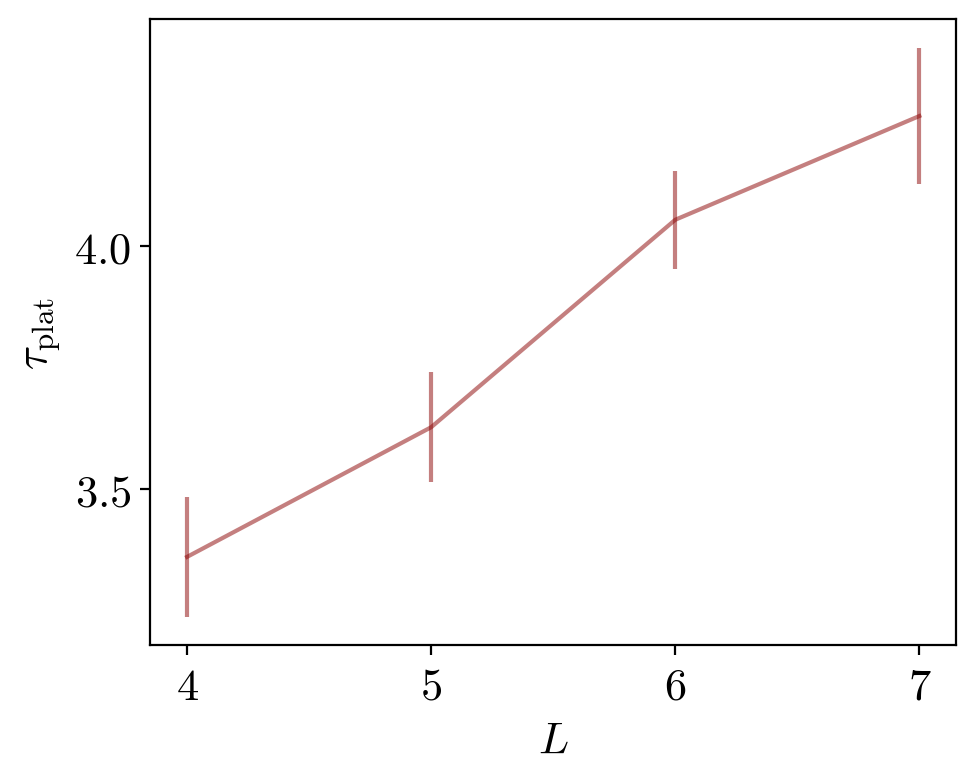}
    \end{minipage}
     \\
\hline
     0D-RL-JO & $z \to z^{1/3}$ & 2D Gaussian
     &
     \begin{minipage}[h]{0.2\textwidth}
    \includegraphics[width=\linewidth,keepaspectratio=true]{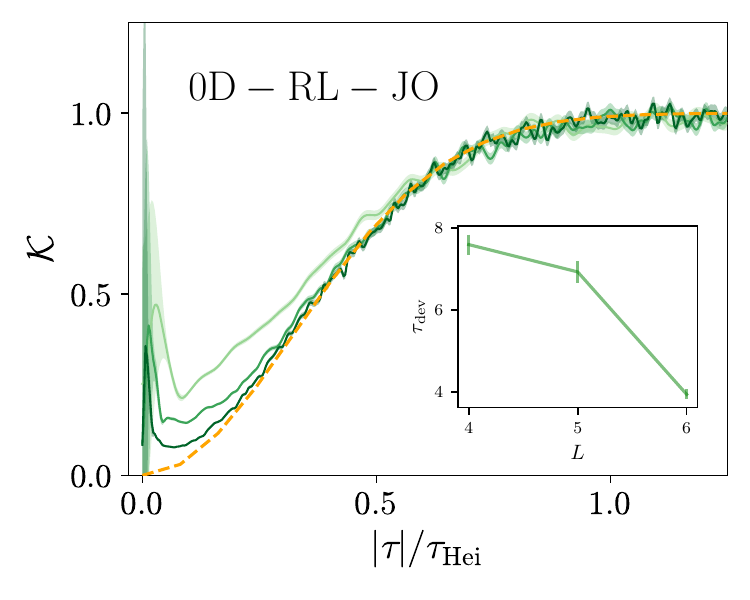}
    \end{minipage}
     &
      \begin{minipage}[h]{0.2\textwidth}
\includegraphics[width=\linewidth,keepaspectratio=true]{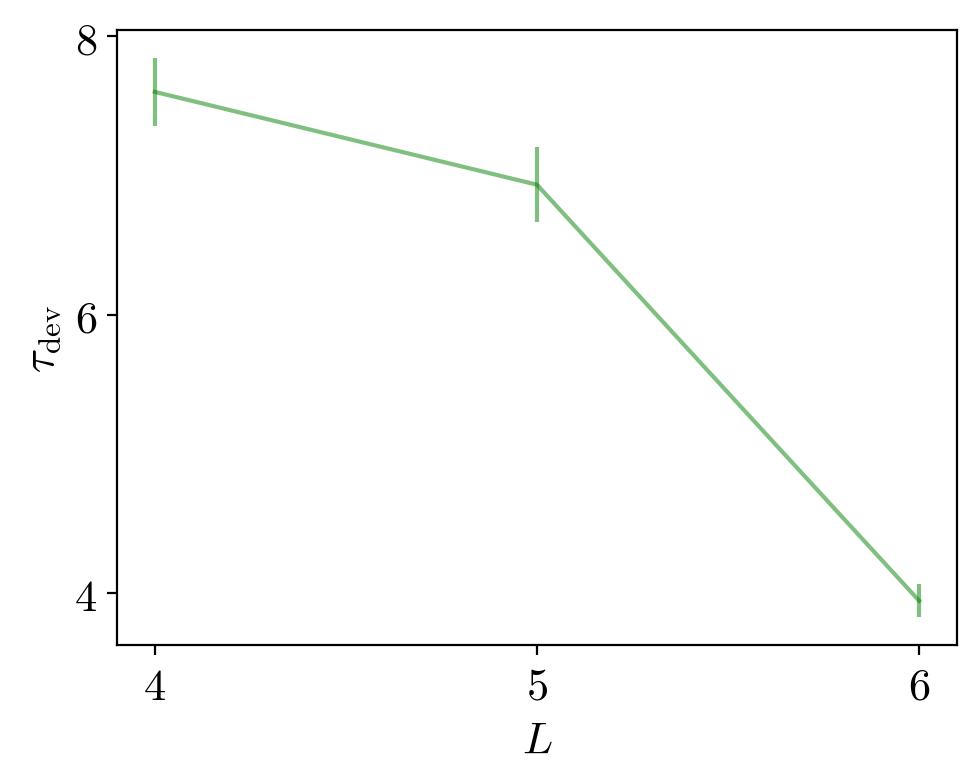}
    \end{minipage}
     \\
\hline
     1D-RL-1JO & $z \to z^{1/2}$ & 2D Gaussian
     &
     \begin{minipage}[h]{0.2\textwidth}
    \includegraphics[width=\linewidth,keepaspectratio=true]{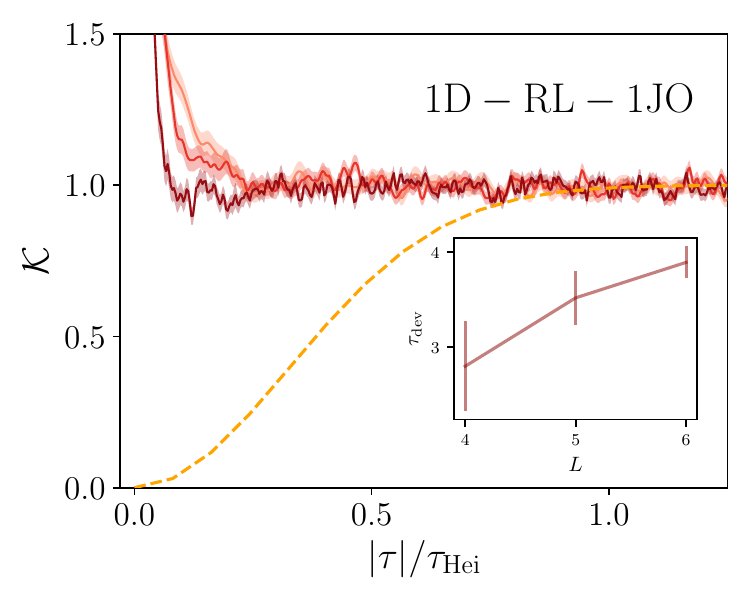}
    \end{minipage}
     &
      \begin{minipage}[h]{0.2\textwidth}
\includegraphics[width=\linewidth,keepaspectratio=true]{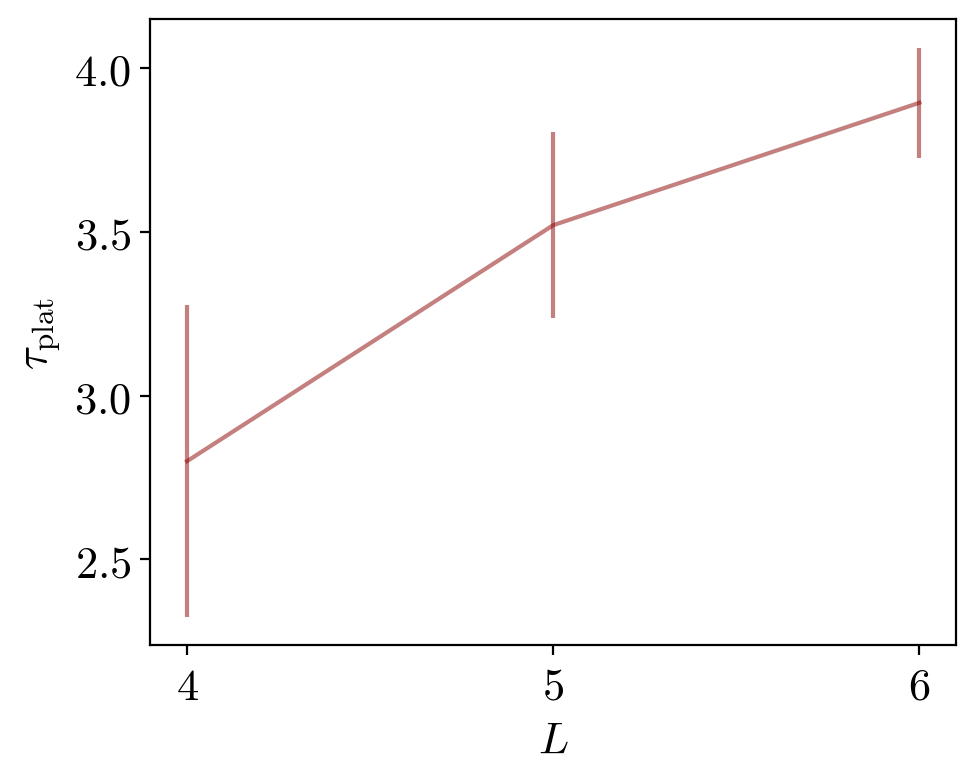}
    \end{minipage}
     \\
\hline
     1D-RL-2JO & $z \to z^{1/2}$ & 2D Gaussian
     &
     \begin{minipage}[h]{0.2\textwidth}
    \includegraphics[width=\linewidth,keepaspectratio=true]{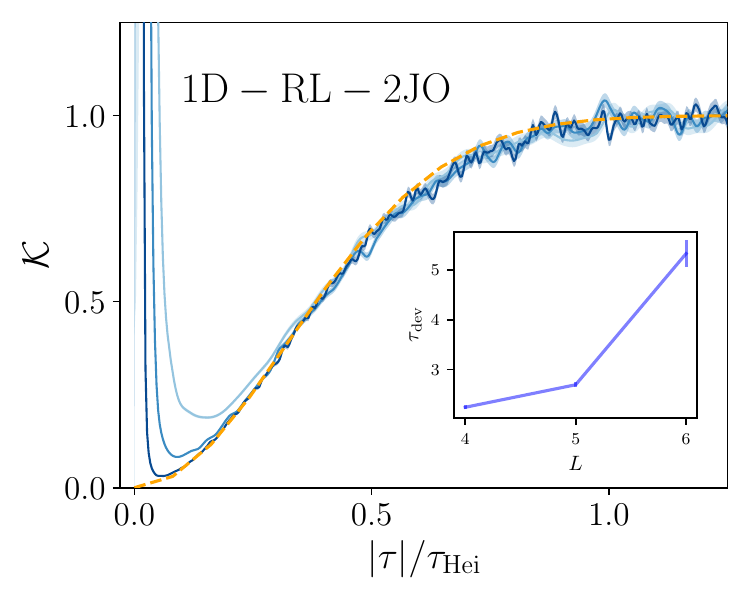}
    \end{minipage}
     &
      \begin{minipage}[h]{0.2\textwidth}
\includegraphics[width=\linewidth,keepaspectratio=true]{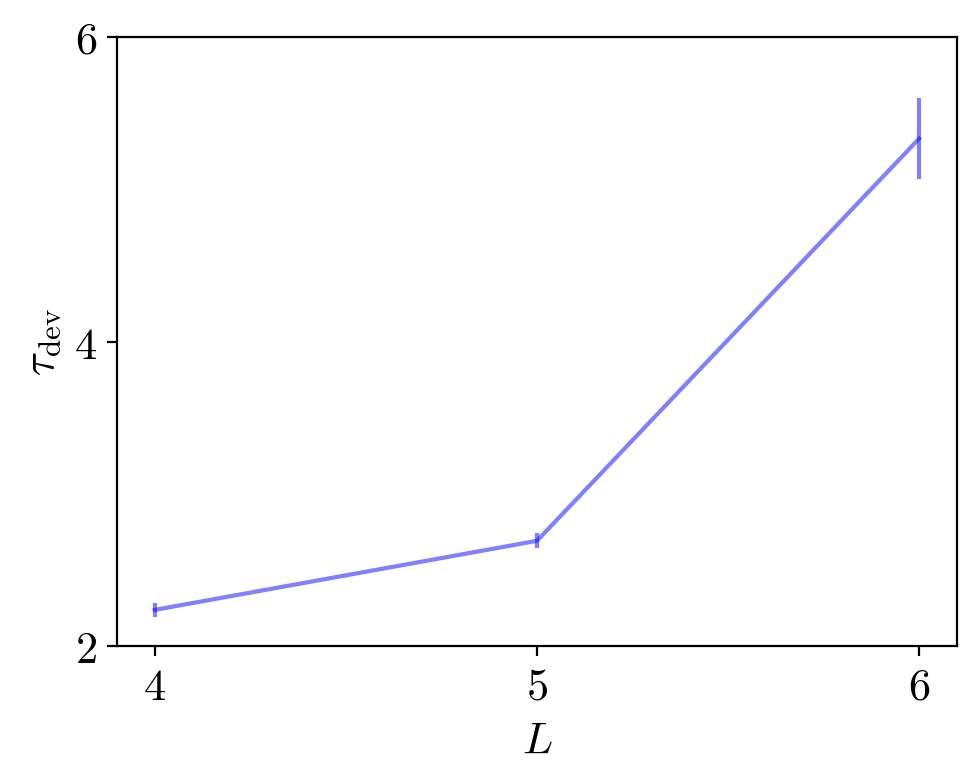}
    \end{minipage}
     \\
\hline
\end{tabular}
\end{center}


\section{Height of bumps}\label{app:peaks}
In this section, for chaotic OQMBS displaying GinUE DSFF behaviours, we determine the system size 
dependence of the height of the bump.
We observe that for non-many-body open quantum systems, e.g. RKO and 0D-RL, the height does not increase with system size. For many-body open quantum systems, e.g. RKC, $\mathbb{Z}_2$-RKC, $U(1)$-RKC, 1D-RL and L=SYK, the height increases.

\begin{figure}[H]
\begin{minipage}[t]{0.46\textwidth}
\includegraphics[width=\linewidth,keepaspectratio=true]{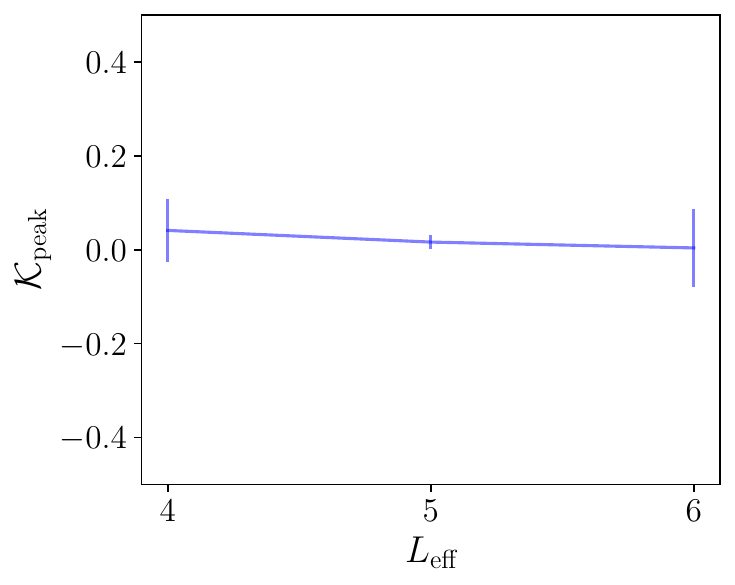}
\end{minipage}
\hspace*{\fill} 
\begin{minipage}[t]{0.46\textwidth}
\includegraphics[width=\linewidth,keepaspectratio=true]{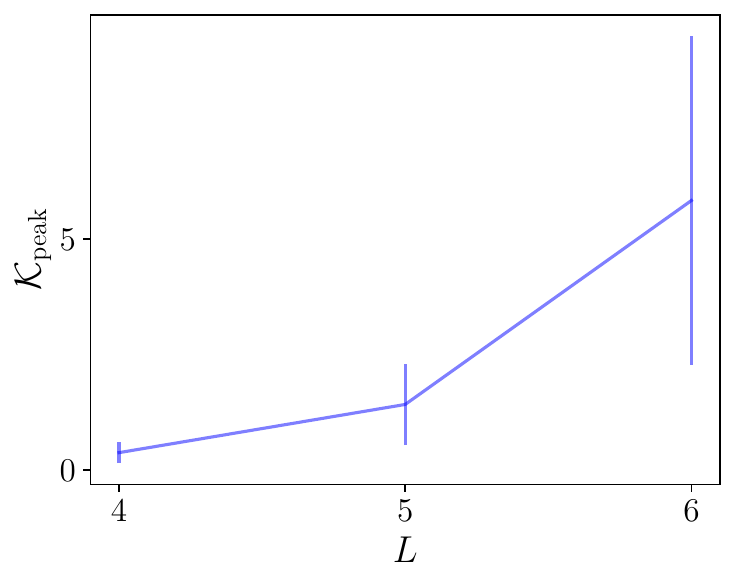}
\end{minipage}
    \caption{ 
    The peak value of the DSFF $\mathcal{K}_\mathrm{peak}$ against the effective system size $L_\mathrm{eff} = \frac{1}{2} \log N$ for RKO (left) and system size $L$ for RKC (right).  
    }
    \label{fig:ko_peak_height_error}
\end{figure}

\begin{figure}[H]
\begin{minipage}[t]{0.32\textwidth}
\includegraphics[width=\linewidth,keepaspectratio=true]{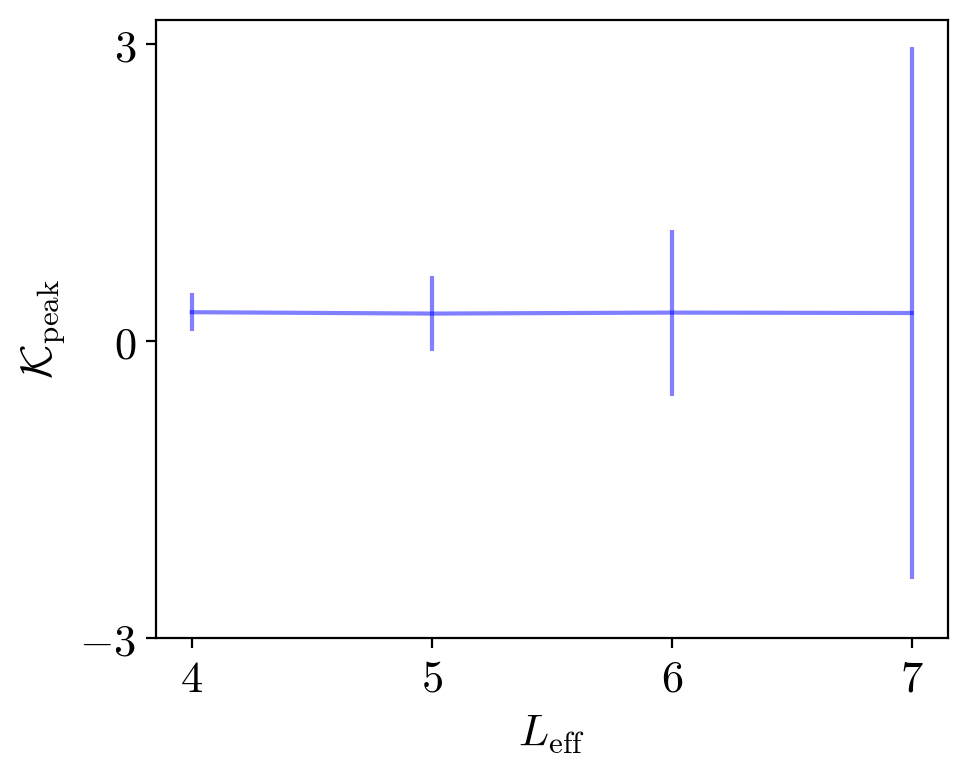}
\end{minipage}
\hspace*{\fill} 
\begin{minipage}[t]{0.32\textwidth}
\includegraphics[width=\linewidth,keepaspectratio=true]{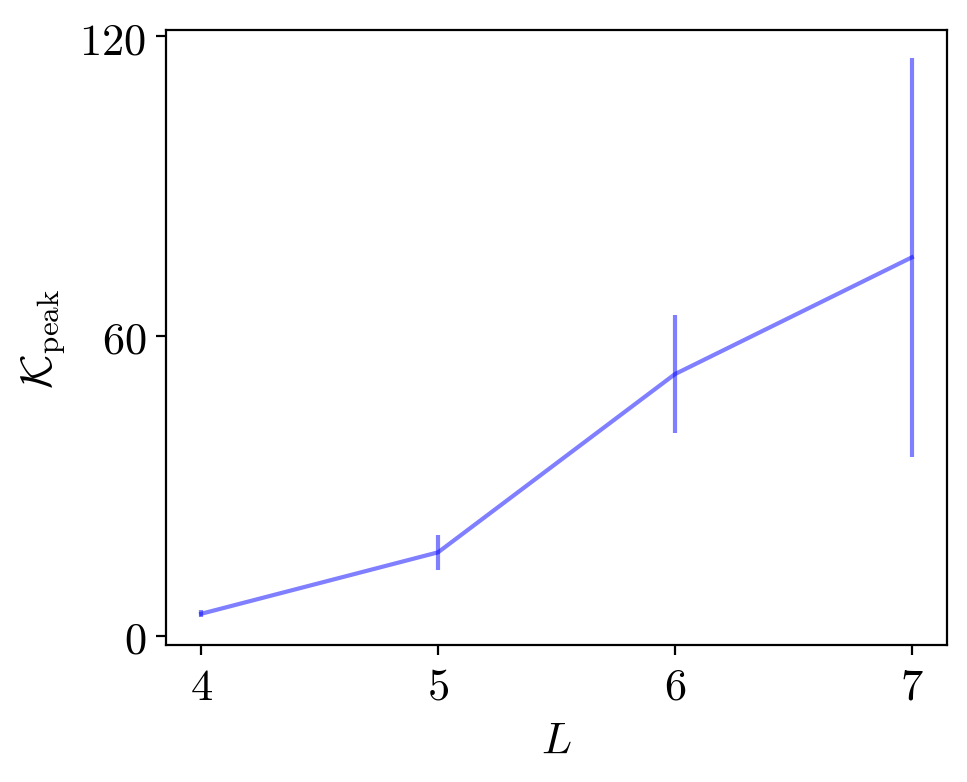}
\end{minipage}
\hspace*{\fill} 
\begin{minipage}[t]{0.32\textwidth}
\includegraphics[width=\linewidth,keepaspectratio=true]{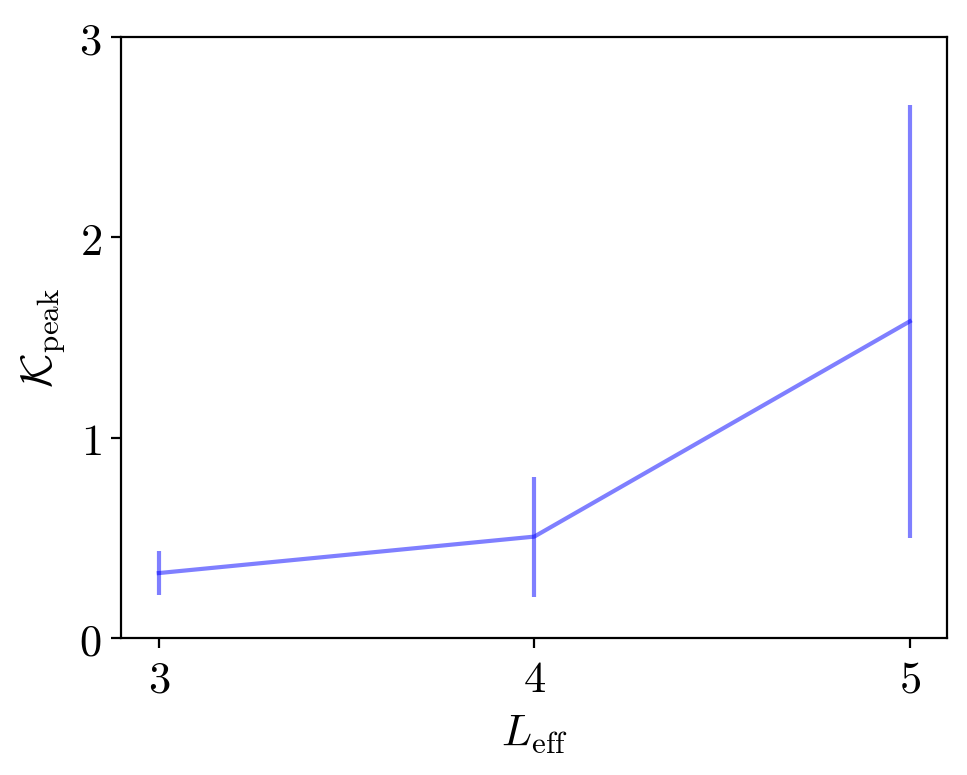}
\end{minipage}
    \caption{ 
    The peak value of the DSFF $\mathcal{K}_\mathrm{peak}$ against the effective system size $L_\mathrm{eff} = \frac{1}{2} \log N$ for 0D-RL (left) and SYK-L (right) and system size $L$ for 1D-RL (middle).  
    }
    \label{fig:rl_peak_height_error}
\end{figure}

Note that the ``bump'' should not to be confused with the ``dip'' in the literature, which refers to the decaying behaviour from $N^2$ to the ramp  due to the \textit{disconnected} part of DSFF~\cite{li2021spectral} or SFF~\cite{Gharibyan_2018}. 
Here we deal with the deviation from RMT behavior in the \textit{connected} part of DSFF.
%
Specifically, we track the dependence of the bump height 
of the DSFF,
normalized by its late-time 
plateau value, defined as follows,
\be\label{eq:tpeak}
\mathcal{K}_{\mathrm{peak}}:=
\frac{
\tilde{K}_\mathrm{c}(|\tau_\mathrm{peak}|,\theta)
}{
\tilde{K}_\mathrm{c}(|\tau|\to \infty,\theta)
} \;, 
\quad \quad 
\tau_{\mathrm{peak}} := 
\mathrm{arg\, max}_{|\tau|
}\left[
\frac{\tilde{K}_{\mathrm{c}}(|\tau|, \theta)}{  \tilde{K}_{\mathrm{c}}(|\tau| \to \infty, \theta)  }
- \frac{ \kcgin(|\tau|, \theta ) }{N}
\right]
\,.
\ee
We see a distinct difference 
between the models with no spatial structure 
(e.g. RKO) and those with spatial structure (e.g. RKC, $\mathbb{Z}_2$-RKC, $U(1)$-RKC),
and we illustrate this difference by comparing, in particular, the RKO and RKC models~\autoref{fig:ko_peak_height_error} and the 0D-RL, 1D-RL and SYK-L~\autoref{fig:rl_peak_height_error}.
Note that in the SYK model, even though the interaction is all-to-all, i.e. there is no spatial structure, the peak of the bump is increasing with system size. This data suggests that the deviation from RMT in DSFF is a generic many-body effect in open quantum many-body chaotic systems. 

%
%
%
%
%
%



\end{document}